\documentclass[default]{aastex701}
\usepackage{natbib}
\usepackage{color} 
\usepackage{ulem}
\usepackage{hyperref} 
\usepackage{amsmath}
\accepted{}
\submitjournal{ApJ}
\shorttitle{Galactic disk viewed by RC}
\shortauthors{Wu et al.}
\begin{document} 
\title{Mapping the Milky Way with Gaia Bp/Rp spectra-\uppercase\expandafter{\romannumeral4}: the broken and asymmetric density profile of the stellar disk traced by a large sample of red clumps}

\author{Wenbo Wu}
\affiliation{Institute of Astronomy and Physics, Inner Mongolia University, Hohhot 010021, People's Republic of China}
\affiliation{National Astronomical Observatories, Chinese Academy of Sciences, Beijing 100101, People's Republic of China}
\affiliation{Instituto de Astrofísica de Canarias, C/ Vía Láctea s/n, E-38205 La Laguna, Tenerife, Spain}
\affiliation{Universidad de La Laguna, Departamento de Astrofísica, 38206 La Laguna, Tenerife, Spain}
\email{wbwu@nao.cas.cn}

\correspondingauthor{Yuqin Chen}
\author{Yuqin Chen}
\affiliation{National Astronomical Observatories, Chinese Academy of Sciences, Beijing 100101, People's Republic of China}
\affiliation{School of Astronomy and Space Science, University of Chinese Academy of Sciences, Beijing 100049, People's Republic of China}
\email[show]{cyq@nao.cas.cn}

\author{Jianhui Lian}
\affiliation{South-Western Institute For Astronomy Research, Astronomy Building, Yunnan University, Chenggong District, Kunming, 650500, People’s Republic of China}
\affiliation{Key Laboratory of Survey Science of Yunnan Province, Yunnan University, Kunming, Yunnan 650500, People’s Republic of China}
\email{jianhui.lian@ynu.edu.cn}

\author{Martín López-Corredoira}
\affiliation{Instituto de Astrofísica de Canarias, C/ Vía Láctea s/n, E-38205 La Laguna, Tenerife, Spain}
\affiliation{Universidad de La Laguna, Departamento de Astrofísica, 38206 La Laguna, Tenerife, Spain}
\email{martin@lopez-corredoira.com}

\author{Chengdong Li}
\affiliation{School of Astronomy and Space Science, Nanjing University, Nanjing 210093, People’s Republic of China}
\affiliation{Key Laboratory of Modern Astronomy and Astrophysics (Nanjing University), Ministry of Education, Nanjing 210093, People’s Republic of China}
\email{chengdong.li@nju.edu.cn}

\author{Xianhao Ye}
\affiliation{National Astronomical Observatories, Chinese Academy of Sciences, Beijing 100101, People's Republic of China}
\affiliation{Instituto de Astrofísica de Canarias, C/ Vía Láctea s/n, E-38205 La Laguna, Tenerife, Spain}
\affiliation{Universidad de La Laguna, Departamento de Astrofísica, 38206 La Laguna, Tenerife, Spain}
\email{yexianhao@nao.cas.cn}

\correspondingauthor{C. Allende Prieto}
\author{C. Allende Prieto}
\affiliation{Instituto de Astrofísica de Canarias, C/ Vía Láctea s/n, E-38205 La Laguna, Tenerife, Spain}
\affiliation{Universidad de La Laguna, Departamento de Astrofísica, 38206 La Laguna, Tenerife, Spain}
\email[show]{carlos.allende.prieto@iac.es}

\author{Xiang-Xiang Xue}
\affiliation{National Astronomical Observatories, Chinese Academy of Sciences, Beijing 100101, People's Republic of China}
\affiliation{School of Astronomy and Space Science, University of Chinese Academy of Sciences, Beijing 100049, People's Republic of China}
\email{xuexx@nao.cas.cn}

\correspondingauthor{Gang Zhao}
\author{Gang Zhao}
\affiliation{National Astronomical Observatories, Chinese Academy of Sciences, Beijing 100101, People's Republic of China}
\affiliation{Institute of Astronomy and Physics, Inner Mongolia University, Hohhot 010021, People's Republic of China}
\affiliation{School of Astronomy and Space Science, University of Chinese Academy of Sciences, Beijing 100049, People's Republic of China}
\email[show]{gzhao@nao.cas.cn}

\author{Jingkun Zhao}
\affiliation{National Astronomical Observatories, Chinese Academy of Sciences, Beijing 100101, People's Republic of China}
\affiliation{School of Astronomy and Space Science, University of Chinese Academy of Sciences, Beijing 100049, People's Republic of China}
\email{zjk@bao.ac.cn}

\author{David S. Aguado}
\affiliation{Instituto de Astrofísica de Canarias, C/ Vía Láctea s/n, E-38205 La Laguna, Tenerife, Spain}
\affiliation{Universidad de La Laguna, Departamento de Astrofísica, 38206 La Laguna, Tenerife, Spain}
\email{david.aguado@iac.es}

\author{Jonay I. González Hernández}
\affiliation{Instituto de Astrofísica de Canarias, C/ Vía Láctea s/n, E-38205 La Laguna, Tenerife, Spain}
\affiliation{Universidad de La Laguna, Departamento de Astrofísica, 38206 La Laguna, Tenerife, Spain}
\email{jonay.gonzalez@iac.es}

\author{Rafael Rebolo}
\affiliation{Instituto de Astrofísica de Canarias, C/ Vía Láctea s/n, E-38205 La Laguna, Tenerife, Spain}
\affiliation{Universidad de La Laguna, Departamento de Astrofísica, 38206 La Laguna, Tenerife, Spain}
\email{rrl@iac.es}
\begin{abstract}
This study explores the density profile of the stellar disk, radially and azimuthally, based on approximately 8.4 million red clump stars selected from Gaia Bp/Rp spectra. After correcting for selection effects and distance uncertainties, we fit the vertical stellar density profile of the Galactic disk with a two-component model consisting of geometrically thin and thick disks. Our derived density profile shows several breaks radially: (1) a steep exponential inside R$\sim3$ kpc; (2) a nearly flat plateau from R$\sim3$ to $\sim7$ kpc; (3) an exponential decline beyond the solar radius to around 13 kpc; (4) a sharper exponential drop-off beyond R$\sim$13 kpc. The parameters of these four main components depend on $\phi$ to some extent. Variation of the termination radius of the first component suggests an interaction with the bar/bulge. Besides the typical flaring at $R>6.4$ kpc, we find that the thin disk also exhibits a similar and smooth thickening/flaring feature toward the Galactic center at $R<6.4$ kpc. The observed inner flaring may indicate heating effects introduced by the Galactic bar, since $R=6.4$ kpc lies close to the co-rotation radius where the bar's dynamical influence becomes significant. Additionally, we identify a localized density bump in the region $5<R<7$ kpc and $-30^\circ<\phi<15^\circ$, where a corresponding metallicity bump is also visible near the Galactic plane. This density/metallicity bump may be related to the recently reported bimodal distribution of the guiding radius of super metal-rich stars in the solar vicinity through radial migration.
\end{abstract}


\section{Introduction} \label{sec:intro}

Disk galaxies are fairly common in the local Universe, but their formation and evolution are still poorly understood. The disk-dominated Milky way (MW) provides a unique opportunity to explore the morphology of typical disk galaxies. Over the last few decades, the spatial distribution of stars in the MW has been well constrained thanks to the advent of large scale photometric surveys, including the Sloan Digital Sky Survey \citep[SDSS,][]{2000AJ....120.1579Y,2004AJ....128..502A,2011ApJS..193...29A}, the Two Micron All Sky Survey \citep[2MASS,][]{2006AJ....131.1163S}, and the Panoramic Survey Telescope and Rapid Response System survey \citep[Pan-STARRS,][]{2016arXiv161205560C}, and spectroscopic surveys including the Sloan Extension for Galactic Understanding and Exploration survey \citep[SEGUE,][]{2009AJ....137.4377Y}, the Large Sky Area Multi-Object Fiber Spectroscopic Telescope \citep[LAMOST,][]{2006ChJAA...6..265Z,2012RAA....12..723Z,2012RAA....12.1197C,2012RAA....12.1243L}, the GALactic Archaeology with HERMES survey \citep[GALAH,][]{2015MNRAS.449.2604D,2021MNRAS.506..150B}, the Apache Point Observatory Galactic Evolution Experiment survey \citep[APOGEE,][]{2017AJ....154...94M} and others. In addition, the data from the Gaia satellite provides proper motions and parallaxes for over 1.4 billion stars \citep{2016A&A...595A...2G,2018A&A...616A...1G,2023A&A...674A...1G}. These massive and accurate data sets allow us to map the MW from the Galactic center to the outer halo in great detail. 

The stellar disk, which comprises approximately three quarters of the Galactic stars, is a crucial component in studying the formation and evolution history of our galaxy. For disk galaxies, the functional form of their light and density profiles has been a long-discussed question since the mid-20th century. In the radial direction, the surface brightness profiles of external disk galaxies generally follow an exponential decline \citep{1970ApJ...160..811F,1981A&A....95..105V,2008MNRAS.387.1099P,2008AJ....135...20E,2011AJ....142..145G,2018ApJS..234...18B}, and the radial stellar density profile of our Milky Way's disk also agrees well with this form \citep{1996A&A...313L..21R,1999MNRAS.308..333H,2001ApJ...556..181D,2001MNRAS.322..426O}. In the vertical direction, early observations suggested that the stellar density profile of a disk galaxy is a squared hyperbolic secant (sech$^2$) function \citep{1981A&A....95..105V,1981A&A....95..116V}, consistent with the theoretical solution of an idealized, self-gravitating isothermal disk \citep{1942ApJ....95..329S,1987gady.book.....B}. However, the R-band studies of some external disk galaxies revealed an excess relative to the sech$^2$ distribution near the Galactic mid-plane, indicating that their vertical stellar density profiles are better described by an exponential form \citep{1994A&AS..103..475B}. For the Milky Way's disk, both exponential \citep{1989MNRAS.239..571K,1992ASPC...32..228K,1998A&A...331..934B,1999MNRAS.308..333H,2002A&A...394..883L,2016ApJ...823...30B} and sech$^2$ \citep{1983MNRAS.202.1025G,1995ApJ...446..646V,2006NewA...12..234B,2012ApJ...750L..41W,2024MNRAS.533L..31W} functions have been adopted to describe its vertical stellar density profile in previous studies. Several works argue that the exponential form provides a significantly better fit than the sech$^2$ function, especially towards the Galactic mid-plane \citep{1983AJ.....88.1476P,1986A&A...157..230V,1999MNRAS.308..333H,2008ApJ...673..864J,2017ApJ...843..141F,2020OJAp....3E...5D,2025ApJ...990L..37L,2026PhR..1163....1J}.   

The Galactic stellar disk is a complex system consisting of multiple components that exhibit different structural properties. Using star counts from photometric surveys, early studies showed that the Galactic stellar disk can be generally split into two components in the solar vicinity, of which one is a geometrically thin disk with a scale height $h_{\mathrm{Z}}\sim$0.3 kpc, and the other one is a thick disk with $h_{\mathrm{Z}}=0.7\sim0.9$ kpc \citep{1982PASJ...34..365Y,1983MNRAS.202.1025G,2001ApJ...553..184C,2008ApJ...673..864J}. The difference in $h_Z$ between these two disk components is explained as a result of the fact that high-$\mathrm{[\alpha/Fe]}$ and/or older stellar populations tend to have a thicker geometric morphology \citep{2016ApJ...823...30B,2016ARA&A..54..529B,2018ApJS..237...33X,2021ApJ...912..106Y,2022MNRAS.513.4130L,2025AJ....169...61Y}. Beyond the solar neighborhood, the scale height grows with increasing Galactocentric radius \citep[e.g.,][]{2014Natur.509..342F,2019ApJ...871..208L,2020A&A...637A..96C,2022A&A...664A..58C,2024MNRAS.527.4863U}. This phenomenon is usually called flaring. The connection between the flare and the stellar age has been studied using mono-abundance and/or mono-age populations. Using 16,944 red clumps selected from the APOGEE survey, \citet{2016ApJ...823...30B} only found a strong evidence of flaring in the low-$\mathrm{[\alpha/Fe]}$ populations. However, by analyzing the stellar sample of the astroNN Value Added Catalog of the APOGEE survey, \citet{2022MNRAS.513.4130L} shows that the high-$\mathrm{[\alpha/Fe]}$ population has the strongest flaring, while the young solar-abundance stellar sample presents the shortest scale height and least flaring. \cite{2021ApJ...912..106Y} find a significant flare both in the low-$\mathrm{[\alpha/Fe]}$ and high-$\mathrm{[\alpha/Fe]}$ populations of red clump stars selected from the LAMOST survey. A similar conclusion is reported by \citet{2025AJ....169...61Y}, who find that all mono-age stellar populations exhibit significant flaring, questioning the role of stellar ages played on the disk flaring. Radially, the scale length $h_R$ of the Galactic disk is largely dependent on the stellar samples, and a wide range of values from 1.6 to 6 kpc has been reported by previous studies \citep[see ][]{2016ARA&A..54..529B}.        

In the radial direction, recent studies find that the density profile of the stellar disk is likely to be more complicated than a single exponential of constant scale height \citep[e.g.,][]{2016ApJ...823...30B,2018MNRAS.478.3367W,2021ApJ...912..106Y,2022MNRAS.513.4130L,2024MNRAS.531.1730T,2024NatAs...8.1302L}. A break in the radial density profile is quite common in external disk galaxies \citep[e.g.,][]{1979A&AS...38...15V,2002A&A...392..807P,2006A&A...454..759P,2014ApJ...782...64K,2016A&A...596A..25L,2024A&A...682L..17X}. There are two main types of broken density profiles: the up-bending form that exhibits a gentler decline beyond the break radius $R_\mathrm{br}$, and the down-bending form which corresponds to a steeper decrease \citep{2006A&A...454..759P,2008AJ....135...20E}. The origin of a broken density profile in disk galaxies remains an unresolved issue, especially for the up-bending form. 

In \citet{2016ApJ...823...30B}, $R_\mathrm{br}$ of the low-$\mathrm{[\alpha/Fe]}$ stellar populations ranges from 7 to 10 kpc according to the their metallicity. Inside $R_\mathrm{br}$, the surface density profile stays almost constant or shows a modest decline towards the Galactic center. The break in the inner disk has been further confirmed by using different stellar samples selected from large spectroscopic surveys \citep{2021ApJ...912..106Y,2022MNRAS.513.4130L,2024A&A...683A.128C,2025AJ....169...61Y}. Notably, \citet{2024NatAs...8.1302L} constructs an age-resolved surface brightness profile of the stellar disk that shows a nearly flat plateau between 3.5 to 7.5 kpc, which results in a much larger estimate of the half-light radius of the Milky Way than in earlier works.    

The presence of a break radius farther out ($R>10\,\mathrm{kpc}$) plays a crucial role in determining the structural boundary of the Milky Way's disk. Using A type stars, \citet{1992ApJ...400L..25R} found a sharp cut-off in the stellar density at around 12 kpc. \citet{2011ApJ...733L..43M} showed a similar edge of the MW disk at around $R=13.9\pm0.5$ kpc by counting clump giants from the Vía Láctea (VVV) survey. However, this truncation is absent in the 3D density distribution of F8V-G5V stars selected from the SDSS survey as demonstrated by \citet{2014A&A...567A.106L}. Disk stars beyond 25 kpc from the Galactic center have been found by \citet{2018A&A...612L...8L}, and a smooth extension of the stellar disk is suggested to continue beyond 20 kpc \citep{2017RAA....17...96L,2020A&A...637A..96C}. Using Red giant branch stars (RGBs) selected from the LAMOST survey, \citet{2018MNRAS.478.3367W} found that the radial density profile of the stellar disk has a sharper decline at $R=11-14$ kpc with a scale length $h_{R}\sim1.18$ kpc, followed by a gentler exponential decline with $h_R\sim2.72$ kpc at $R=16-19$ kpc. In \citet{2024MNRAS.531.1730T}, the radial density profile traced by M giants from LAMOST DR5 presents two possible breakpoints at $R=14$ and 20 kpc. Their results support a gentler decline of the surface density beyond these breakpoints for the thin disk. However, in \citet{2024NatAs...8.1302L}, the surface brightness profile traced by APOGEE stars has a steeper decrease at $R>14$ kpc compared to the region at $R=7.5-14$ kpc. A possible complex radial density profile of the outer disk is suggested by these studies, but its shape remains a subject of debate and needs to be further confirmed by more investigations.

An additional complexity in the stellar density distribution of the Galactic disk is manifested through its dependence on azimuthal angle $\phi$. The warp is a well-known asymmetric feature in the outer stellar disk, which represents a large-scale distortion that warps upward in the north and downward in the south. The first detection of it was made through the 21-cm H-\uppercase\expandafter{\romannumeral1} in the gaseous disk \citep{1957AJ.....62...90B,1958MNRAS.118..379O}. Since then, its morphology and kinematics have been extensively studied \citep[e.g.,][]{1993AJ....105.2127C,1998AJ....115.2384D,2002A&A...394..883L,2009A&A...495..819R,2017A&A...602A..67A,2018MNRAS.478.3809S,2019ApJ...871..208L,2019NatAs...3..320C}. 

The formation mechanism of the galactic warp remains a topic of ongoing debate, with no consensus reached yet. Several hypotheses have been proposed to explain its origin, including the accretion of intergalactic matter \citep{2002A&A...386..169L}, the presence of a misaligned dark matter halo \citep{1989MNRAS.237..785O,2023ApJ...957L..24H,2023NatAs...7.1481H}, gravitational interactions with satellite galaxies \citep{2024MNRAS.533.2997W,2024MNRAS.535.1898B}, and the major merger event associated with the Gaia-Sausage-Enceladus \citep{2024ApJ...975...28D}. 

In the solar vicinity, recent studies suggest an asymmetric density profile of the stellar disk, which is quantified by variations of the scale length, scale height, and surface density change with $\phi$ \citep{2024MNRAS.531..495T,2024RAA....24f5005L}. However, their star samples are constrained to a relatively small local volume spanning $6<R<12$ kpc and $|\phi|<10^\circ$, which prevents them checking asymmetries across the entire Galactic disk. Recently, the residuals between an axisymmetric model and observed star counts in \cite{2025A&A...701A.270K} suggest the possible presence of a two-armed perturbation in the Galactic stellar disk. 

\textit{Gaia} Data Release 3 \citep[DR3;][]{2023A&A...674A...1G} has released around 220 million low-resolution spectra obtained from the blue ($330\leq\lambda\leq680\,\mathrm{nm}$) and red ($640\leq\lambda\leq1050\,\mathrm{nm}$) \textit{Gaia} slitless spectrophotometers \citep{2021A&A...652A..86C,2023A&A...674A...3M,2023A&A...674A...2D}. These spectra (hereafter referred to as `Bp/Rp spectra' or `XP spectra') have a variable resolving power ranging from 20 to 90 as a function of wavelength. The extensive data provide a unique opportunity to map the stellar spatial distribution of the Milky Way from its Galactic center to the distant outer halo \citep{2022ApJ...941...45R,2024ApJ...972..112C,2024arXiv241022250K,2025arXiv250114089C}. 

In Paper-\uppercase\expandafter{\romannumeral2} of this series \citep{2025A&A...700A.244W}, we explore the density shape of the stellar halo using Blue horizontal branch stars selected from Gaia Bp/Rp spectra. As an another type of standard candles, red clump (RC) stars are widely used to trace the spatial distribution of the stellar disk. The low resolution of the Gaia Bp/Rp spectra brings uncertainties in deriving the stellar atmospheric parameters, but a clear RC structure is still apparent in the $T_\mathrm{eff}-\log g$ diagram as shown by \citet{2023ApJS..267....8A} and \citet{2024ApJS..272....2L}. 

In this study, we aim to map the density profile of the Galactic disk and investigate its azimuthal dependence by RCs selected from Gaia XP spectra. The paper is organized as follows. In Section~\ref{sec:identification} we introduce the selection methodology and the distance derivation for the RC sample. The reconstruction of the selection function is illustrated in Section~\ref{sec:selectfunction}. We describe the derivation and the modeling of the spatial distribution of the Galactic stellar disk in Section~\ref{sec:densityfitting}. The broken and asymmetric density profile of the disk is presented in Section~\ref{sec:results}. We discuss the possible relationship between the density bump and the recent found bimodal distribution in the guiding radius of super metal-rich stars in Section~\ref{sec:discussion}. Finally, a summary is given in Section~\ref{sec:summary}.

\section{Data}\label{sec:identification}
In this study, we selected RCs from Gaia XP spectra based on their distributions in $T_\mathrm{eff}-\log g-\mathrm{[M/H]}$ diagram. To reduce the effect of Galactic dust reddening, we cross-matched the RC catalog with the 2MASS survey, and derived the photometric distance $d_\mathrm{rc}$ based on the dereddened near-infrared $K_s$ band magnitude. To reduce possible contaminants, we compared the derived $d_\mathrm{rc}$ with the photometric distance of StarHorse Catalog v.1.1 \citep{2022A&A...658A..91A}, and retained only the RC candidates that have a consistent distance estimates between us and \citet{2022A&A...658A..91A}. 

\subsection{Identification of the RCs}\label{sub:identification}
The most accurate way of separating red giant branch stars (RGB) and RCs is through the asteroseismic data, since these two types of stars are located in different regions of the $\Delta\pi_{1}$ (asymptotic period spacing) versus $\Delta\nu$ (mean frequency difference) diagram \citep{2013ApJ...765L..41S,2016A&A...588A..87V}. However, the small number of stars with high-precision light curves from the CoRoT \citep{2009A&A...506..411A} and Kepler \citep{2010PASP..122..131G} space mission satellites prevents us from obtaining a full-sky coverage and a large RC catalog. Although less accurate than the asteroseismic data analysis, previous studies have shown that it is still possible to achieve an accurate selection of RCs from large spectroscopic surveys in $T_\mathrm{eff}-\log g-\mathrm{[M/H]}$ diagram \citep{2014ApJ...790..127B,2020ApJS..249...29H}. Inspired by these results, it is worthwhile to make an attempt on constructing a RC catalog from the low-resolution XP spectra survey considering its large volume and full-sky coverage.         

The identification of RCs requires an accurate estimation of the stellar atmospheric parameters, which is challenging to do based on Gaia XP spectra, considering the low resolution ($\mathrm{R}\sim60$) and the presence of systematic errors in the data. Previous studies have adopted different data-driven and model-driven methods to solve this problem, including machine learning \citep{2023MNRAS.524.1855Z,2023ApJS..267....8A,2024MNRAS.531.2126F,2024A&A...691A..98K,2024ApJS..272....2L}, traditional spectral matching \citep{2024ApJS..272...20A,2025A&A...695A..75Y}, and synthetic photometry \citep{2023A&A...674A.194B,2024A&A...692A.115M}. The first publication of this series fits the flux-corrected XP spectra with a theoretical spectral model and estimates the stellar atmospheric parameters for around 68 million stars \citep{2025A&A...695A..75Y}. However, the derived $\log g$ is not precise enough for us to exclude contaminants from RGBs with low surface gravity. Therefore, we explored other available catalogs of Gaia XP spectra and finally decided to adopt the parameters from \cite{2023ApJS..267....8A} (\texttt{An23}). The \texttt{An23} catalog contains the vast majority of sources in the XP spectra and achieves a precision of about 0.08 dex in estimating $\log g$. This precision is substantially better than our own results in paper-\uppercase\expandafter{\romannumeral1}, with a precision of about 0.4 dex, thanks to the consideration of the Gaia parallaxes and the stellar luminosity in the analysis.  

\begin{figure}
	\centering
	\includegraphics[width = 8.8cm]{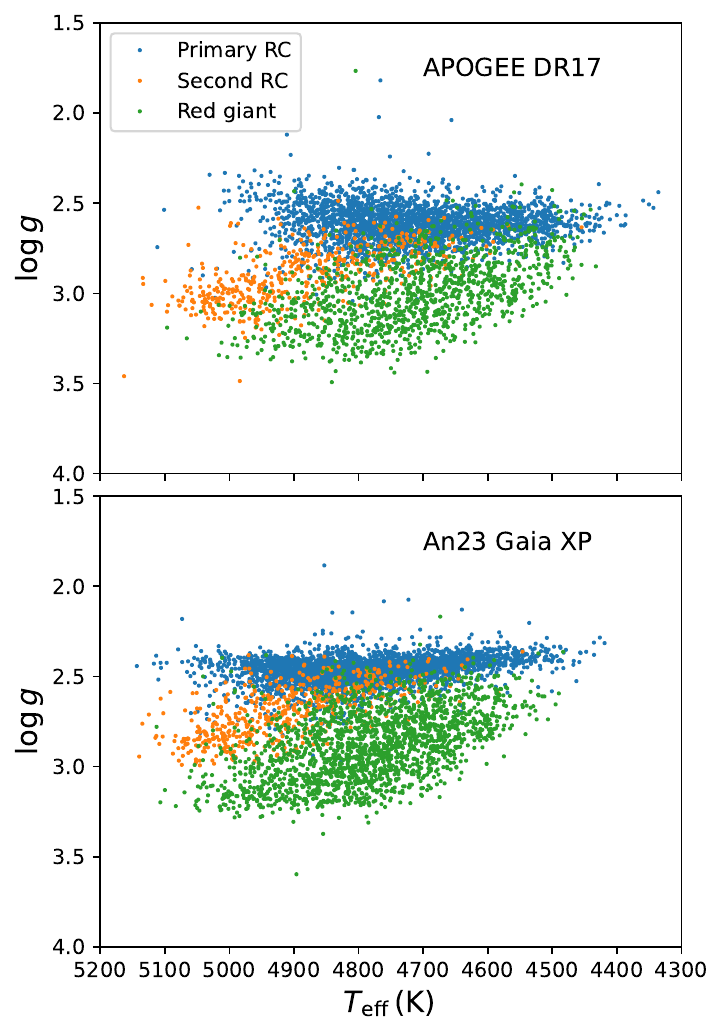}
	\caption{Distributions of the primary RC (blue points), secondary RC (orange points), and RGB (green points) identified by \cite{2016A&A...588A..87V} in the $T_\mathrm{eff}-\log g$ diagram. The stellar atmospheric parameters are provided by the APOGEE DR17 (top panel) and the \texttt{An23} Gaia XP catalog (bottom panel).}
	\label{fig:apogeeandxp}
\end{figure}

\cite{2016A&A...588A..87V} provide an estimation of $\Delta\pi$ and $\Delta\nu$ for around 6,100 Kepler stars by applying a Fourier analysis to their light curves. They classify 6,100 Kepler stars into three groups of primary RCs, second RCs, and RGBs in the $\Delta\pi$ versus $\Delta\nu$ diagram. Among them, 4,296 stars are available in the APOGEE DR17 catalog and 5,943 stars have Gaia XP spectra. In Figure~\ref{fig:apogeeandxp}, we can see a clear separation of these three groups in the $T_\mathrm{eff}-\log g$ diagram both for the APOGEE parameters and for those derived from Gaia XP spectra. \citet{2014ApJ...790..127B} provided an empirical $T_\mathrm{eff}-\log g-\mathrm{[M/H]}$ relationship to select RCs from the APOGEE survey using the asteroseismic identified RCs as a reference. Although the \texttt{An23} catalog is trained on stellar parameters from the APOGEE survey, the existence of a small offset between these two catalogs is apparent, as we can see in Figure~\ref{fig:apogeeandxp} that the distribution of $\log g$ for the Primary RCs is slightly narrower in the \texttt{An23} catalog. 

A previously identified RC catalog is needed to derive an empirical relationship to select RCs from Gaia XP spectra. We chose a machine-learning constructed RGB/RC catalog of \cite{2018ApJ...858L...7T} (\texttt{Ting18}) as reference. \texttt{Ting18} selected 210,371 RCs from LAMOST and APOGEE surveys with a contamination rate of 9\%. They also provided a pristine subsample of 92,249 RCs with 3\% contamination by requiring a spectra signal-to-noise ratio S/N $>$ 75. There are two reasons for our choice: (1) the Ting18 sample has a more completed coverage in [M/H] than the asteroseismic catalog, especially in the relatively metal-poor end of [M/H]$<-0.5$; (2) it is hard to separate RCs from RGBs that have a similar $\log g$ by the empirical relationship, while the machine-learning method can cleanly differentiate them by some additional information hidden in the single-epoch spectra such as C/N ratios. 

\begin{figure*}
	\centering
	\includegraphics[width =0.8\textwidth]{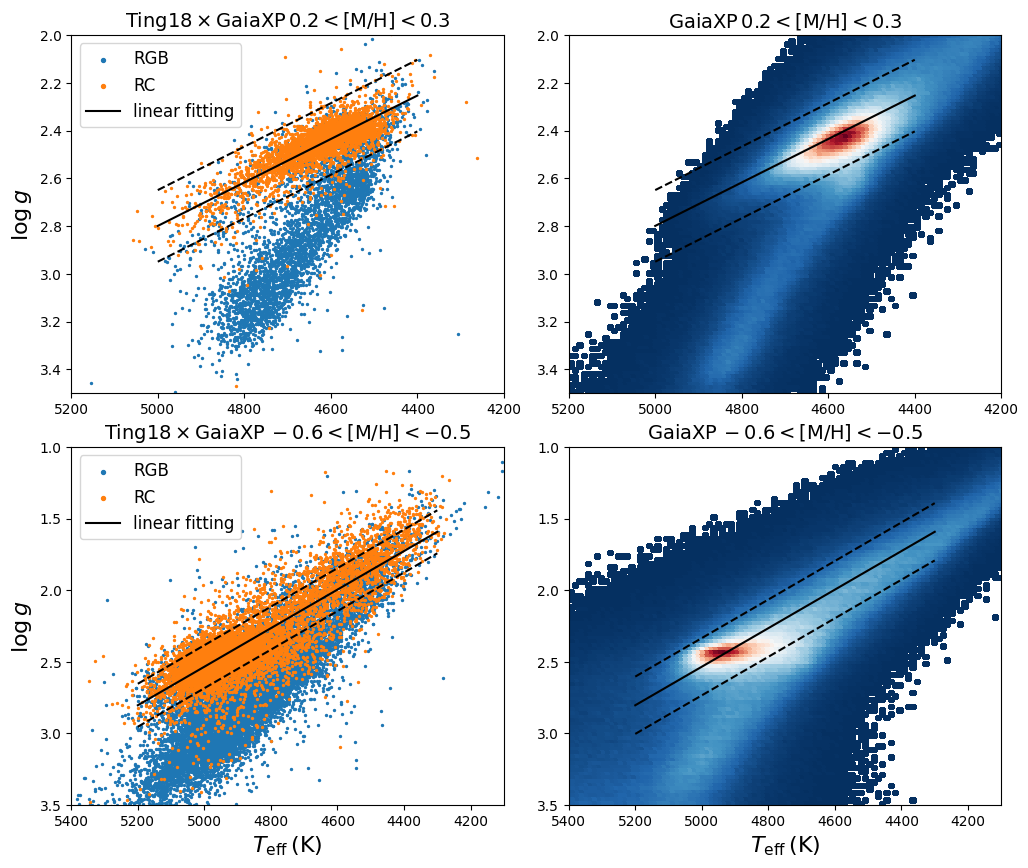}
	\caption{Selection of RCs in two specific metallicity bins. Top left panel shows the distributions of RGBs (blue points) and RCs (orange points) identified by \citet{2018ApJ...858L...7T} in a bin of $0.2<\mathrm{[M/H]}<0.3$. A linear fit is applied to derive the $T_\mathrm{eff}$–$\log g$ relation for these RC stars. Two dashed lines, offset by $\pm0.15$ dex in $\log g$ from the fitting line (shown in black), are used to include the majority of RCs within the temperature range $4400<T_\mathrm{eff}<5000$ K. We applied the two dashed lines to stars satisfying $0.2<\mathrm{[Fe/H]}<0.3$ in the \texttt{An23} catalog to make a rough selection of the RCs, and their density distributions in the $T_\mathrm{eff}-\log g$ diagram is exhibited in the top right panel. The bottom panel shows the results of another metallicity bin of $\mathrm{-0.6<\mathrm{[Fe/H]}<-0.5.}$ All the stellar atmospheric parameters are provided by the \texttt{An23} catalog.}
	\label{fig:comparetwobins}  
\end{figure*}

\begin{figure}
	\centering
	\includegraphics[width =8.8cm]{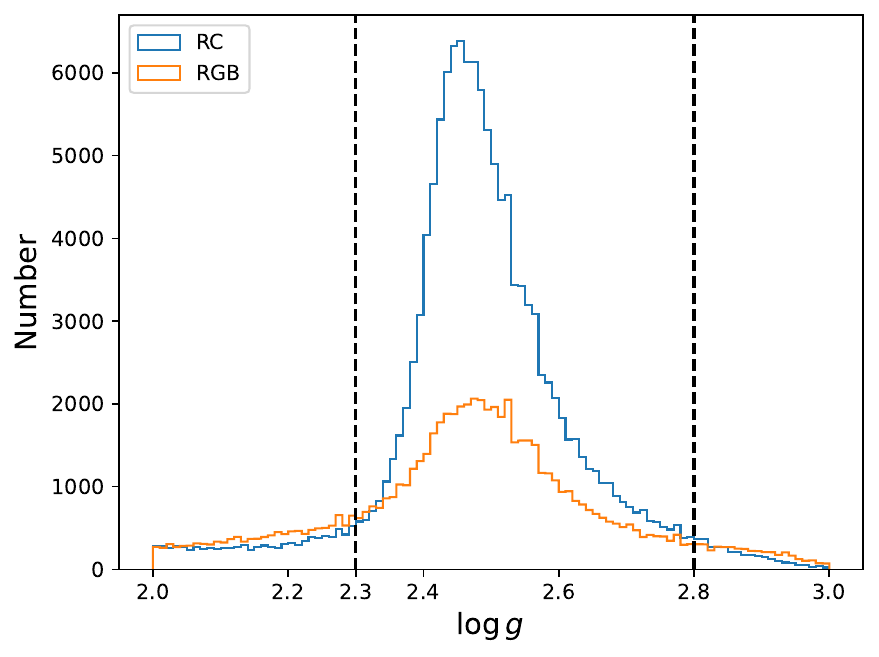}
	\caption{Density distributions of $\log g$ for the non-duplicated sources satisfying the $T_\mathrm{eff}-\log g$ selection criteria. According to the Ting18 catalog, some of them are identified as RCs (blue hist), while others are possibly RGBs (orange hist). We only retain the region where RCs dominate by requiring $2.3<\log g<2.8$.}
	\label{fig:loggRC}  
\end{figure}

A cross-match between the \texttt{Ting18} and the \texttt{An23} catalog returns us 353,970 non-duplicated sources, among them 155,522 are identified as RCs (128,007 from the LAMOST survey and 52,464 from the APOGEE survey, with 5,259 duplicated sources removed). We divided these previously identified RCs into 13 metallicity bins with edges of [M/H]= -1.0, -0.8, -0.7, -0.6, -0.5, -0.4, -0.3, -0.2, -0.1, 0, 0.1, 0.2, 0.3, and 0.5. For each bin, we fit the $T_\mathrm{eff}-\log g$ relationship of these identified RCs with a linear line, and apply it to the whole \texttt{An23} catalog to make a rough selection of RCs as Equation~\ref{eq:selection}: 
\begin{equation}
	\begin{aligned}
		&|\log g-(a\times10^{-3}T_\mathrm{eff}+b)|<\delta\log g\\
		&T_\mathrm{eff;min}<T_\mathrm{eff}<T_\mathrm{eff;max},
		\label{eq:selection}
	\end{aligned}
\end{equation} 
where $a$ and $b$ are the derived parameters of the $T_\mathrm{eff}-\log g$ relationship. $\delta\log g, T_\mathrm{eff;min}$, and $T_\mathrm{eff;max}$ are used to include the region that RC population dominates. The specific values of these parameters for all bins are given in Table~\ref{tab:rcselection}. The selection of RCs for two specific metallicity bins is provided in Figure~\ref{fig:comparetwobins}. Figure~\ref{fig:loggRC} shows the distributions of $\log g$ for the mentioned non-duplicated sources that pass our selection criteria. To improve the purity, an additional cut of $2.3<\log g<2.8$ is applied to the \texttt{An23} catalog and this returns a sample of around 10 million RC candidates.           

\begin{table}
	\centering
	\caption{Parameters of selecting RCs based on the $T_\mathrm{eff}-\log g-\mathrm{[M/H]}$ relationship.}
	\tabcolsep=0.11cm
	\begin{tabular}{ccccccc}
		\hline
		$\mathrm{[M/H]_{min}}$&$\mathrm{[M/H]_{max}}$&a&b& $\delta\log g$&$T_\mathrm{eff;min}$&$T_\mathrm{eff;max}$ \\
		\hline
		-1.00 & -0.80 & 1.31 & -4.21 & 0.20 & 4300 & 5200 \\
		-0.80 & -0.70 & 1.23 & -3.77 & 0.20 & 4300 & 5200 \\
		-0.70 & -0.60 & 1.27 & -3.91 & 0.20 & 4300 & 5200 \\
		-0.60 & -0.50 & 1.28 & -3.89 & 0.20 & 4300 & 5200 \\
		-0.50 & -0.40 & 1.28 & -3.83 & 0.20 & 4300 & 5200 \\
		-0.40 & -0.30 & 1.24 & -3.59 & 0.20 & 4300 & 5200 \\
		-0.30 & -0.20 & 1.19 & -3.31 & 0.20 & 4300 & 5200 \\
		-0.20 & -0.10 & 1.15 & -3.03 & 0.20 & 4300 & 5200 \\
		-0.10 &  0.00 & 1.07 & -2.59 & 0.20 & 4300 & 5200 \\
		0.00 &  0.10 & 1.03 & -2.35 & 0.20 & 4400 & 5200 \\
		0.10 &  0.20 & 1.07 & -2.52 & 0.15 & 4400 & 5000 \\
		0.20 &  0.30 & 1.09 & -2.57 & 0.15 & 4400 & 5000 \\
		0.30 &  0.50 & 1.25 & -3.23 & 0.10 & 4400 & 4800 \\
		\hline
	\end{tabular}
	\label{tab:rcselection}
\end{table}

\subsection{Distance estimation of the RCs}\label{sub:distance}
The heliocentric distance $d_\mathrm{rc}$ (in the unit of pc) can be obtained from Equation~\ref{eq:distance},
\begin{equation}
	d_\mathrm{rc}=10^{0.2(m_\lambda-A_\lambda-M_{\lambda}+5)},
	\label{eq:distance}
\end{equation}
where $m_\lambda$ is the apparent magnitude, $A_\lambda$ is the interstellar extinction, and $M_\lambda$ is the absolute magnitude. RCs are known as a type of standard candle that has an almost constant absolute magnitude due to their similar helium-burning core mass \citep{2016ARA&A..54...95G}. Although many previous studies have made great efforts to construct a complete and accurate Galactic dust map, there are still some uncertainties in the extinction especially towards low latitude regions. To reduce the influence of interstellar extinction, we will derive $d_\mathrm{rc}$ through the $K_s$ band, which is fairly insensitive to the ISM absorption. We cross-matched our RC sample with the 2MASS survey by the \texttt{gaiadr3.tmass\_psc\_xsc\_best\_neighbour} catalog \citep{2017A&A...607A.105M,2019A&A...621A.144M} and obtained the ${K_s}$ band magnitude for these RC candidates. 

Two types of the dust map are frequently used in the astronomical researches: (1) a two-dimensional (2D) map where the extinction only depends on the sky positions; (2) a three-dimensional (3D) map that both the sky positions and the distances matter in deciding the extinction. A fraction of our obtained RCs are in the solar neighborhood, and a two-dimensional map might overestimate the dust extinction for these stars \citep{2003A&A...409..205D}. Therefore, we chose the dust map \texttt{mwdust.Combined19} from the Python package \texttt{mwdust} \citep{2016ApJ...818..130B}, which is a combination of various 3D maps including \cite{2003A&A...409..205D}, \cite{2006A&A...453..635M}, and \cite{2019ApJ...887...93G}. The usage of \texttt{mwdust.Combined19} requires a prior estimation of the heliocentric distance. At first we try the value derived from the Gaia DR3 parallax, but it is not accurate for relatively distant stars, and there are a small fraction of stars having a negative parallax. Therefore, we turned to the photo-geometric distance dist50 provided by the \texttt{gaiaedr3\_contrib.starhorse}\footnote[1]{through an ADQL query interface of \href{https://gaia.aip.de}{https://gaia.aip.de}} catalog of \citet{2022A&A...658A..91A}, and used it to obtain the extinction $A_{K_s}$ from the \texttt{mwdust.Combined19} map. Our private check finds many RC candidates in the sky regions of the Magellanic clouds. \texttt{mwdust.Combined19} is only a description of the Galactic dust map, which might not be suitable for stars in the Magellanic clouds. Therefore, following \citet{2021A&A...654A.107C}, we remove stars in the region of the Magellanic clouds where the local apparent stellar population density $>50,000$ per $\deg^2$: (1) the Large Magellanic Cloud $274.5^\circ<l<286.5^\circ$ and $-37.9^\circ<b<-27.9^\circ$. (2) the Small Magellanic Cloud $299.8^\circ<l<305.8^\circ$ and $-46.3^\circ<b<-42.3^\circ$.  

Previous studies investigate the absolute magnitude of RCs in different passbands, and their results point to a mean value of $ M_{K_s}=-1.60$ for the ${K_s}$ band \citep[e.g.,][]{2000ApJ...539..732A,2007A&A...463..559V,2017MNRAS.471..722H,2018A&A...609A.116R,2020ApJ...893..108P}. Several recent studies suggest that $M_\mathrm{K_s}$ actually varies with multiple factors such as ages, metallicity, and stellar colors \citep{2017ApJ...840...77C,2019ApJ...872...95M,2019MNRAS.486.5600O,2021ApJ...923..145W}. To trace the slight change of $M_{K_s}$ with [M/H], we derived $M_{K_s}$ for RC candidates that have an accurate estimation of the parallax (parallax\_over\_error$>$5) after compensating for a zero-point offset of -0.021 mas provided by \citet{2021A&A...654A..20G}. To reduce the influence of the ISM extinction, we only kept stars with a high Galactic latitude ($|b|>20^\circ$) and low $K_s$ band extinction ($A_{K_s}<0.03$). As mentioned before, a fraction of these selected RC candidates may actually belong to the RGB population. Therefore, we will only use the candidates that have already been previously identified as RCs or RGBs in the \texttt{Ting18} catalog. 

\begin{figure}
	\centering
	\includegraphics[width = 8.8cm]{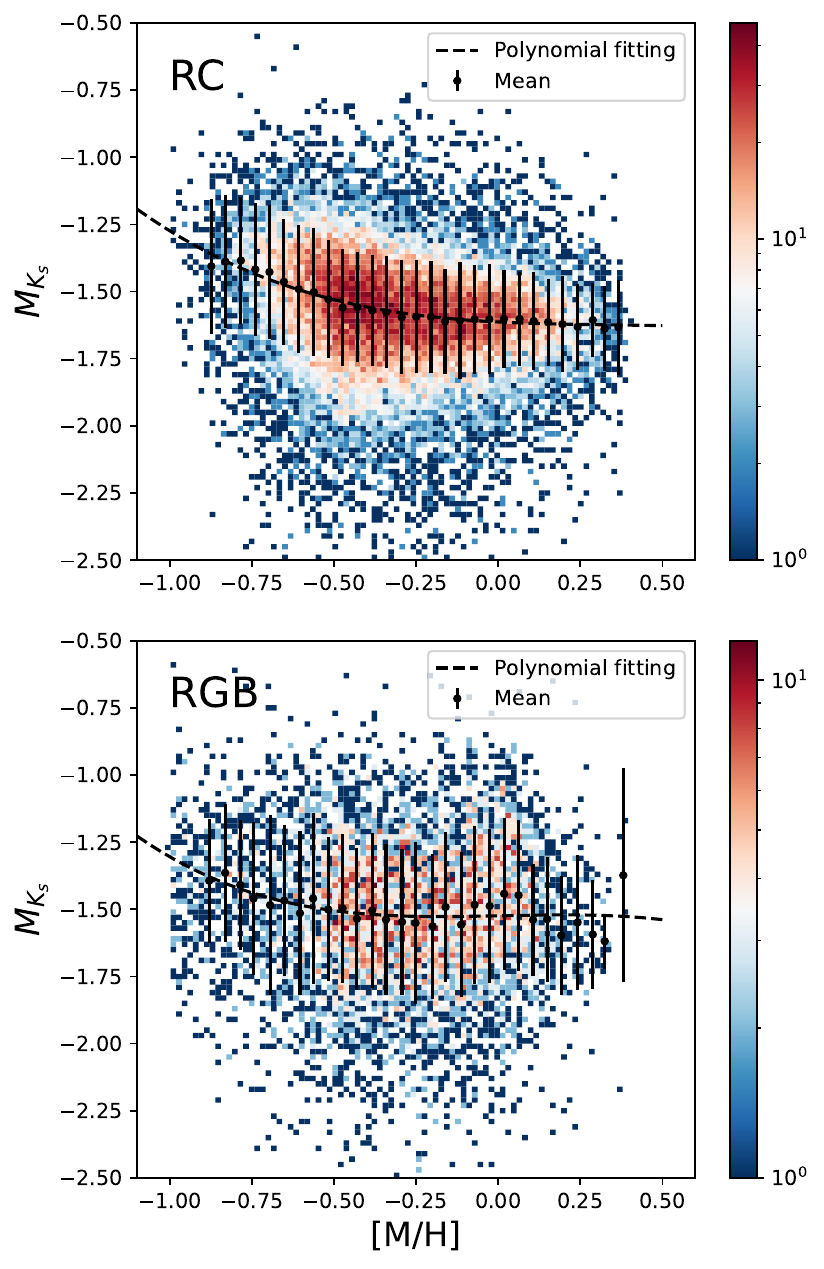}
	\caption{Density distribution of the RCs (top panel) and RGBs (bottom panel) on a logarithmic scale in [M/H]-$M_\mathrm{K_s}$ diagram. The black point represents the mean value, and the dashed black line is a cubic polynomial that describes $M_{K_s}$ as a function of [M/H].}
	\label{fig:MKsandfeh}
\end{figure}

Figure~\ref{fig:MKsandfeh} shows the density distribution of RCs and RGBs in the reddening corrected $\mathrm{[M/H]}-M_{K_s}$ diagram. For RCs, $M_{K_s}$ declines slightly with [M/H], while the RGBs have a broader distribution of $M_{K_s}$ that shows a weaker dependence on [M/H]. A similar declining tendency in RCs has already been theoretically proved \citep{2002MNRAS.337..332S} and empirically observed \citep{2022MNRAS.512.1710H,2025AJ....169...61Y} before. $M_{K_s}$ of the previously identified RCs does not have a substantial difference from that of the RGBs, since these selected RGBs occupy the same part of the Hertzsprung-Russell (HR) diagram as the RCs. We divided these stars into different metallicity bins with a step size of 0.05 dex, and obtain the mean values and standard deviations of $M_{K_s}$ in every [M/H] bins. The standard deviations of RCs differ slightly but all are around 0.2, which is a typical value of the intrinsic dispersion of $M_{K_s}$ according to previous studies \citep[e.g.,]{2017MNRAS.471..722H,2018A&A...609A.116R,2020ApJ...893..108P,2025A&A...701A.270K}. We fit the relationship between $M_{K_s}$ and [M/H] for RCs with a cubic polynomial as follows
\begin{equation}
	M_{K_s} = -0.137\mathrm{[M/H]^3}+0.139\mathrm{[M/H]^2}+0.062\mathrm{[M/H]}-1.614.\label{eq:MksFeh}
\end{equation}

For each RC, its distance module is described by a one-dimensional Normal distribution as Equation~\ref{eq:distancemodule},
\begin{equation}
	m_\mathrm{K_s}-M_\mathrm{K_s}\sim\mathcal{N}(\langle m_\mathrm{K_s}\rangle-\langle M_\mathrm{K_s}\rangle,{{\sigma_{m_\mathrm{K_s}}}^2+{\sigma_{M_\mathrm{K_s}}}^2})\,
	\label{eq:distancemodule}
\end{equation}
where $\langle m_{K_s}\rangle$ is the apparent ${K_s}$ band magnitude, $\langle M_{K_s}\rangle$ is obtained from Equation~\ref{eq:distancemodule} based on the stellar metallicity, $\sigma_{m_{K_s}}$ is the measurement uncertainty in the apparent magnitude that provided by the 2MASS catalog, and $\sigma_{M_{K_s}}=0.2$ is the intrinsic dispersion of $M_{K_s}$. By substituting Equation~\ref{eq:distancemodule} to ~\ref{eq:distance}, we can obtain the 16th, 50th ,and 84th percentile distributions of the photometric distance. Hereafter, $d_\mathrm{rc}$ refers to the 50th percentile distribution of the photometric distance.

\begin{figure}
	\centering
	\includegraphics[width = 8.8cm]{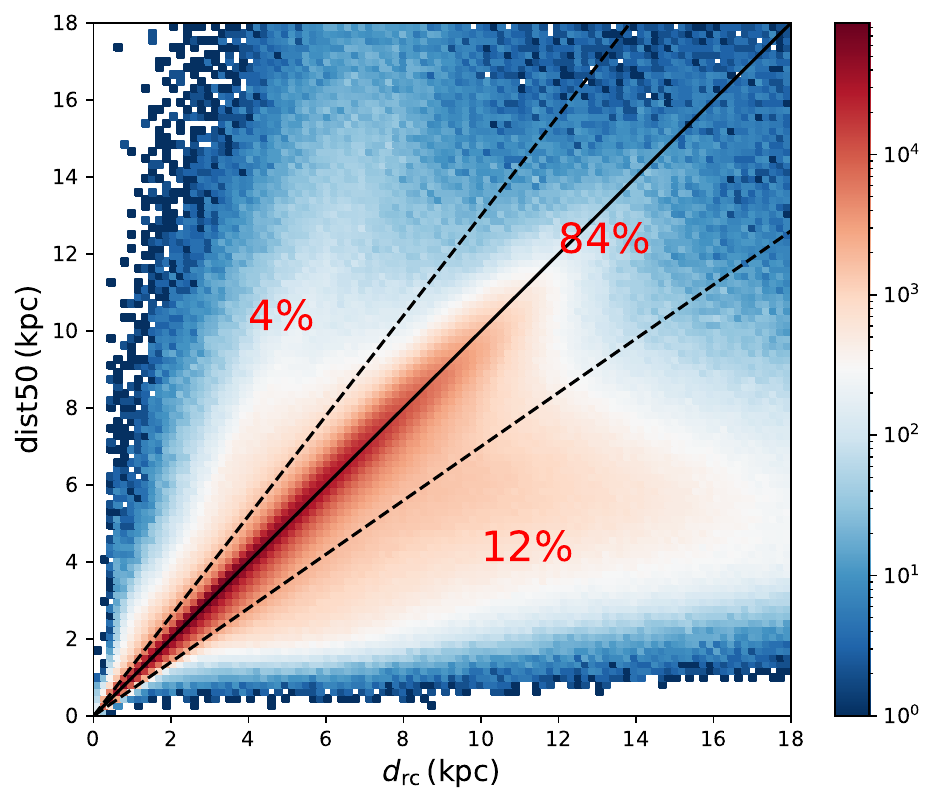}
	\caption{Comparison between our calculated heliocentric distance ($d_\mathrm{rc}$) and the dist50 from \citet{2022A&A...658A..91A} for the selected RC sample. The two dashed lines indicates the boundaries of a 30\% difference between the two types of distance ($\frac{|d_\mathrm{rc}-\mathrm{dist50}|}{d_\mathrm{rc}}=0.3$). We only kept stars where the two distances agree, referring to the 84\% sources located between the two boundaries.}
	\label{fig:comparedistance}
\end{figure}

As shown in Figure~\ref{fig:comparedistance}, our obtained $d_\mathrm{rc}$ is consistent with the StarHorse distance dist50 for most stars, and a small portion ($16\%$) of them shows a relatively larger dispersion with $|d_\mathrm{rc}-\mathrm{dist50}|/d_\mathrm{rc}>0.3$. It is hard to distinguish which distance performs better for these outliers, but we think that the inconsistency may indicate a possible larger contaminant rate. To ensure the reliability, we require that the two types of distances should agree within 30\% ($|d_\mathrm{rc}-\mathrm{dist50}|/d_\mathrm{rc}<0.3$), which finally results in a catalog containing 8.4 million RCs.  

\subsection{Spatial distribution of the RC sample}
We transform the sky position and the heliocentric distance into Cartesian ($X,Y,Z$) and Cylindrical ($R,Z,\phi$) Galactocentric coordinates by the conventions in \texttt{astropy} \citep{2018AJ....156..123A}. We use the default values of the Solar Galactocentric distance $r_{\mathrm{gc},\odot}$ = 8.122 kpc \citep{2018A&A...615L..15G} and height $Z_{\odot}$ = 20.8 pc \citep{2019MNRAS.482.1417B}. 

\begin{figure}
	\centering
	\includegraphics[width = 8.8cm]{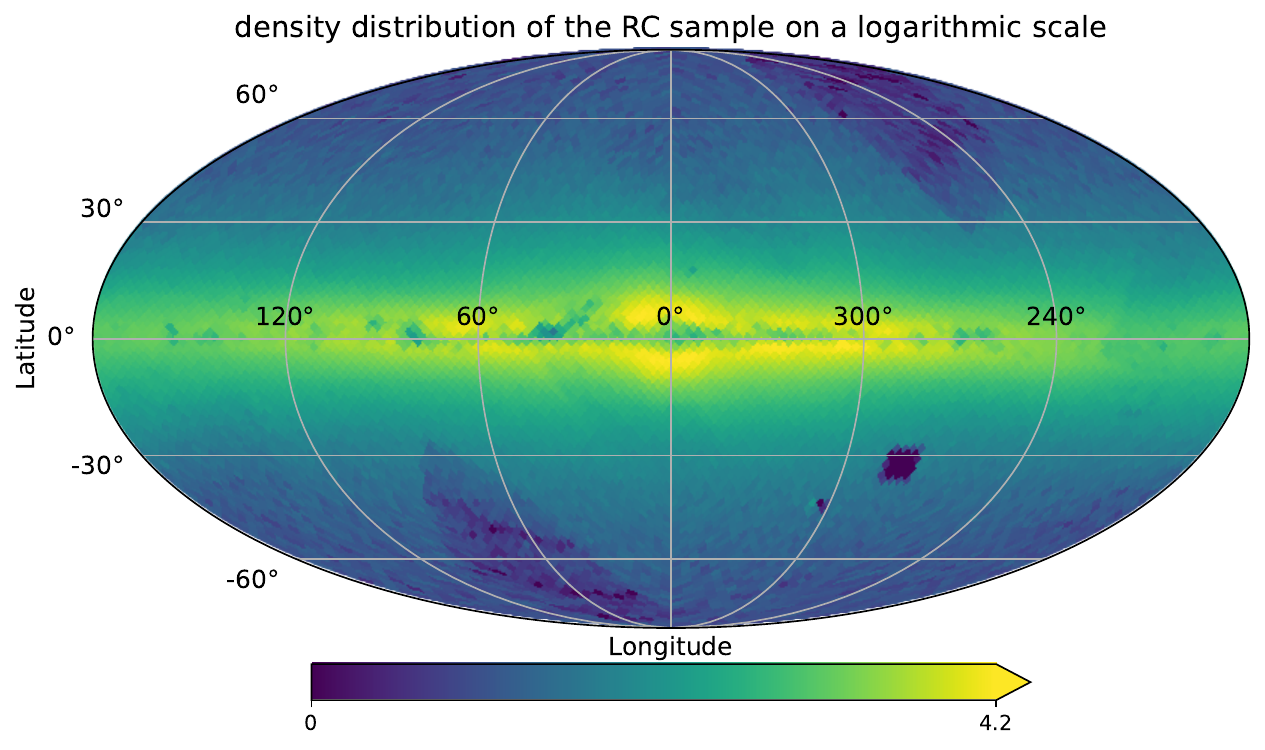}
	\caption{Density distribution of the RC sample in the Sky positions $(l,b)$ on a logarithmic scale with base 10.}
	\label{fig:lb}
\end{figure}

\begin{figure}
	\centering
	\includegraphics[width = 8.8cm]{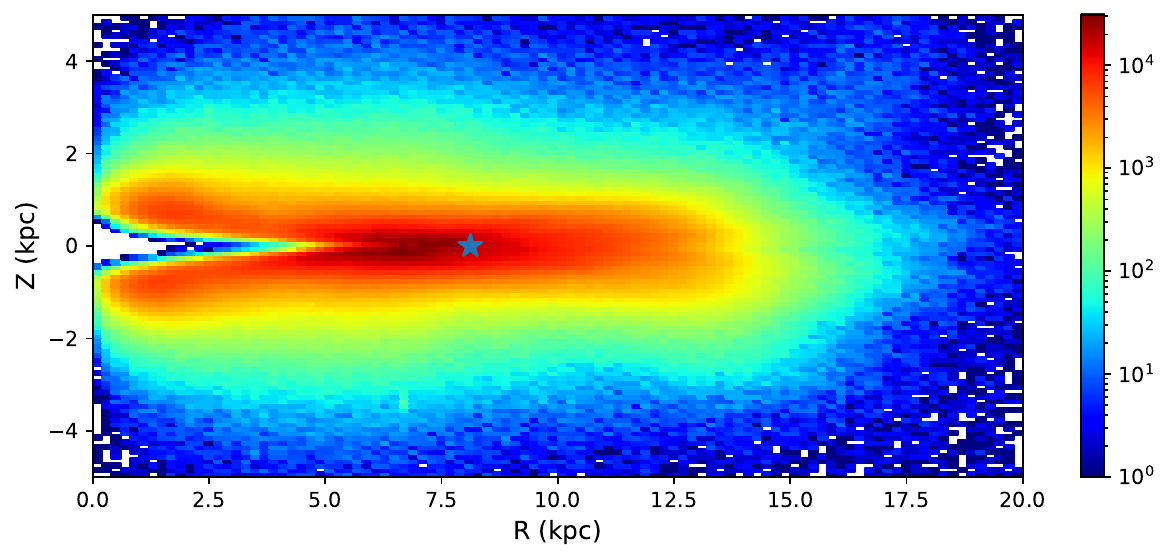}
	\caption{Density distribution of the RC sample in the $R-Z$ diagram on a logarithmic scale. The location of the Sun is marked by a blue star.}
	\label{fig:RZdensity}
\end{figure}

\begin{figure}
	\centering
	\includegraphics[width = 8.8cm]{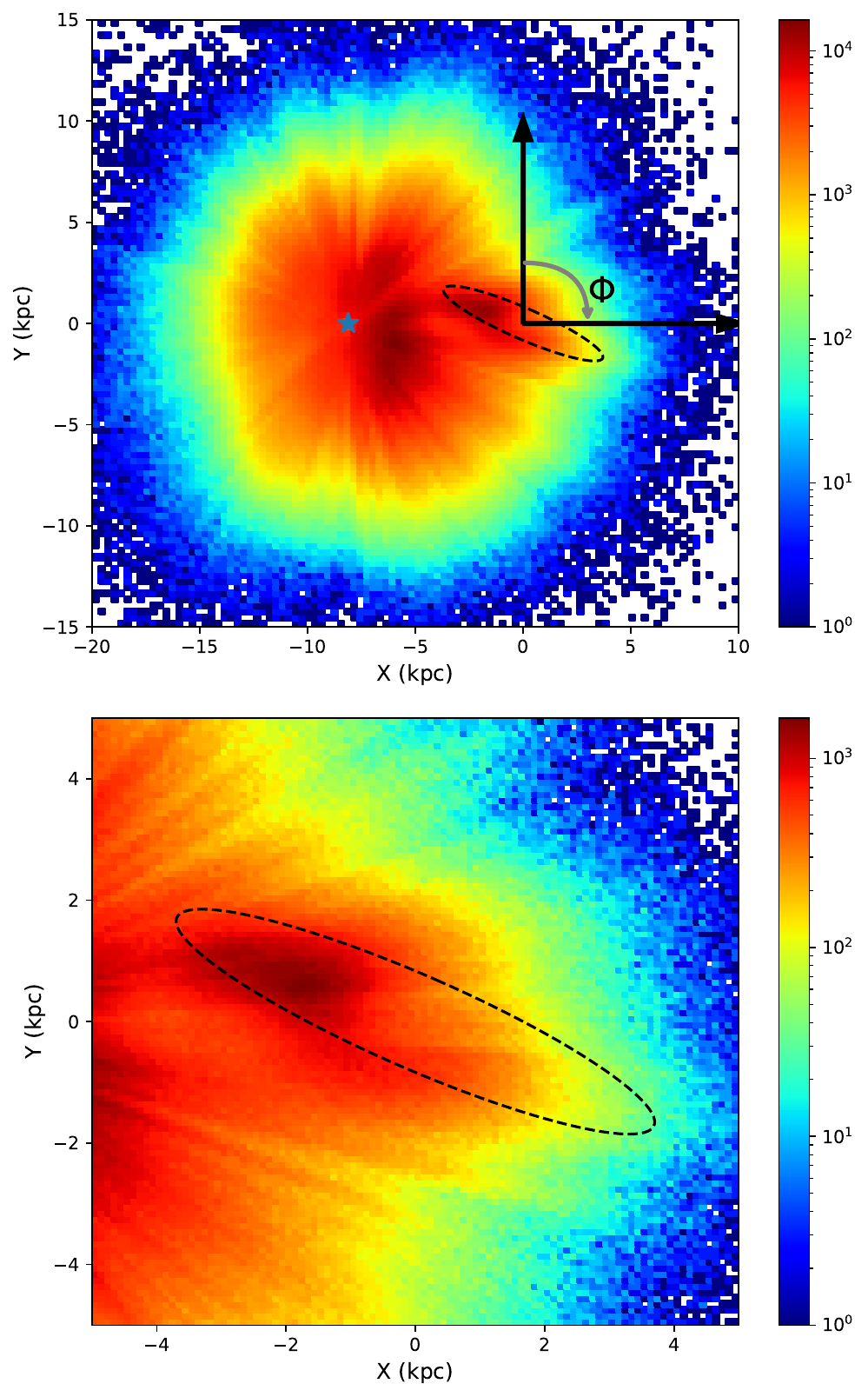}
	\caption{Density distribution of our RC candidates in the Galactocentric $X-Y$ diagram, where the position of the Sun is indicated by a blue star. In this right-handed coordinate system, the azimuthal angle $\phi$ is measured from the center-Sun-anticenter direction and rotates clockwise like the Milky Way. The dashed ellipse indicates the orientation angle (25 deg with respect to the Sun-Galactic center line) and extent (semi-major axes a=4.07 kpc and semi-minor axes b=0.76 kpc, same as \citet{2022A&A...658A..91A}) of the Galactic bar. The bottom panel is a zoom in of the Galactic center. We can see an overdensity around $X,Y=-2,1$ kpc, which is well included in the dashed ellipse.}
	\label{fig:xydensity}
\end{figure}

Our RC sample has full sky coverage, but most of the stars are confined within $|b|<10^\circ$ as shown in Figure~\ref{fig:lb}, which is in line with the fact that RCs are mainly found in the disk and bulge. From Figure~\ref{fig:RZdensity}, we can see that our RC sample has good coverage from the Galactic center to R=15 kpc. Even inside $R=5$ kpc, we still have a large number of RC stars near the Galactic plane, which will help us obtain an accurate mid-plane density. Figure~\ref{fig:xydensity} shows the density distribution of the RC sample on a logarithmic scale in $X-Y$ diagram. An overdensity of RCs is found around $(X,Y)=(-2,1)$ kpc, which is well included by the ellipse that indicates the region of the Galactic bar. Therefore, the influence of the Galactic bar/bulge should be noted when deriving the density profile of the inner disk. Since our sample has good coverage in the region of $R < 15$ kpc and $|\phi| < 75^\circ$, we can divide these RCs into different bins of $\phi$ to examine the influence of the bar/bulge, which will be discussed in more detail in Section~\ref{sec:results}.  

\begin{figure}
	\centering
	\includegraphics[width=8.8cm]{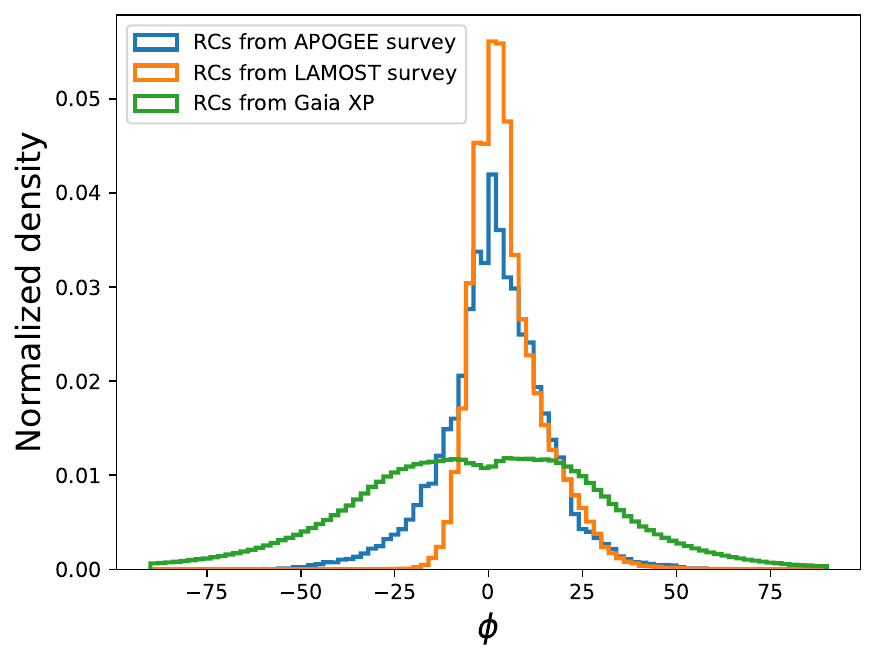}
	\caption{Azimuthal distribution ($\phi$) of RC samples selected from the APOGEE survey (blue hist, \citet{2016ApJ...823...30B}), the LAMOST survey (orange hist, \citet{2020ApJS..249...29H}), and Gaia XP spectra (green hist). RC sample obtained from Gaia XP spectra has a more completed coverage in $\phi$ than the other two samples.}
	\label{fig:phi}
\end{figure}

\section{Selection function of the RC sample}\label{sec:selectfunction}

The wide spatial distribution of our RC sample enables us to explore the density profile of the Galactic disk from R=2 kpc to R=17 kpc. The selection function of our RC sample is defined as Equation~\ref{eq:SFall}:
\begin{equation}
	\begin{aligned}
		S_\mathrm{RC}&=\frac{n_\mathrm{RC;sub}(l,b,c,m)}{n_\mathrm{RC;ph}(l,b,c,m)}\\&=
		\frac{n_\mathrm{RC;\mathrm{combined}}(l,b,c,m)}{n_\mathrm{RC;XP}(l,b,c,m)}\times\frac{n_\mathrm{RC;XP}(l,b,c,m)}{n_\mathrm{RC;ph}(l,b,c,m)}\\&\times\frac{n_\mathrm{RC;sub}(l,b,c,m)}{n_\mathrm{RC;\mathrm{combined}}(l,b,c,m)},\label{eq:SFall}
	\end{aligned}
\end{equation}
where $n_\mathrm{RC;sub}$, $n_\mathrm{RC;ph}$, $n_\mathrm{RC;XP}$, and $n_\mathrm{RC;An23}$ are RC star counts in our sample, a complete photometric data, Gaia XP spectra, and the combined catalog in the color-magnitude diagram, respectively. We use the subscript ‘sub’ to indicate that our sample is only a subsample of all real RCs hidden in Gaia XP spectra. The combined catalog refers to a merger of the \texttt{An23}, 2MASS, and \texttt{gaiaedr3\_contrib.starhorse} catalogs used in this study. The selection function $S_\mathrm{RC}$ describes the probability that a true RC star can be included in the catalog according to their Galactic coordinates $(l, b)$, colors ($c$), and apparent magnitudes ($m$). We choose the Gaia DR3 photometric data as the reference, since it is almost complete for stars of $G<19$ except for some extremely crowded regions, such as the center of globular clusters or the Galactic bulge \citep{2023A&A...669A..55C}. We adopt the $G$ band apparent magnitude and the color $G-G_\mathrm{RP}$ in the calculation of $S_\mathrm{RC}$. Following \citet{2023A&A...677A..37C}, we use $G-G_\mathrm{RP}$ instead of $G_\mathrm{BP}-G_\mathrm{RP}$ because of some known calibration issues affecting $G_\mathrm{BP}$ for faint red sources \citep{2021A&A...649A...3R}.

The first factor ${n_\mathrm{RC;\mathrm{combined}}(l,b,c,m)}/{n_\mathrm{RC;XP}(l,b,c,m)}$ accounts for the ratio of RCs between the combined catalog and Gaia XP spectra. However, ${n_\mathrm{RC;\mathrm{combined}}(l,b,c,m)}$ and ${n_\mathrm{RC;XP}(l,b,c,m)}$ are unknown factors since we are unable to perfectly extract all RCs from Gaia XP spectra. Therefore, we adopt the ratio of all stellar objects ${n_\mathrm{\mathrm{combined}}(l,b,c,m)}/{n_\mathrm{XP}(l,b,c,m)}$ as an approximation of the first factor, assuming that the \texttt{An23} catalog does not have any particular favor or unfavor of deriving the parameters for RCs in a certain $(l,b,c,m)$ bin. Because the \texttt{An23} catalog includes most of the stellar objects of Gaia XP spectra, we think that this approximation will not largely change the results. We generated the first factor at the resolution of the HEALPix\footnote[1]{HEALPix, an acronym for Hierarchical Equal Area isoLatitude Pixelisation of a sphere, is an algorithm that describes the Sky positions $(l,b)$. Stars with similar coordinates $(l,b)$ have the same HEALPix index number. Further information about HEALPix is available in the Gaia archive documentation.} level 5 in the Sky positions (12,288 equally sized areas, referred to as the 12,288 base pixels), $G-G_\mathrm{RP}\in[0.5, 1.5]$ in bins of 0.1, and $G\in[6, 18]$ in steps of 0.2. 

\begin{figure}
	\centering
	\includegraphics[width = 8.8cm]{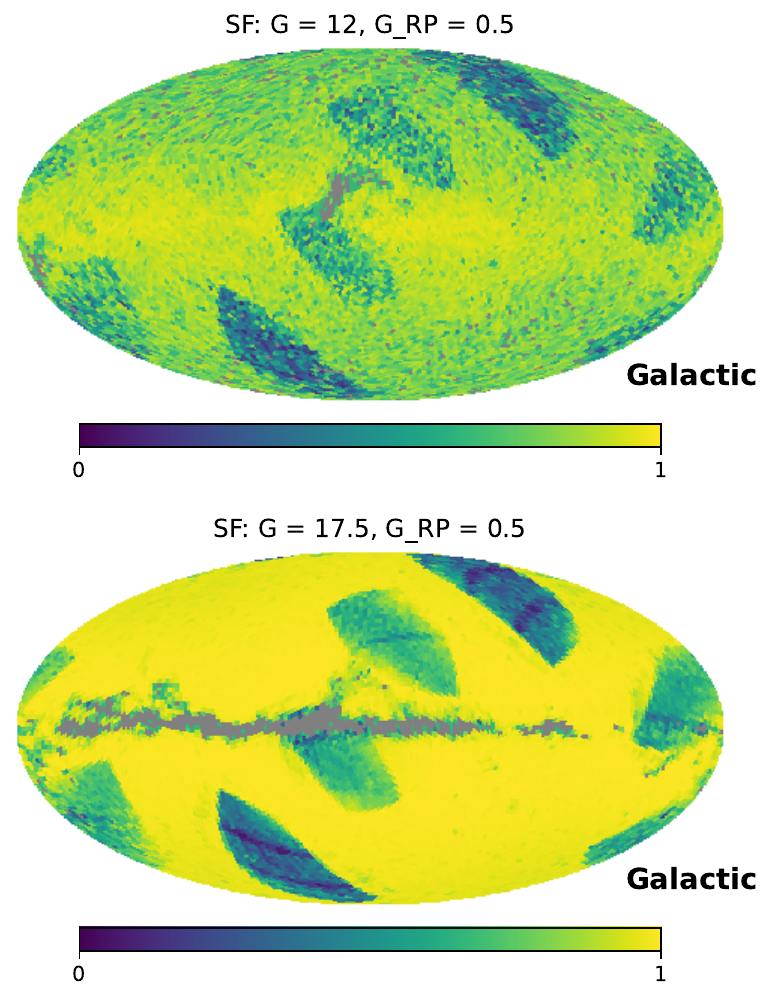}
	\caption{Map of the selection function for Gaia XP spectra using Gaia DR3 photometry as the reference for a bright (top panel) and a faint (bottom panel) magnitude. In the bottom panel we can see a lack of faint stars within $|b|<10^\circ$ for XP spectra, which causes the deficiency of RCs in the disk mid-plane towards the Galactic center ($R<3.5$ kpc).}
	\label{fig:secondfactor}
\end{figure}

The second factor ${n_\mathrm{RC;XP}(l,b,c,m)}/{n_\mathrm{RC;ph}(l,b,c,m)}$ can be written as Equation~\ref{eq:SFXP},
\begin{equation}
	\begin{aligned}
		\frac{n_\mathrm{RC;XP}(l,b,c,m)}{n_\mathrm{RC;ph}(l,b,c,m)}&=\frac{n_\mathrm{XP}(l,b,c,m)\times r_\mathrm{RC;XP}(l,b,c,m)}{n_\mathrm{ph}(l,b,c,m)\times{r_\mathrm{RC;ph}(l,b,c,m)}}\\&=S_\mathrm{XP}(l,b,c,m)\times\frac{r_\mathrm{RC;XP}(l,b,c,m)}{r_\mathrm{RC;ph}(l,b,c,m)}
		,\label{eq:SFXP}
	\end{aligned}
\end{equation}
where $r_\mathrm{RC;XP}(l,b,c,m)$ and $r_\mathrm{RC;ph}(l,b,c,m)$ are the ratio of RCs to all stellar objects in Gaia XP spectra and Gaia DR3 photometric survey. The stars found in a certain $(l,b,c,m)$ bin might be our needed RC or contaminants such as RGBs/dwarfs. However, we do not know a prior what their true stellar types are before observation. Therefore, during target selection, we assumed that Gaia XP spectra do not favor observing one type of stars over another. The random target selection means that $r_\mathrm{RC;XP}(l,b,c,m)$ will be almost equal to $r_\mathrm{RC;ph}(l,b,c,m)$. In this case, the only factor concerned in the second factor is $S_\mathrm{XP}(l,b,c,m)$. \citet{2023A&A...677A..37C} develop a Python class \texttt{gaiaunlimited.SubsampleSelectionFunction} to compare the number of stars in a given subsample to that in the overall Gaia catalog. They provide an estimate of the membership probability of the subsample as a function of sky position ($l,b$), magnitude ($G$), and color ($G-G_\mathrm{RP}$). We generated the selection function $S_\mathrm{XP}(l,b,c,m)$ by \texttt{gaiaunlimited.SubsampleSelectionFunction} at the same resolution as the first factor. Figure~\ref{fig:secondfactor} shows $S_\mathrm{XP}$ for both a bright magnitude and a faint magnitude. The second factor is larger than 0.9 for around 80\% of our selected RCs.

The third factor $n_\mathrm{RC;sub}(l,b,c,m)/n_\mathrm{RC;{combined}}(l,b,c,m)$ describes the probability of a RC that can be properly selected from the combined catalog using our methodology. We adopted several cuts to reduce contaminant RGBs/dwarfs, which will inevitably exclude some real RCs and lower our completeness. However, $n_\mathrm{RC;combined}$ is an unknown factor because we are unable to identify all RCs from Gaia XP spectra. Although we cannot perfectly recover all RCs, a subset of them has been reliably identified from higher-resolution spectroscopic surveys, which is the 155,522 common sources obtained from a cross-match between the \texttt{Ting18} and Gaia XP spectra mentioned in Section~\ref{sub:identification}. We required that these common sources are also available in the 2MASS survey and the \texttt{gaiaedr3\_contrib.starhorse} catalog, which returns us a total of 151,797 identified RCs. Among these common sources, 117,196 are also identified as RCs by Gaia XP spectra in this study, and the rest are excluded during the identification. The probability of correctly identifying RC stars within this subset is used as an approximation for the third factor.   

\begin{figure}
	\centering
	\includegraphics[width = 8.8cm]{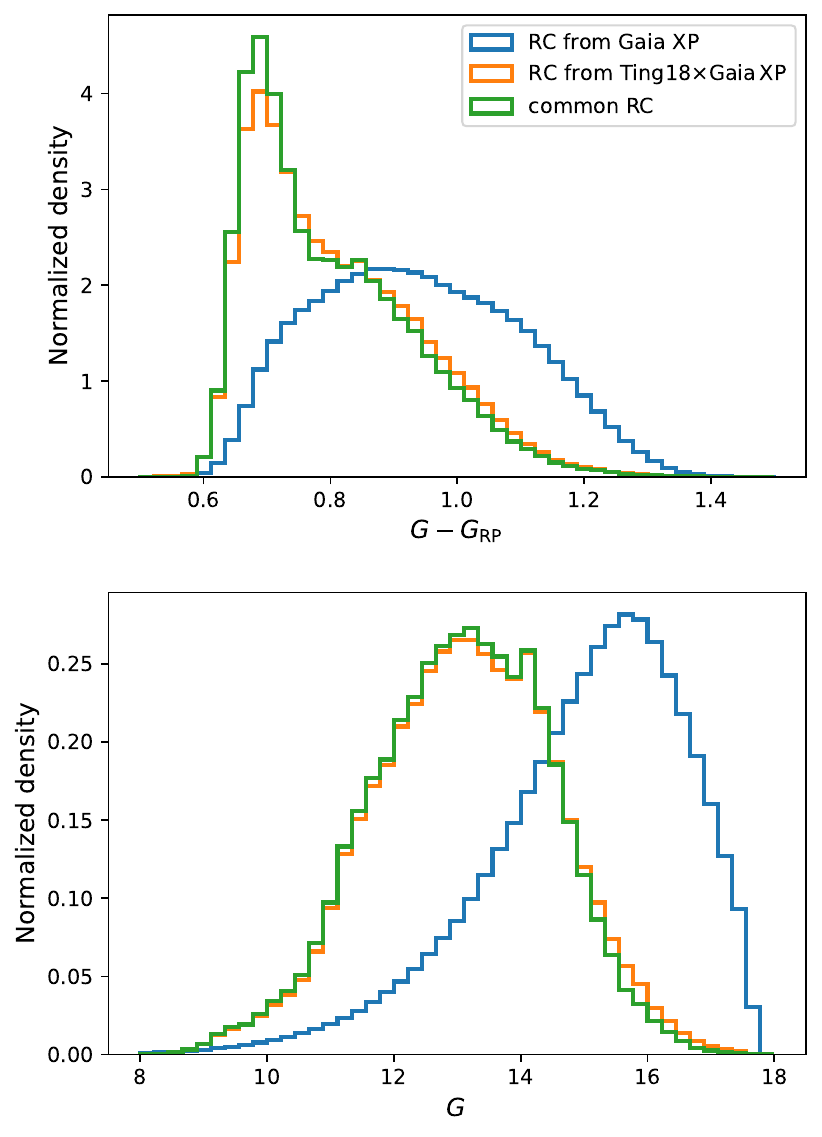}
	\caption{Normalized density distributions of the RCs selected from XP spectra (blue normalized histogram, by us) and the combined LAMOST/APOGEE survey (orange histogram, by \citet{2018ApJ...858L...7T}) in $G-G_\mathrm{RP}$ (top panel) color and $G$ (bottom panel) band magnitude. The green histogram represents the RCs in common between the two catalogs.}
	\label{fig:color_dis}
\end{figure}

\begin{figure}
	\centering
	\includegraphics[width = 8.8cm]{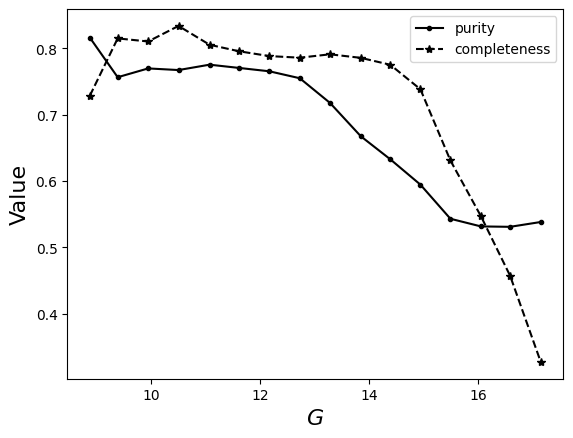}
	\caption{Completeness (black dashed lines) and purity (black solid lines) as a function of $G$ estimated from a direct comparison with the Ting18 catalog. A one-dimensional quadratic polynomial fit is applied to both completeness and purity, and the resulting functions are used in modeling the stellar disk density.}
	\label{fig:thirdfactor}
\end{figure}

\begin{figure}
	\centering
	\includegraphics[width = 8.8cm]{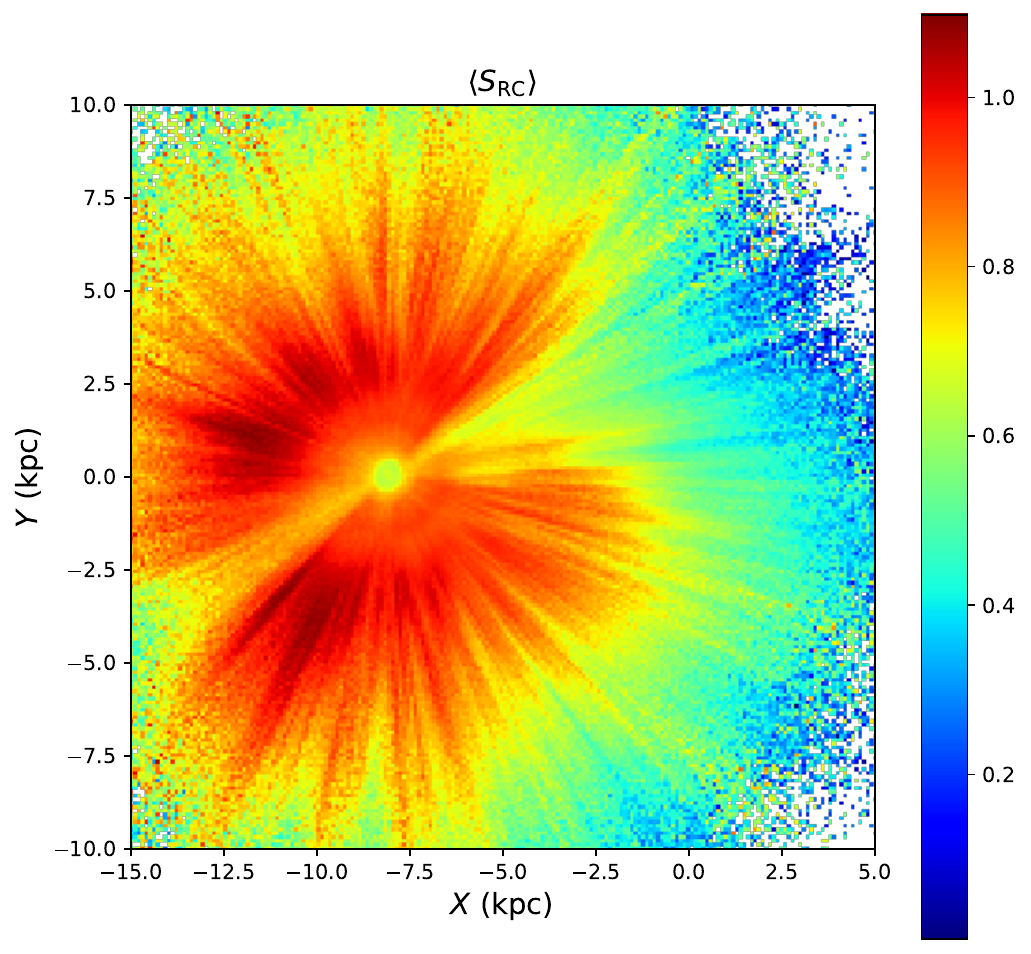}
	\caption{Averaged $S_\mathrm{RC}$ in $X-Y$ panel of our RC sample. The pixel size is $0.1\times0.1$. }
	\label{fig:sfxy}
\end{figure}

Because this subsample does not cover the full sky like Gaia XP spectra, we have to neglect the influence of sky positions on the third factor. From Figure~\ref{fig:color_dis}, we can see that our identified RCs are more weighted in the red end of $G-G_\mathrm{RP}$ than the sample of \texttt{Ting18}. Since Gaia XP spectra are more concentrated toward the Galactic plane and bulge region than the other two surveys, it is expected that our selected RCs suffer more from dust reddening. The small number of the common RCs makes the exploration of completeness hard in the red end of $G-G_\mathrm{RP}$. Therefore, we only focus on the variation of completeness with $G$ in this study, even though this simplification will inevitably introduce some uncertainties in our selection function. Figure~\ref{fig:thirdfactor} shows the variation of completeness derived from this subset with $G$, which keeps almost constant as 0.8 at $G<15.5$ and drops quickly to around 0.35 at $G\sim17.5$. We adopt the completeness estimated from the combined RC catalog as the third factor and apply it to all RCs in our sample. Figure~\ref{fig:sfxy} exhibits the averaged final selection function in the $X-Y$ (Galactocentric Cartesian coordinates) diagram, where more distant stars tend to have a smaller $S_\mathrm{RC}$. Although the construction of $S_\mathrm{RC}$ is accompanied by several assumptions, it works well in deriving the density profile of the Galactic disk and brings results compatible with previous studies in the following sections. 

Besides the identified RCs, the \texttt{Ting18} catalog also includes a large number of RGBs from the LAMOST DR3 and APOGEE DR14 data. A cross match between our RC sample and Ting18 catalog returns us a total of 167,869 common stars. Among them, 117,196 are also identified as real RCs by \texttt{Ting18}, while the rest 50,673 are classified as RGBs according to predicted asteroseismic parameters. We define the purity as the proportion of real RCs in these 167,869 candidates, and show its dependence on $G$ magnitude in Figure~\ref{fig:thirdfactor}. The purity keeps almost constant as 0.75 at $G<13.5$ and then drops gradually to around 0.55 at $G\sim17.5$.

\section{Method}\label{sec:densityfitting}
Following \cite{2017RAA....17...96L}, we derived the stellar density at the position of every RC after correcting the selection function and the uncertainty of the distance estimation.

As several candidates may actually be RGBs overlapping with genuine RCs in the Hertzsprung–Russell diagram, we first corrected the estimated distance based on the purity obtained in section~\ref{sec:selectfunction}. The corrected distance is described as follows:
\begin{equation}
		\begin{aligned}
	&m_\mathrm{K_s}-M_\mathrm{K_s}\sim\mathrm{P}(G)\mathcal{N}(\langle m_\mathrm{K_s}\rangle-\langle M_\mathrm{K_s}\rangle,{{\sigma_{m_\mathrm{K_s}}}^2+{\sigma_{M_\mathrm{K_s}}}^2})\\&
	+(1-\mathrm{P}(G))\mathcal{N}(\langle m_\mathrm{K_s}\rangle-\langle M_\mathrm{K_s;RGB}\rangle,{{\sigma_{m_\mathrm{K_s}}}^2+{\sigma_{M_\mathrm{K_s;RGB}}}^2})\,
	\label{eq:corrected}
		\end{aligned}
\end{equation}
where $\mathrm{P}(G)$ is the purity as a function of $G$ magnitude. In Figure~\ref{fig:MKsandfeh}, we find that $M_\mathrm{K_s}$ of the RGB contaminants in the RC catalog depends weakly on $\mathrm{[M/H]}$. Therefore, in Equation~\ref{eq:corrected}, we adopt a simple Gaussian function to describe the distribution of $M_\mathrm{K_s}$ for the RGB part, where $\langle M_\mathrm{K_s;RGB}\rangle=-1.52$ and $\sigma_{M_\mathrm{K_s;RGB}}=0.29$ are obtained by applying a Gaussian fitting to the RGB sample in Figure~\ref{fig:MKsandfeh}. The 16th, 50th ,and 84th percentile distributions of the photometric corrected distance are derived by a random sampling of Equation~\ref{eq:corrected}. This correction only causes a slight change in the estimated distance $d_\mathrm{rc}$, since the absolute magnitude of the overlapped RGB contaminant does not differ significantly from that of the real RC. However, due to the larger dispersion of the absolute magnitude in the RGB part, the estimated distance uncertainty ($\sim10\%$) is slightly larger than the one ($\sim7.5\%$) obtained in Section~\ref{sub:distance}, where only the dispersion of the RC component is considered. In this study, the density profile of the stellar disk is derived from the corrected distance.  

\cite{2017RAA....17...96L} describe a method to derive the stellar density profile along each line of sight (an observed $\textit{l}-\textit{b}$ plate) of a given spectroscopic survey. In their work, the ratio between the stellar density obtained from the photometric ($\nu_\mathrm{ph}$) (a complete sample) and spectroscopic ($\nu_\mathrm{sp}$) (an uncompleted subsample) survey is defined as:
\begin{equation}
	\nu_\mathrm{ph}(D|\textit{l}, \textit{b}, \textit{c}, \textit{m}) = \nu_\mathrm{sp}(D|\textit{l}, \textit{b}, \textit{c}, \textit{m}){S^{-1}(\textit{l}, \textit{b}, \textit{c}, \textit{m})},
	\label{eq:nu}
\end{equation}
where $D$ is the distance along a given line of sight, and $S$ is the selection function of the RC sample ($S_\mathrm{RC}$) in this study. 

After integrating over $\textit{c}$ and $\textit{m}$, the stellar density profile for a given line of sight is defined as:
\begin{equation}
	\nu_\mathrm{ph}(D|l, b) = \iint{\nu_\mathrm{sp}(D|\textit{l}, \textit{b}, \textit{c}, \textit{m})S^{-1}(\textit{l}, \textit{b}, \textit{c}, \textit{m})\mathrm{d}c\mathrm{d}m}.
	\label{eq:ph}
\end{equation}

In a given light of sight, the contribution $p_i$ of an identified RC $i$ at distance $D$ is weighted by a Gaussian kernel with width of the distance uncertainty, and $p_i$ is defined as,
\begin{equation}
	p_i(D) = \frac{\mathcal{N}(D|D_i, \sigma_{D_i}^2)}{\int_{D_\mathrm{min}}^{D_\mathrm{max}}\mathcal{N}(D_x|D_i, \sigma_{D_i}^2)dD_x}\times{\mathrm{P}(G)_i},
	\label{eq:pi}
\end{equation}
where $\mathcal{N}$ is a normal function, and $\mathrm{P}(G)_i$ is the purity of star $i$. $D_i$ is the estimated distance of star $i$, and $\sigma_{D_i}$ is the uncertainty in the distance estimation. In this study, $\sigma_{D_i}$ is defined as,
\begin{equation}
	\sigma_{D_i} = \frac{D_i(84\%)-D_i(16\%)}{2}
\end{equation}
where $D_i(16\%)$ and $D_i(84\%)$ are the $16\%$ and $84\%$ percentiles distribution of the corrected distance. In this calculation, We impose $D_\mathrm{min} = 0$ kpc and $D_\mathrm{max} = 80$ kpc.

We take into account the contribution of all stars in a certain color-magnitude bin. The density profile $\nu_\mathrm{sp}$ is defined as
\begin{equation}
	\nu_\mathrm{sp}(D|\textit{l}, \textit{b}, \textit{c}, \textit{m}) = \frac{1}{{\Omega}D^2}\sum_{i}^{n_{sp}(\textit{l}, \textit{b}, \textit{c}, \textit{m})}p_i(D),
	\label{eq:sp}
\end{equation}
where $\Omega$ is the solid angle of the given line of sight. We divided the sky positions into 12,288 $l-b$ plates at the resolution of the HEALPix level 5, and all RCs are distributed into different plates according to their Galactic coordinates. Since $\Omega$, $dc$, and $dm$ are constant values which does not influence the derived density shape, we normalize them to 1 for computational convenience. To correct the selection function, we substitute Equation~\ref{eq:sp} into ~\ref{eq:ph} and obtain the intrinsic density $\nu_\mathrm{ph}$. 

In principle, we can derive $\nu_\mathrm{ph}$ for all $l-b$ plates at any arbitrarily selected distance $D$. However, as emphasized by \citet{2018MNRAS.473.1244X}, the derived $v_\mathrm{ph}$ at the positions without any stars sampled by the spectroscopic survey suffers from a large uncertainty. Therefore, we only use $\nu_\mathrm{ph}$ at the positions where we can find a star. In other words, every RC in our sample is assigned a $\nu_\mathrm{ph}$ that represents an estimation of the stellar density $\nu$ at its position.

\section{Results}\label{sec:results}           
\subsection{Modeling of the Galactic disk}\label{sub:modeldisk}
The vertical number density profile $\nu(Z|R)$ of the stellar disk is described by a combination of two exponential disks as follows:
\begin{equation}
	\begin{aligned}
		\nu_\mathrm{model}(Z|R)=f_\mathrm{thin}(R)\nu_0(R)\exp(-\frac{|Z-Z_0(R)|}{h_{\mathrm{thin}}(R)})&\\+(1-f_\mathrm{thin}(R))\nu_0(R)\exp(-\frac{|Z-Z_0(R)|}{h_{\mathrm{thick}}(R)}),\label{eq:towdisk}
	\end{aligned}
\end{equation}
where $f_\mathrm{thin}$ is the fraction of the geometric thin disk component, $\nu_0$ is the mid-plane density at cylindrical Galactocentric radial coordinate $R$, and $Z_0$ is used to describe the warp structure. $h_{\mathrm{thin}}$ and $h_{\mathrm{thick}}$ are the scale heights of the geometric thin and thick disks, respectively. All of these free parameters vary with $R$. During the fitting, we exclude stars in sky regions that are strongly affected by the dust extinction requiring $E(B-V)<2$ \citep{1998ApJ...500..525S}

\begin{figure}
	\centering
	\includegraphics[width = 8.8cm]{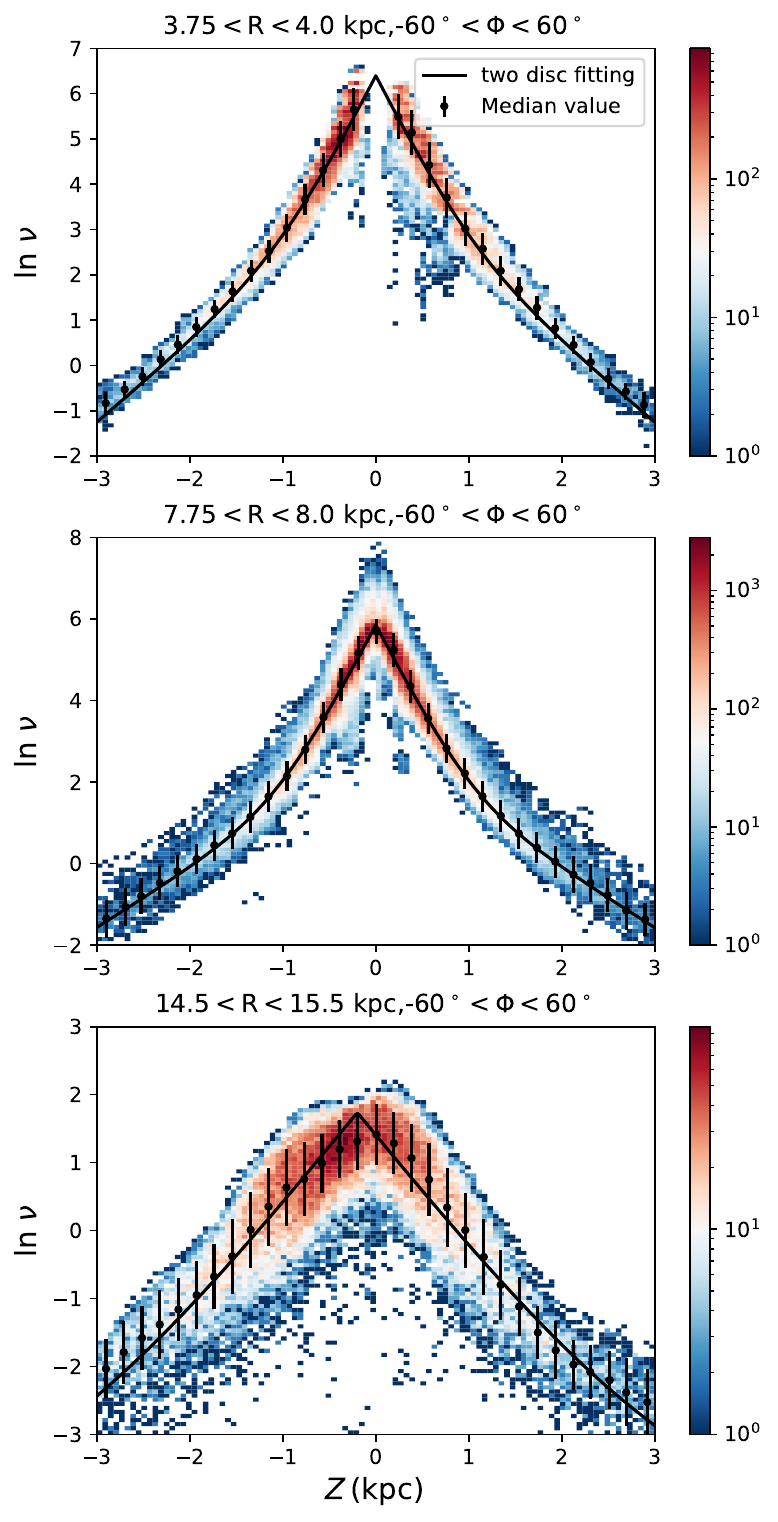}
	\caption{Density distribution of RCs in three certain bins of $3.75<R<4.0\,\mathrm{kpc}$ (top panel, near the bar/bulge), $7.75<R<8.0\,\mathrm{kpc}$ (middle panel, in the solar neighborhood),and $14.5<R<15.5\,\mathrm{kpc}$ (bottom panel, in the outer disk) in $Z-\ln\nu$ panel. The black points represent the median values of $\ln\nu$ in these vertical bins, and the black lines are the best fitting results based on the two disk model.}
\label{fig:anexample}
\end{figure}

Figure~\ref{fig:anexample} presents the density distributions of the RCs in three certain bins of $3.75<R<4.0\,\mathrm{kpc}$, $7.75<R<8.0\,\mathrm{kpc}$, and $14.5<R<15.5\,\mathrm{kpc}$ in $\log\nu-Z$ panel. We excluded RC stars outside the edges of the azimuthal distribution in Figure~\ref{fig:phi} by applying the criterion $|\phi|<60^\circ$. We divided these stars into 31 bins along the vertical direction ($|Z|\leq3$ kpc). For each $Z$ bin, the median value of $\ln\nu$ is obtained as the observed vertical stellar density $\ln\nu_\mathrm{obs}$ and the standard deviation is provided as $\sigma_{\ln\nu_\mathrm{obs}}$. The likelihood of $\ln\nu_\mathrm{obs}(Z_i|R)$ is defined as: 
\begin{equation}
	\begin{aligned} \mathcal{L}_i=\mathcal{N}(\ln\nu_\mathrm{obs}(Z_i|R)|\ln\nu_\mathrm{model}(Z_i|R),\sigma^2_{\ln\nu_\mathrm{obs}(Z_\mathrm{i}|R)}),
	\end{aligned}
	\label{eq:single} 
\end{equation}
where $\mathcal{N}$ is a one-dimensional normal distribution with $\ln\nu_\mathrm{model}(Z_i|R)$ as the mean value and $\sigma_{\ln\nu_\mathrm{obs}(Z_\mathrm{i}|R)}$ as the standard deviation. $Z_i$ is the $i$th bin along the $Z$ direction, and $\nu_\mathrm{obs}(Z_\mathrm{i}|R)$ represents the average observed vertical stellar density in this specific bin. $\theta$ is a combination of the 5 free parameters of $(f_\mathrm{thin},\nu_0,Z_0,h_{\mathrm{thin}},h_{\mathrm{thick}})$. The likelihood function for the vertical density profile in a certain $R$ bin is defined as a combination of all $Z_i$ bins as follows:
\begin{equation}
	\mathcal{L}(\ln\nu_\mathrm{obs}(Z|R)|\theta)=\prod_{i=1}^{N}\mathcal{L}_i(\ln\nu_\mathrm{obs}(Z_i|R)|\theta)
\end{equation}

\begin{figure}
\centering
\includegraphics[width = 8.8cm]{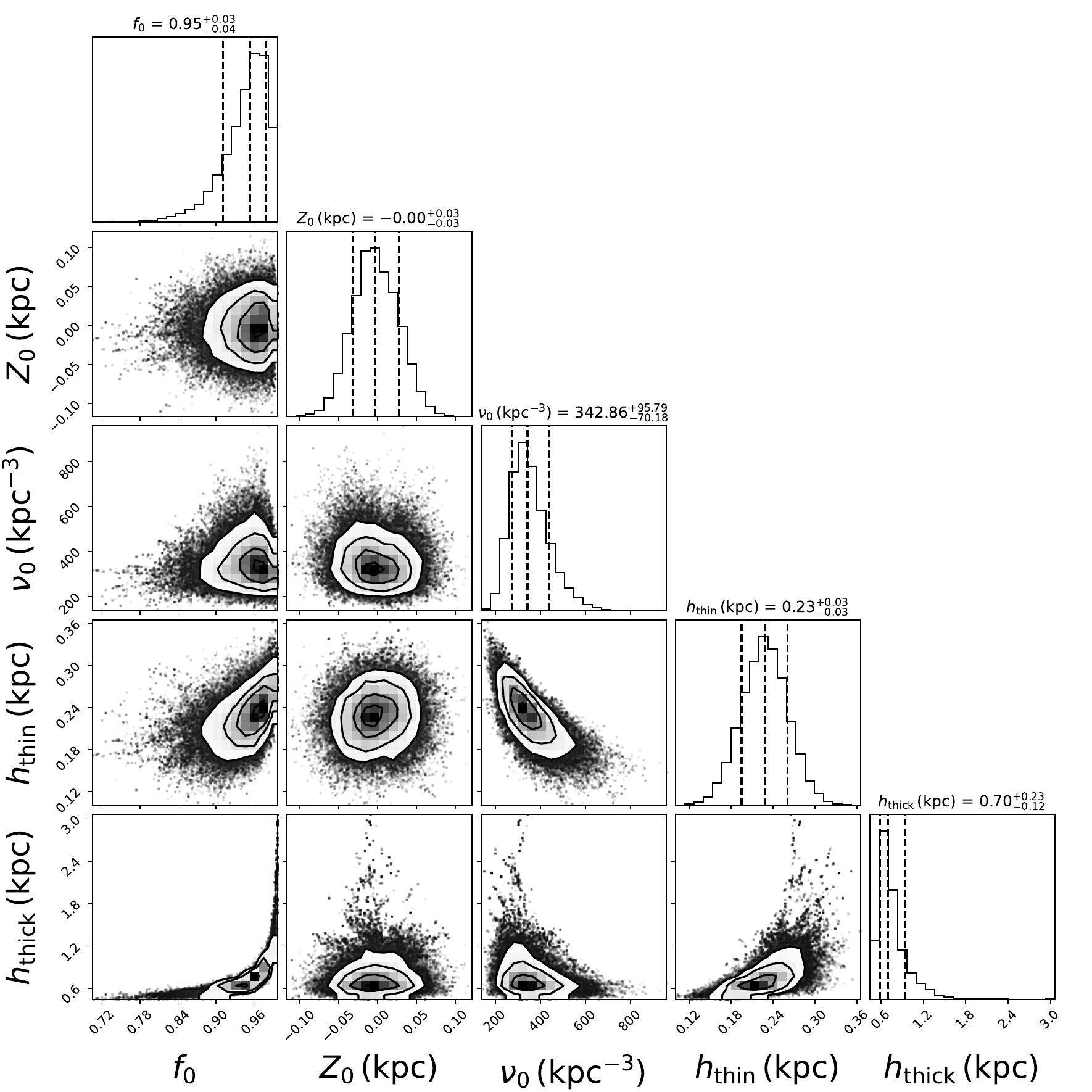}
\caption{Corner plot of the five free parameters in the regions of $7.75<R<8.0$ kpc. The dashed lines show the 16th, 50th, and 84th percentiles of the marginalized distribution of each parameter.}
\label{fig:cornerbin}
\end{figure}

We perform the Markov Chain Monte Carlo (MCMC) method to derive the maximum value of the posterior probability for the 5 free parameters $\theta=(f_\mathrm{thin},\nu_0,Z_0,h_{\mathrm{thin}},h_{\mathrm{thick}})$ using the Python package \texttt{emcee} of \citet{2013PASP..125..306F}. We use 50 walkers and 500 steps as the burn-in, then followed by 3,500 steps to get the posterior distributions of these free parameters. We use the 50th percentile of the marginalized posterior distributions as the best estimated value, and the 16th and 84th percentiles as the uncertainties. Figure~\ref{fig:cornerbin} shows the corner plot of the fit for the bin $7.75<R<8.0\,\mathrm{kpc}$. As shown in Figure~\ref{fig:anexample}, the vertical density profile obtained from these best estimated values generally agrees well with the distribution of RCs in $Z-\ln\nu$ panel.

\subsection{Stellar density profile with varied radius}\label{sub:variationR}
\begin{figure*}
\centering
\includegraphics[width = \textwidth]{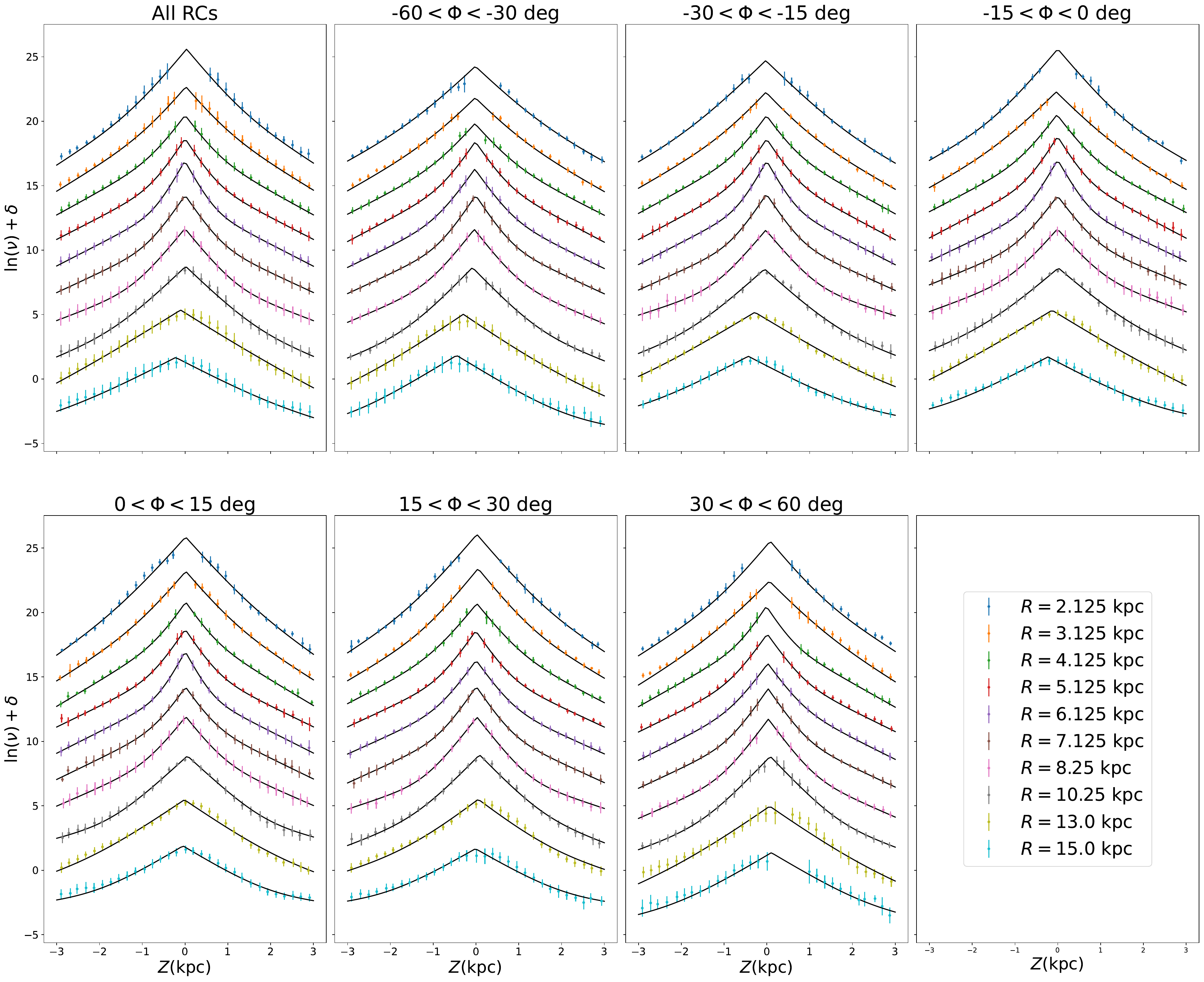}
\caption{Vertical density distribution (points) and best fitting results (black lines) for RC sample of different bins of $\phi$. In each sub figure, the differently colored points represent the vertical density distributions at increasing radii from the inner to outer disk: $R=$2.125 (blue), 3.125 (orange), 4.125 (green), 5.125 (red), 6.125 (purple), 7.125 (brown), 8.25 (pink), 10.25 (gray), 13 (yellow), and 15 (cyan) kpc. To better distinguish these differently colored points, a constant $\delta$ of 18, 16, 14, 12, 10, 8, 6, 4, 2, and 0 is added to $\ln\nu$, respectively. In general, the two disk model works well in describing the vertical density distributions in most cases except for several outermost bins, such as $R=13$ kpc (yellow points) in the bin of $30^\circ<\phi<60^\circ$.}
\label{fig:allfitting}
\end{figure*}

We divided our RCs into different bins of $R$ to obtain a radial density profile of the stellar disk. From $R=2$ to 8 kpc, we adopt a smaller bin with a size of 0.25 kpc to explore the tiny variation of the stellar density distribution with $R$ thanks to the large number of stars inside the solar radius. From $R=8$ to 11 kpc this size is set to 0.5 kpc and increases to 1 kpc beyond $R=11$ kpc. We fit the vertical density profile with the two-disk model for all 35 $R$ bins. Generally, the vertical density profile is well fitted for most bins and the fitting results of several bins are displayed in Figure~\ref{fig:allfitting}. We can see a clear tendency of a slower decline of $\ln(\nu)$ over $|Z|$ in the outer disk, which indicates a phenomenon of flare that has been well identified by many previous studies.

\begin{figure}[!htp]
\centering
\includegraphics[width = 8.8cm]{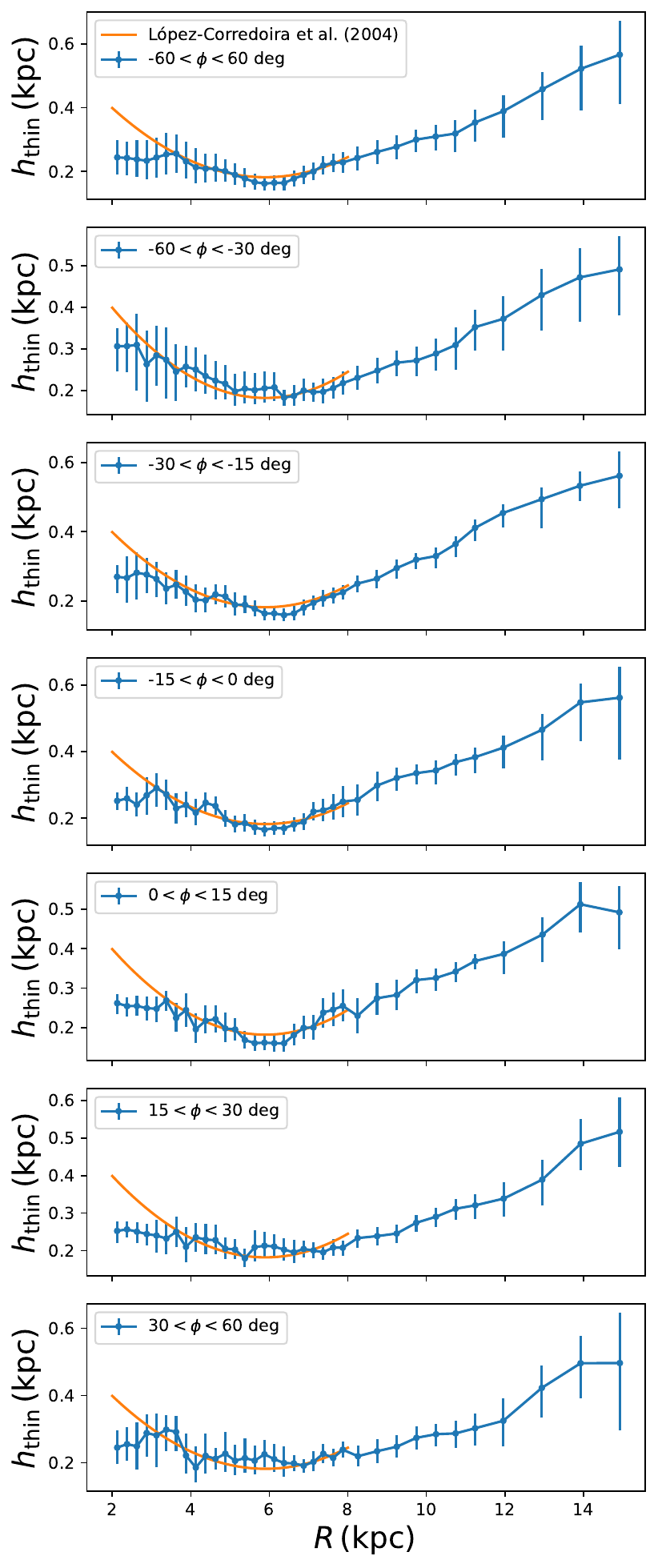}
\caption{Scale height of the thin disk as a function of $R$ for different bins of $\phi$, which presents a V-shaped pattern flaring both inside and outside. Yellow lines are results of \citet{2004A&A...421..953L} with an systematic offset of -0.04 corrected, which also display a flaring morphology inside and outside $R\sim6$ kpc.}
\label{fig:thindiskhz}
\end{figure}

Many previous studies have investigated the variation of scale heights of the thin disk beyond the solar radius using different tracers, and their results suggest that older stellar populations usually flare more. Therefore, we will focus on the comparison with studies using stellar populations of a similar age distribution to ours, like RCs or giants. In Figure~\ref{fig:thindiskhz}, the scale height increases from $h_\mathrm{thin}\sim0.23$ kpc in the solar neighborhood, to around 0.6 kpc in the outer disk of $R=15$ kpc. It is a typical value of $h_\mathrm{thin}$ for the solar neighborhood, which varies from around 0.17 to 0.38 kpc using RCs or Giants as tracers \citep[e.g.,][]{2021ApJ...912..106Y,2024MNRAS.527.4863U,2024MNRAS.531..495T,2025AJ....169...61Y,2025A&A...701A.270K}. However, a much larger range of $h_\mathrm{thin}$ in the outer disk is seen in different studies. Several results provide a value of $h_\mathrm{thin}$ from around 0.5 to 0.9 kpc \citep{2006A&A...451..515M,2014A&A...567A.106L,2021ApJ...912..106Y,2022A&A...664A..58C,2024MNRAS.531.1730T}, while some other studies tend to obtain a much larger result of $h_\mathrm{thin}>1.5$ kpc \citep{2002A&A...394..883L,2024MNRAS.527.4863U,2025A&A...701A.270K}. 

Inside the solar radius, $h_\mathrm{thin}$ is less studied by previous studies due to the small volume of stellar samples and the influence of Galactic bar/bulge, especially for $R<6$ kpc. An advantage of Gaia XP spectra is that a large number of stars towards the inner Galaxy are included in this survey, which enables us to study the stellar spatial distributions in the inner disk. As shown in Figure~\ref{fig:thindiskhz}, $h_\mathrm{thin}$ gradually decreases from 0.23 kpc at $R\sim8$ kpc to 0.16 kpc as the radius decreases to $R\sim6.4$ kpc, and then steadily increases to approximately 0.28 kpc at $R\sim3$ kpc where the Galactic bar's stars can not be neglected during the density fitting. According to \citet{2015MNRAS.450.4050W}, the long bar has two scale heights of 0.18 kpc for the thin bar component and 0.045 kpc for the super thin bar component. Therefore, the mixture of bar's stars in the observational data is likely to lead to an underestimation of $h_\mathrm{thin}$ in the very inner region. Inside $R\sim3$ kpc we can see a slight decline of $h_\mathrm{thin}$ towards the Galactic center, as the bar's stars fraction becomes larger with decreasing $R$.   

Although less significant than the outer flaring at $R>7$ kpc, the inner disk thickening between $R=3$ to 7 kpc shows a steadily and smoothly tendency. In this study, we refer to the variation of $h_\mathrm{thin}$ with $R$ as a V-shaped pattern since it flares both inside ($R<6.4$ kpc) and outside ($R>6.4$ kpc). The V-shaped pattern of $h_\mathrm{thin}$ observed in this study differs somewhat from several previous studies, which think the scale heights of the thin disk is nearly constant inside $R\sim7$ kpc \citep{2016ARA&A..54..529B,2021MNRAS.507.5246M,2025A&A...701A.270K,2025ApJ...984L..48L}. The increase of $h_\mathrm{z}$ towards the Galactic center is consistent with \citet{2004A&A...421..953L}, and they interpreted the observed deficiency of stars in the in-plane central disk as a consequence of a flare in the inner disk. Their expression of $h_\mathrm{thin}$ is obtained by a combination of results from the outer disk ($R>6$ kpc) and the inner disk ($4>R>2.5$ kpc), while our results fill in the observation gap between $6>R>4$ kpc and presents a more detailed morphology, owing to the larger number of RC stars in this study. \citet{2004A&A...421..953L} attributed the inner disk thickening to the dynamical heating introduced by an in-plane bar, which heats the near-plane stars and bring them to higher $|Z|$ positions.   

The increase of $h_\mathrm{thin}$ towards the Galactic center starts from $R\sim6.4$ kpc, which is very near to the co-rotation radius according to the bar's current pattern speed of $\sim30\,\mathrm{km\,s^{-1}\,kpc^{-1}}$ \citep{2017MNRAS.465.1621P,2019A&A...626A..41M,2021MNRAS.500.4710C}. The dynamical heating introduced by the Galactic bar is likely to be more efficient in the inner regions, and the bar-introduced thickening in the inner disk has been shown previously by analyzing the time evolution of Milky Way analogues in hydro-dynamical \textit{N}-body simulations \citep{2012A&A...548A.127M,2019MNRAS.482.1983F}. However, besides the Galactic bar, the vertical density distribution of the stellar disk is also influenced by many other factors like the star formation history \citep{2025NatAs...9..101X}, non-isothermal velocity dispersion \citep{2020MNRAS.499.2523S}, gravitational potential \citep{2018A&A...617A.142S}, and the heating from other asymmetric structures of spiral arms \citep{2015ApJ...802..109M}. Therefore, an accurate test-particle or \textit{N}-body simulation is needed to further check the bar's influence, which is beyond the scope of this work and we leave it to a further study of Wu et al. (2025, in preparation). 

\begin{figure}[!htp]
\centering
\includegraphics[width = 8.8cm]{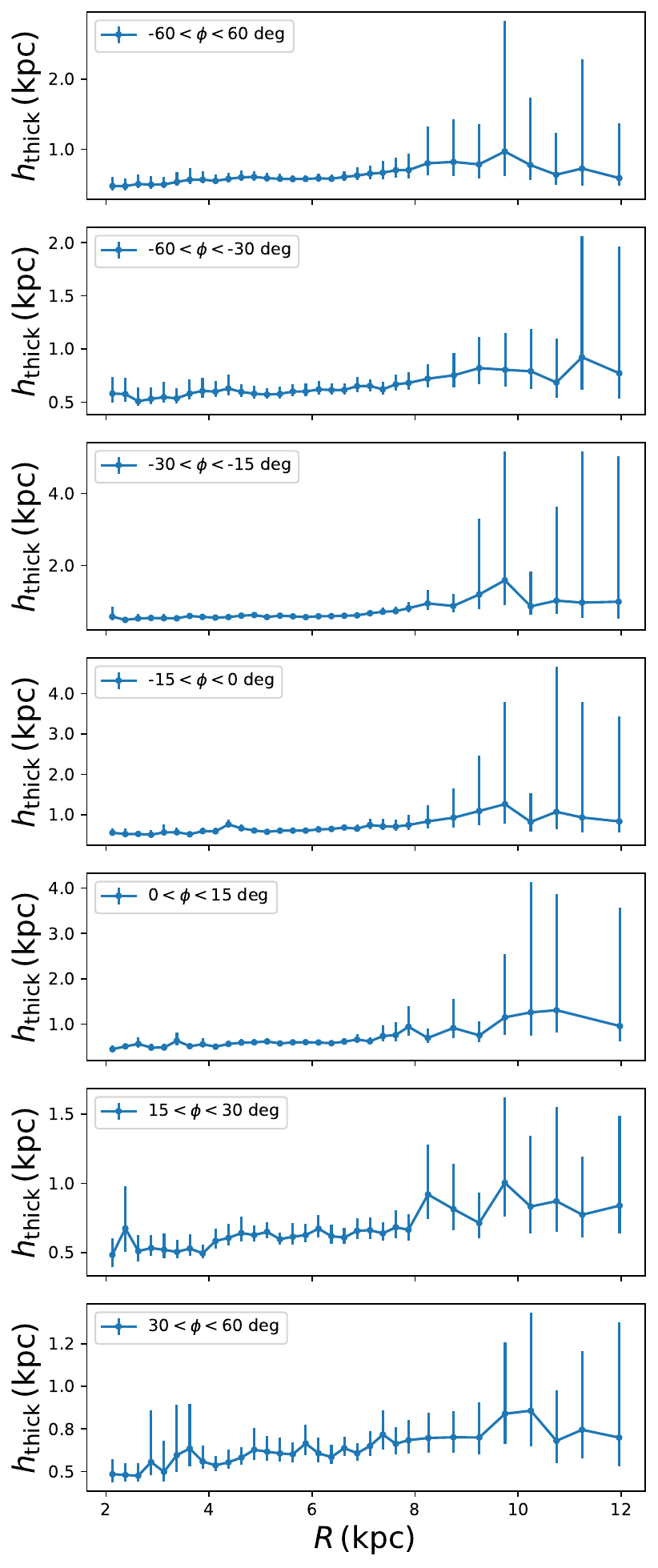}
\caption{Scale height of the thick disk as a function of $R$ for different bins of $\phi$. It increases steadily from $\sim$ 0.50 kpc at $R=2$ kpc to $\sim$ 0.75 kpc in the solar neighborhood.}
\label{fig:thickiskhz}
\end{figure}

\begin{figure}[!htp]
\centering
\includegraphics[width = 8.8cm]{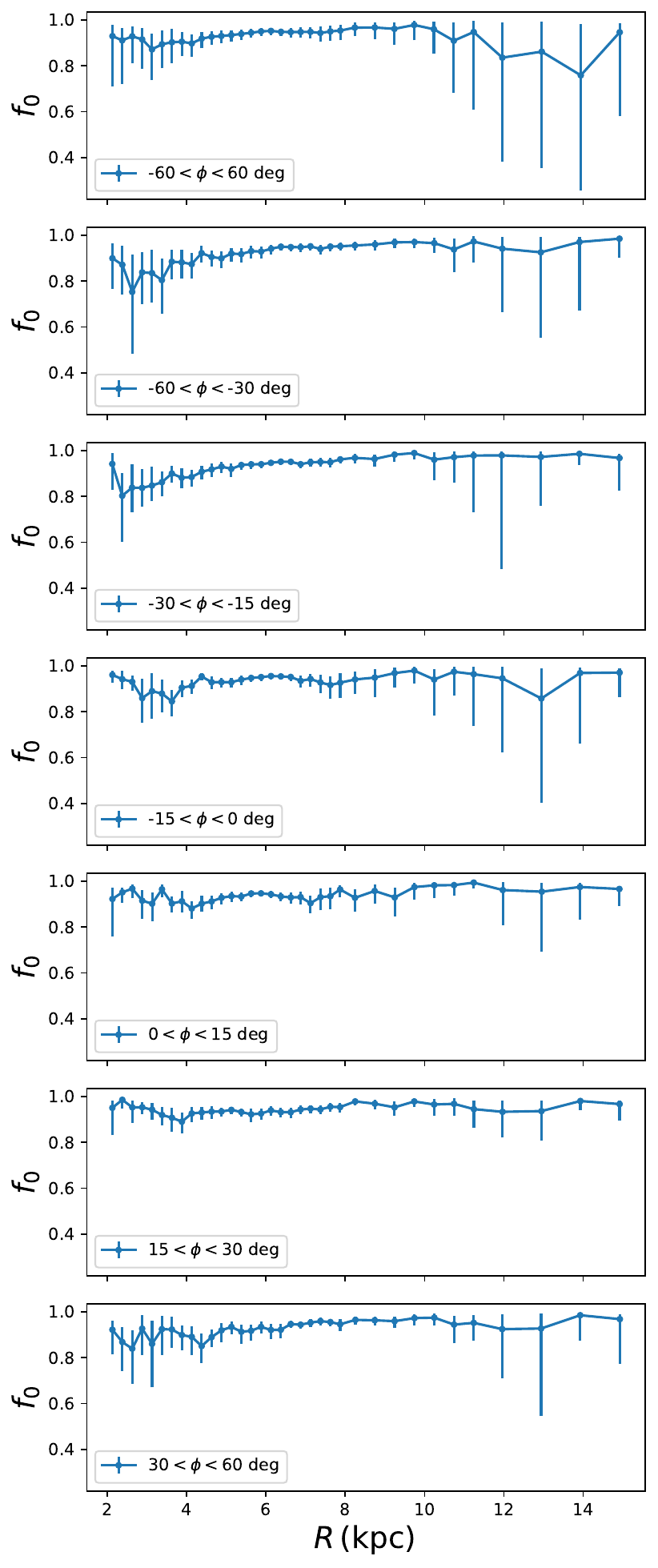}
\caption{Fraction of the geometric thin disk as a function of $R$ for different bins of $\phi$. However, we note that $f_0$ has a large uncertainty at $R>12$ kpc for several bins, which might be caused by the worse fitting in the remoter disk as shown in Figure~\ref{fig:RZdensity}.}
\label{fig:f0}
\end{figure}

\begin{figure}[!htp]
	\centering
	\includegraphics[width = 8.8cm]{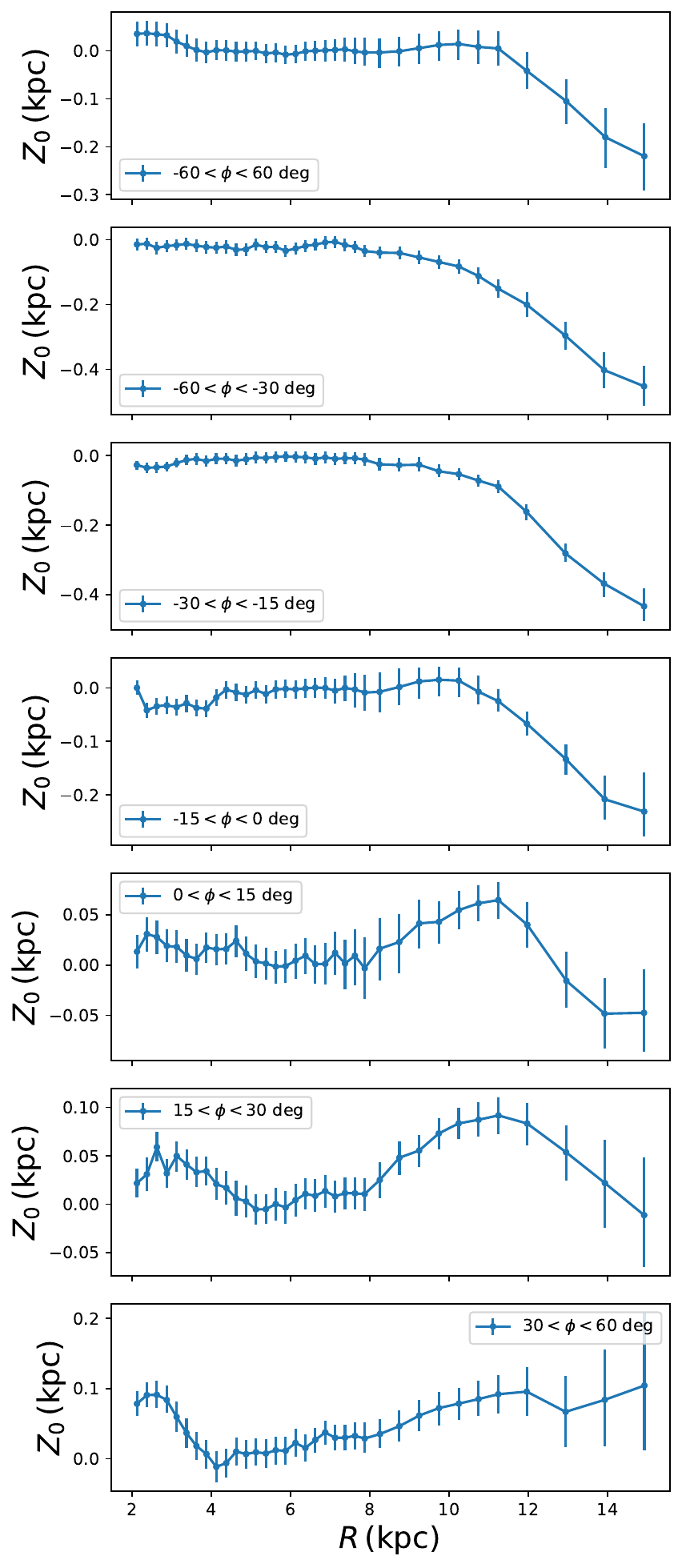}
	\caption{Warp parameter $Z_0$ as a function of $R$. We can see a clear down warp at bins of $\phi<15^\circ$ beyond $R\sim8$ kpc. However, at bins of $\phi>15^\circ$, the upward warp is absent at $R>12$ kpc, which might be introduced by the lack of near-plane stars and the bad fitting of the vertical density profile as shown in Figure~\ref{fig:allfitting}.}
	\label{fig:Z0}
\end{figure}

Figure~\ref{fig:thickiskhz} shows the variation of $h_\mathrm{thick}$, which increases steadily from around 0.50 kpc at $R=2$ kpc to 0.75 kpc at the solar radius. As shown in Figure~\ref{fig:f0}, our derived $f_0$ is larger than 95\% at $8.25<R<11\,\mathrm{kpc}$, which means that the thick disk component could be neglected during the fitting. Therefore, the observation data could be over fitted by the two disk model and we are not sure about the accuracy of our estimated $h_\mathrm{thick}$. At $R>12\,\mathrm{kpc}$, there is a large uncertainty of the derived $f_0$, which may cause the large uncertainty in determining $h_\mathrm{thick}$. Therefore, we only exhibits $h_\mathrm{thick}$ inside $R=12$ kpc to maintain the readability of this image. A similar situation is also seen in \citet{2024MNRAS.531.1730T}, and they attribute it to a large degree of overlap between the thin and thick disks in the outer regions. Figure~\ref{fig:Z0} exhibits the variation of $Z_0$, which remains approximately constant at zero within R$=11$ kpc, and displays a downward warp ($Z_0<0$) beyond this radius.   

\begin{figure}[!htp]
\centering
\includegraphics[width = 8.8cm]{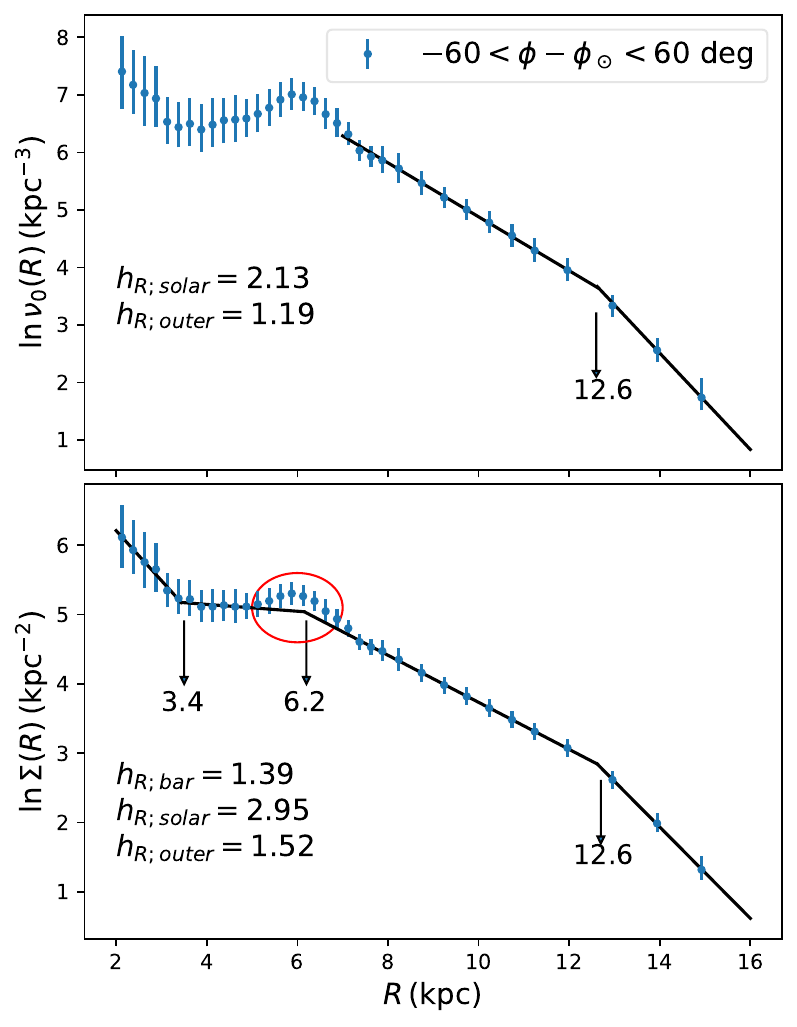}
\caption{Mid-plane ($\ln\nu_0$, top panel) and surface ($\ln\,\Sigma$, bottom panel) density profiles as a function of $R$. Both of them can be well described by a combination of four main components with an additional density bump at $5<R<7$ kpc. $h_{R;\mathrm{bar}}$ is the scale length of the first component at $R<3.5$ kpc. $h_{R;\mathrm{solar}}$ is estimated from the density profile at $8<R<12$ kpc, and $h_{R;\mathrm{outer}}$ represents the scale length of the outer disk at $R>12.6$ kpc.}
\label{fig:surfacedensity}
\end{figure}

An integral of Equation~\ref{eq:towdisk} returns us the surface density ($\Sigma(R)$) of the stellar disk defined as follows,
\begin{equation}
	\begin{aligned}
\Sigma(R)&=\int_{-\infty}^{\infty}{\nu_\mathrm{model}(Z|R)dZ}\\&=2\nu_0(f_0{h_\mathrm{thin}+(1-f_0){h_\mathrm{thick}}}).\label{eq:surface}
\end{aligned}
\end{equation} 
Figure~\ref{fig:surfacedensity} shows the variation of the mid-plane ($\ln\,\nu_0$) and surface ($\ln\,\Sigma$) density profiles with $R$. As shown, both of them can be roughly described by four components like \citet{2024NatAs...8.1302L}, but with an additional density bump at $5<R<7$ kpc. From $R\sim6.2$ to 13 kpc, both two density profiles decline with $R$, exhibiting a typical exponential with a scale length of $h_{R;\mathrm{solar}}=2.14$ kpc and 2.95 kpc for $\ln\,\nu_0$ and $\ln\,\Sigma$, respectively. A steeper decrease of the density is seen beyond $R=13$ kpc, which is consistent with the sharper drop identified by \cite{2024NatAs...8.1302L} beyond $R=14$ kpc. Between R$\sim3.5$ to 6.2 kpc, the two density profiles show a slight decrease towards the Galactic center, and $\ln\,\Sigma$ keeps almost constant as a flat plateau. Inside $R$=3.5 kpc, this flat plateau disappears and $\ln\,\Sigma$ increases sharply towards the Galactic center. It is likely caused by the inclusion of the Galactic bar/bulge stars in the fitting considering the bar's length ($3<R_\mathrm{b}<5\,\mathrm{kpc}$). 

There are three main break radius ($R_\mathrm{br}$) in $\ln\,\Sigma$: (1) the first at $R=3.5$ kpc which is also corresponds to the point where $h_\mathrm{thin}$ ceases its increasing tendency towards the Galactic center; (2) the second at $R\sim6.2$ kpc, near the location where $h_\mathrm{thin}$ begins to increase towards the Galactic center; (3) the third at $R\sim13$ kpc, beyond which the stellar disk presents a down-bending density profile. However, we need to note that our completeness drops sharply beyond $G=15.5$ in Figure~\ref{fig:thirdfactor}. Although selection function is corrected, the low completeness at faint stars may still introduce some uncertainties in the derived density profile of the outer disk. Another interesting thing is the density bump between $R\sim5$ to 7 kpc. The location of this density bump is similar to a slight bump in the observational data at around 6 kpc found by \citet{2025A&A...701A.270K}, which will be further discussed in the following section.  

\subsection{Stellar density profile with varied azimuthal angle}\label{sub:variationphi}              
Our obtained break radius at $R=6.2$ kpc is a little smaller than results from previous studies at around $R\sim7-9$ kpc, which might be caused by the different spatial distribution in these stellar samples. \cite{2024MNRAS.531..495T} found that the surface density profile changes evidently with the azimuthal angle in the solar neighborhood. A similar asymmetry may also exist in the innermost stellar disk considering its complex environment introduced by the Galactic bulge and/or bar. Exploring the variation of the density profile with $\phi$ will also help determine whether the observed bump is a local feature or a global characteristic of the stellar disk. 

We divide the RC sample into six bins of $-60^\circ<\phi\leq-30^\circ, -30^\circ<\phi\leq-15^\circ, -15^\circ<\phi\leq0^\circ, 0^\circ<\phi\leq15^\circ, 15<\phi\leq30$, and $30^\circ<\phi\leq60^\circ$. The Galactic bar is expected to have a different influence on these parts due to its approximately elliptical density shape. Because the major axes of the Galactic bar is located at $\phi\sim25^\circ$, the bin of $30^\circ<\phi\leq15^\circ$ is more likely to be heavily influenced. For each bin, the division of $R$ and the fitting process are the same with that we mentioned in section~\ref{sub:variationR}. Figure~\ref{fig:allfitting} shows the fitting results of several $R$ bins from $R=$ 2 to 15 kpc for all parts. In general, the two disk model works well in describing $\nu(Z|R)$ except for several outermost bins where the stellar disk is heavily distorted by the warp structure.

In Figure~\ref{fig:thindiskhz} and ~\ref{fig:thickiskhz}, we can see the general pattern of the scale heights changes little with $\phi$. For the thin disk, $h_\mathrm{thin}$ exhibits a typical flare feature beyond $R\sim6.4$ kpc, and also begins to increase towards the Galactic center inside $R\sim6.4$ kpc. This increasing tendency ceases at larger $R$ in $\phi$ bins closer to the major axes of the Galactic bar. As shown in Figure~\ref{fig:f0}, the overall behavior of $f_0$ is similar across all $\phi$ bins, including large uncertainties seen at several bins of $R>12$ kpc. A down warp is clearly shown in Figure~\ref{fig:Z0} for parts with $\phi<0^\circ$ at $R>10$ kpc. For parts with $0^\circ<\phi$, we can see signs of an upward warp at $8<R<12$ kpc. However, at $R>12$ kpc, the upward warp disappears and even changes into a downward form at $0^\circ<\phi<30^\circ$. For the outer disk ($R>13$ kpc) in bins of $\phi>15^\circ$ in Figure~\ref{fig:allfitting}, we can see a deficiency and bad fit of the data points near the Galactic plane. Therefore, this change is more likely a result of data deficiency rather than a reflection of the actual situation.  
 
\begin{figure*}
\centering
\includegraphics[width = \textwidth]{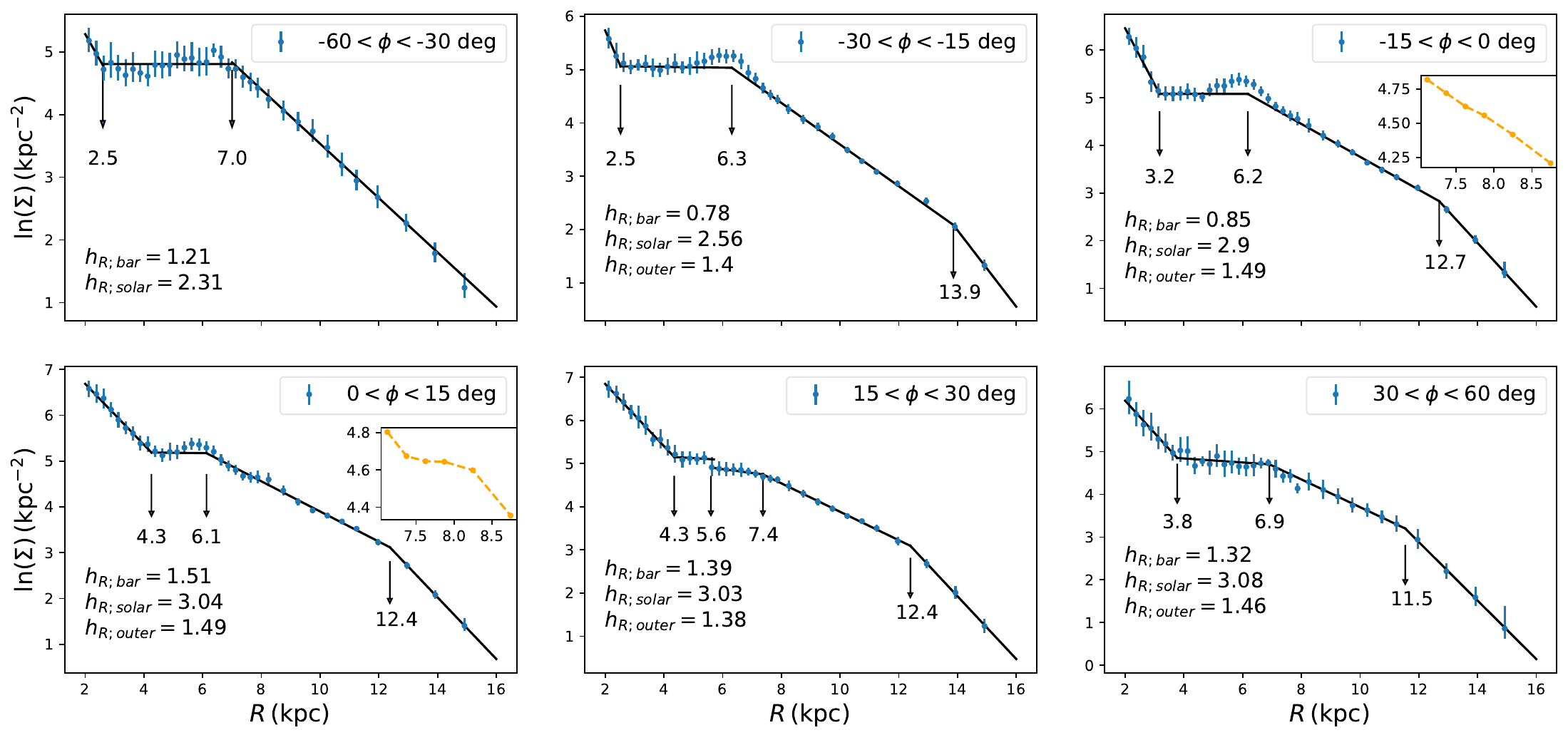}
\caption{Surface density profile $\ln\,\Sigma$ as a function of $R$ for six different bins of $\phi$. Except for the bin of $-60^\circ<\phi<-30^\circ$, the conclusion that $\ln\,\Sigma$ is composed of four main components remains unchanged, but the specific parameters exhibit some degree of dependence on the azimuthal angle. The density bump at $5<R<7$ kpc is a local feature that mainly exists in the bins of $-15^\circ<\phi<0^\circ$ and $-30^\circ<\phi<-15^\circ$. An additional density break is found at $R=5.6$ kpc in the bin of $15^\circ<\phi<30^\circ$, which splits the second component into two different flat plateaus at $4.3<R<5.6$ kpc and $5.6<R<7$ kpc. For the bin of $0^\circ<\phi<15^\circ$, there is an additional flat plateau between R=7.5 to 8.5 kpc as shown in the zoomed-in figure. Our results show that the stellar disk is not only radially broken but also exhibits significant azimuthal variation.}
\label{fig:surfaceplane}
\end{figure*}

Figure~\ref{fig:surfaceplane} presents the change of $\ln\,\Sigma$ as a function of $R$ for all parts. Overall, $\ln\,\Sigma$ can be well described by four components, but the parameters (scale length, break radius, initial and end radii of the flat plateau) associated with these components vary to some extent with $\phi$. The first $R_\mathrm{br}$, which corresponds to the Galactic radius that the bar's influence cannot be neglected inside it, changes significantly with $\phi$. In the bin of $15^\circ < \phi < 30^\circ$, where the bar's major axis (at $\sim25^\circ$)is located, the first $R_\mathrm{br}$ appears at $\sim4.3$ kpc. This is consistent with the thin bar length $R_\mathrm{b}=4.6\pm0.3$ kpc derived from the stellar count fitting in \citet{2015MNRAS.450.4050W}, but slightly larger than the recently reported dynamical bar length of $R_\mathrm{b}\sim3.5$ kpc \citep{2023MNRAS.520.4779L,2024MNRAS.528.3576V}. The variation of the first $R_\mathrm{br}$ with $\phi$ can be well fit by an ellipsoid with a similar orientation angle (25 deg with respect to the Sun-Galactic center line) and extent (semi-major axes a=4.3 kpc and semi-minor axes b=2 kpc) as the Galactic bar. The second $R_\mathrm{br}$, which is the ending radius of the flat plateau, varies from around 6 to 7.5 kpc in different bins of $\phi$. The change of the second $R_\mathrm{br}$ indicates that the inner disk is broken and asymmetric. In the part of $15^\circ<\phi<30^\circ$, we find an additional break point at $R=5.6$ kpc, and $\ln\,\Sigma$ exhibits two different flat plateaus at $R=4.3-5.6$ kpc and $R=5.6-7.0$ kpc respectively. 

Beyond the inner disk, we find a slight dependence of its scale length $h_{R;\mathrm{solar}}$ on $\phi$ in the third component. For the four parts of $\phi>-15^\circ$, the values of $h_{R;\mathrm{solar}}$ show minimal variation and remain nearly constant at approximately 2.9 kpc. After that, $h_{R;\mathrm{solar}}$ shows a gradual decrease as $\phi$ decreases, reaching values of 2.52 kpc and 2.30 kpc in the part of $-30^\circ<\phi<-15^\circ$ and $-60^\circ<\phi<-30^\circ$, respectively. Previous studies also found some dependence of the scale length on $\phi$, but their results are not completely consistent \citep{2020A&A...637A..96C,2022A&A...664A..58C,2024MNRAS.527.4863U,2024RAA....24f5005L}. The nearly constant $h_{R;\mathrm{solar}}$ at the three parts of $\phi>0^\circ$ agrees with \citet{2024MNRAS.527.4863U}, while the smaller $h_{R;\mathrm{solar}}$ at the two parts of $\phi<-15^\circ$ is consistent with \citet{2020A&A...637A..96C}, in which their scale length reaches the lowest value at $\phi=-30^\circ$ among $-30^\circ,\,0^\circ\,\mathrm{and}\,30^\circ$. In the part of $0^\circ<\phi<15^\circ$, we find an additional flat plateau between R=7.5 to 8.5 kpc. As a comparison, in the part of $-15^\circ<\phi<0^\circ$, the density profile follows a typical exponential without any additional flat plateau from R=7 to 13 kpc. This is consistent with a recent study of the surface density profile in the solar neighborhood \citep{2024MNRAS.531..495T}, in which their obtained $\ln\,\Sigma$ changes little from $R=7.6$ to 8.4 kpc at $\phi=2.5^\circ$ and $5^\circ$. The third $R_\mathrm{br}$ shows a gradual and continuous decrease with $\phi$. It declines slowly from 13.8 kpc at $-30^\circ<\phi<-15^\circ$ to around 11.7 kpc at $<30^\circ<\phi<60^\circ$. Except for the bin of $60^\circ<\phi<30^\circ$, the rest five bins exhibit a fourth component that supports a down-bending form in the stellar density profile of the outer disk. The scale length $h_{R;\mathrm{outer}}$ of the fourth component presents no dependence on $\phi$ and keeps almost constant as around 1.50 kpc.

In Figure~\ref{fig:surfaceplane}, we can see that the density bump is a local feature that mainly exists in the two bins of $-15^\circ<\phi<0^\circ$ and $-30^\circ<\phi<-15^\circ$. The density bump is also visible but less prominent at $5<R<6.5$ kpc in the bin of $0^\circ<\phi<15^\circ$. By analyzing the simulated Milky Way, \citet{2025MNRAS.537.1620D} find that the azimuthal variation of the mean metallicity closely follows the surface density variation, especially in these radii ($5.75\leq R\leq10.25$ kpc) where the spiral arms dominant. Inspired by their results, Figure~\ref{fig:Metaldistribution} shows the mean metallicity $\langle \mathrm{[M/H]}\rangle$ in the $X-Y$ diagram for $|\phi|<60^\circ$. We can clearly see a metallicity bump at around $-7<X<-5$ kpc and $-4<Y<2$ kpc, which is consistent with the location of the density bump. A similar metallicity bump is also seen in the chemical cartography of the stellar disk obtained from the Gaia RVS spectroscopy survey \citep{2023A&A...674A...1G}. The azimuthal metallicity map of the stellar disk in \citet{2025A&A...693A...3B} also shows an excess of [M/H] around $R\sim6$ kpc at $-20^\circ<\phi<10^\circ$ for both young and old stellar populations.

By constraining the vertical heights $|Z|$ of our RC sample, we find that the metallicity bump appears most clearly near the Galactic plane of $|Z|<0.15$ kpc. Therefore, we removed the data points at $|Z|<0.15$ kpc in Figure~\ref{fig:allfitting} and repeated the fitting of the vertical density profile. Figure~\ref{fig:surfaceplane_z02} shows the variation of $\ln\,\Sigma$ after excluding data points of $|Z|<0.15$ kpc. We can clearly see that the density bump disappears, along with the additional break at $R\sim4.3$ kpc in the bin of $15^\circ<\phi<30^\circ$. The density bump is possibly introduced by an over-accumulation of super metal-rich $\mathrm{[M/H]>0.1}$ stars near the Galactic plane. Using high-resolution $N$-body simulations, recent study reveals that the spiral perturbation can introduce azimuthal [M/H] variations with peaks at the locations of the density peaks for both young and old stellar populations \citep{2018A&A...611L...2K,2025MNRAS.537.1620D}. The coincidence of the metallicity and density bump found in this study may suggest a possible relationship with the spiral perturbation. However, we need to note that not all $R$ bins within the range of $5<R<7$ kpc have a good measurement of $\nu$ inside $|Z|=0.15$ kpc, especially for the bins of $15^\circ<\phi<30^\circ$ and $30^\circ<\phi<60^\circ$. For these bins, although the absence of a metallicity bump is clear in Figure~\ref{fig:Metaldistribution}, we still can not rule out the possible existence of a density bump.  

The removal of low $|Z|$ data points makes the obtained second $R_\mathrm{br}$ a little larger than before and more consistent with previous studies. $h_\mathrm{thin}$ at $R\sim6$ kpc also increases from 0.16 to around 0.18 for the two $\phi$ bins influenced by the density bump, while $\ln\,\Sigma$ beyond the solar radius changes little with low $|Z|$ data points excluded. Our results indicate that the density profile of the inner disk is complicated, and an accurate understanding of it relies more on the near-plane stars compared to the density profile of the outer disk. 

\section{Discussion: density bump and bimodal distribution in the guiding radius for super metal rich stars}\label{sec:discussion}    
Recently, \citet{2024A&A...681L...8N} report a bimodal spatial distribution of super metal-rich stars ([M/H]$>$0.1) in the solar neighborhood, where their guiding radius ($R_g$) distribution show two peaks at 6.9 and 7.9 kpc, respectively. They interpret the observed bimodality as possible imprints of the Galactic bar. Notably, the inner $R_g$ peak at 6.9 kpc is very close to the outer edge of the density bump, and they are both related to the super metal rich stars. 

The radial migration, specifically 'churning', can increase/decrease a star's guiding radius by changing its angular momentum due to co-rotation resonances, but still maintain its circular orbit after migrating inwards or outwards \citep{2002MNRAS.336..785S,2009MNRAS.396..203S}. Since the position of the density bump is near the co-rotation radius of the bar, it is possible that stars inside it could be heavily influenced by churning. Therefore, although the density bump is nearly invisible beyond $R = 6.9$ kpc, radial migration may cause several stars originating from this bump to be redistributed to outer regions, eventually becoming concentrated around $R_g \sim 6.9$ kpc.        

The density bump found in this study is a localized feature that mainly exists in $-30^\circ<\phi<0^\circ$. The radial migration mainly changes $R_g$, while the azimuthal angle is less influenced by churning. If the inner $R_g$ peak is partly related to the density bump, it might also present a similar localized feature in the azimuthal angle. We cross-matched our RC sample with the Gaia RVS catalog and obtained around 4.1 million sources with a measurement in the radial velocity. We use the conventions from Python package astropy \citep{2018AJ....156..123A}, adopting a solar motion of $(+12.9, +245.6,
+7.78)\,\mathrm{km\,s^{-1}}$ \citep{2004AJ....128..502A,2018A&A...615L..15G}. The guiding radius $R_g$is defined as $L_z/V_0$, where $L_z$ is the star's angular momentum and $V_0=235\,\mathrm{km\,s^{-1}}$ is the circular velocity of local standard of rest (LSR) \citep{2015ApJS..216...29B}.

\begin{figure*}
\centering
\includegraphics[width = \textwidth]{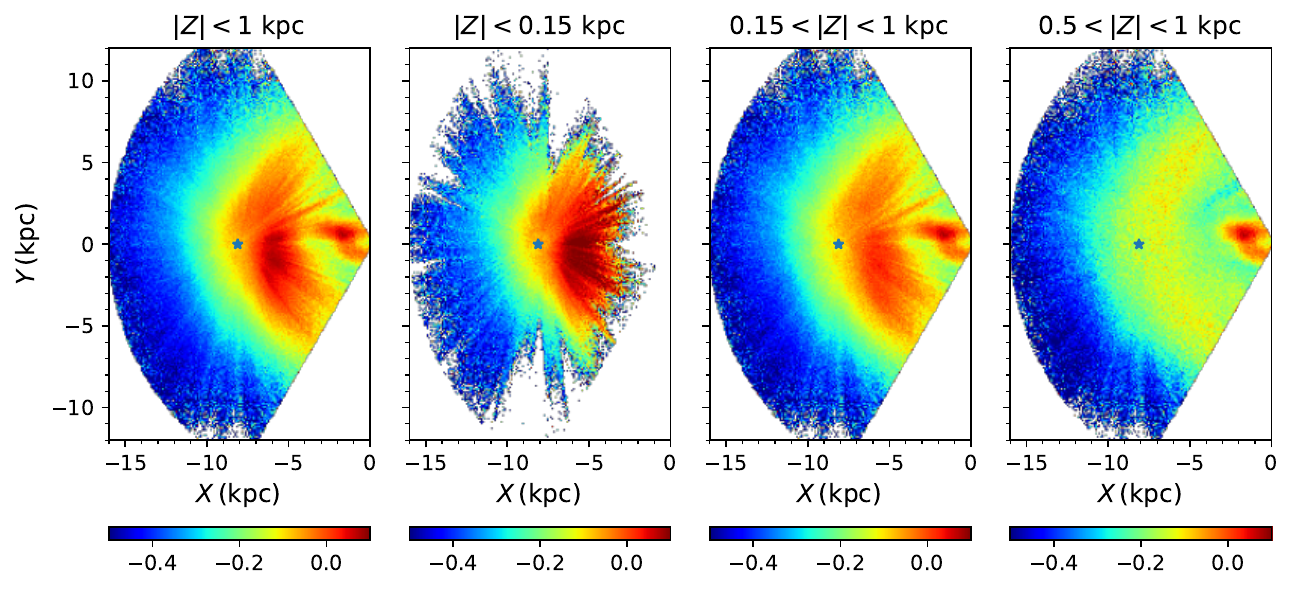}
\caption{Distributions of mean metallicity $\langle\mathrm{[M/H]}\rangle$ in the $X-Y$ diagram for RCs with $|Z|<1$ kpc, $|Z|<0.15$ kpc, $0.15<|Z|<1$ kpc, and $0.5<|Z|<1$ kpc (in each row, respectively). We can notice a metallicity bump at around $-7<X<-5$ kpc and $-4<Y<2$ kpc in the left two rows which include near-plane stars. The position of this metallicity bump is consistent with the location of the density bump shown in Figure~\ref{fig:surfaceplane}.}
\label{fig:Metaldistribution}
\end{figure*}

\begin{figure*}
\centering
\includegraphics[width = \textwidth]{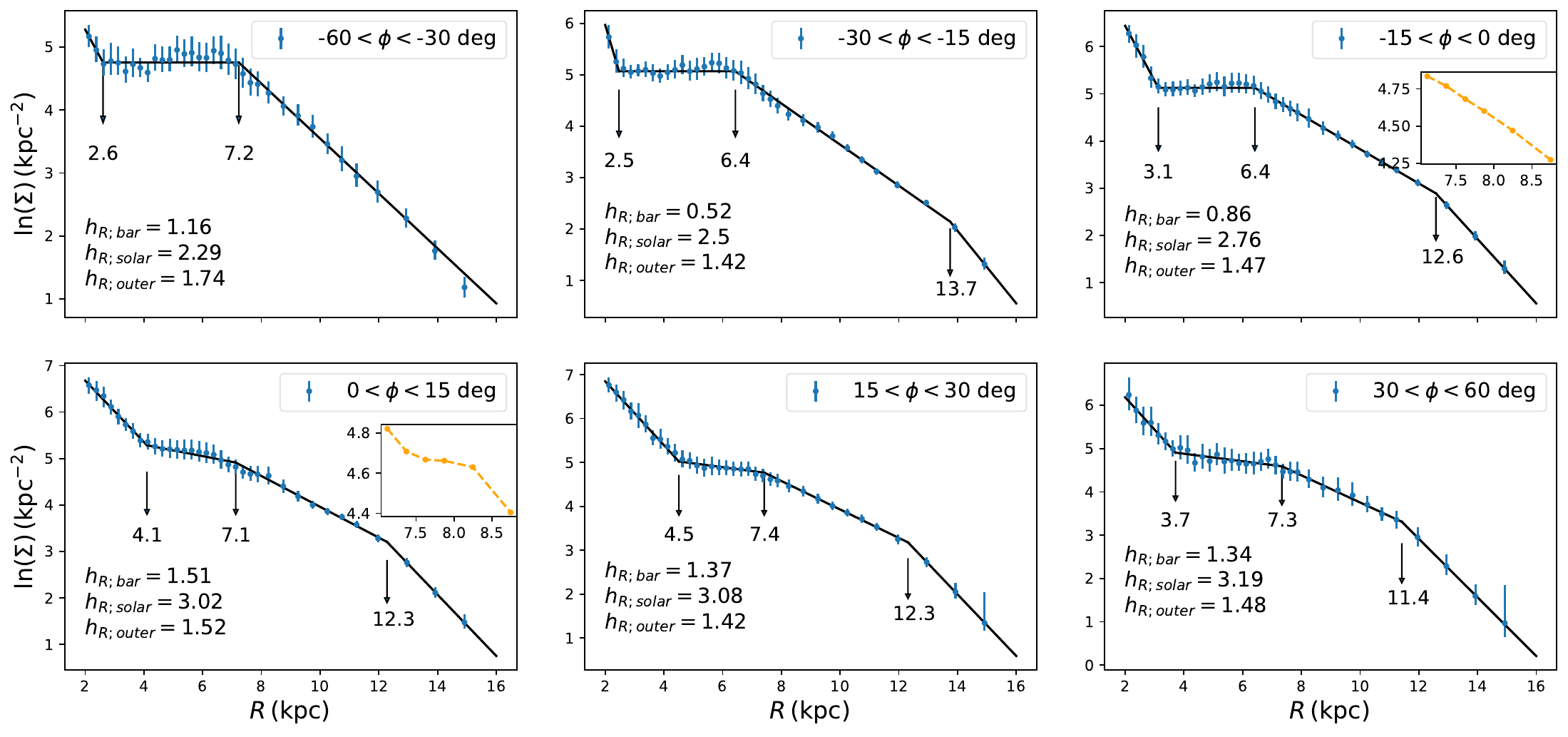}
\caption{Surface density profile $\ln\,\Sigma$ as a function of $R$ for six different bins of $\phi$ after removing data points of $|Z|<0.15$ kpc during the fitting process. The density bump and the additional break mentioned in Figure~\ref{fig:surfaceplane} disappear, which means that these two local features are mainly caused by near-plane stars like the metallicity bump in Figure~\ref{fig:Metaldistribution}.}
\label{fig:surfaceplane_z02}
\end{figure*}

\begin{figure*}
\centering
\includegraphics[width = \textwidth]{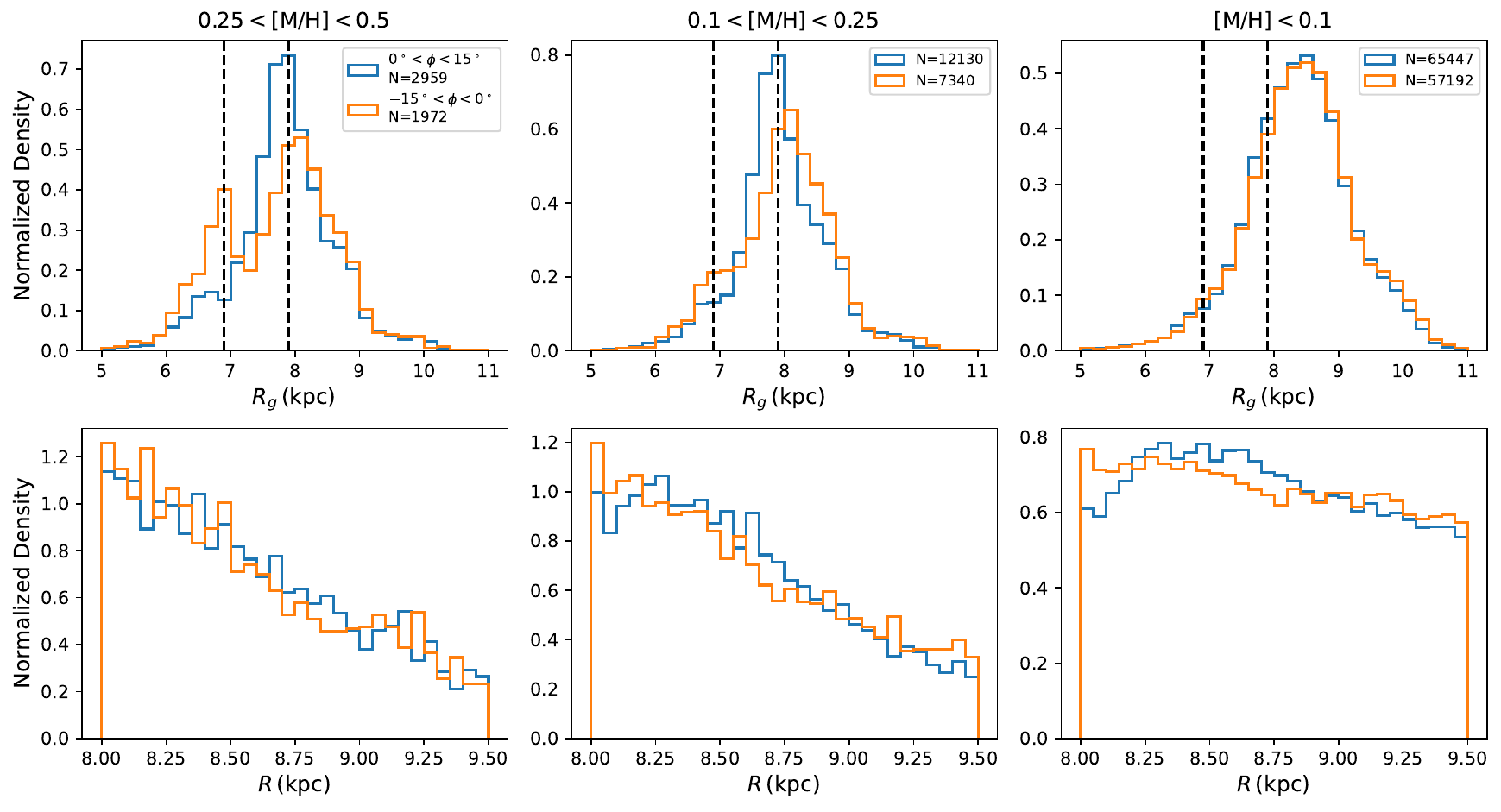}
\caption{Normalized density distributions of $R_g$ (top panel) and $R$ (bottom panel) for stars of $\mathrm{0.25<[M/H]<0.5}$ (left panel), $\mathrm{0.1<[M/H]<0.25}$ (middle panel), and $\mathrm{[M/H]<0.1}$ (right panel) in two different bins of $0^\circ<\phi<15^\circ$ (blue hist) and $-15^\circ<\phi<0^\circ$ (orange hist). Two vertical dashed lines represent the two $R_g$ peaks at 6.9 and 8.1 kpc found by \citet{2024A&A...681L...8N}, respectively. The bimodal $R_g$ distribution only exists in super metal-rich stars of $\mathrm{[M/H]}>0.1$ and become much less pronounced in the bin of $0^\circ<\phi<15^\circ$. The consistency in the $R$ distributions means that stars in these two bins are likely affected by similar selection effects, indicating that the bimodal $R_g$ distribution is an intrinsic feature rather than a result of artificial effects.}
\label{fig:guidingradius}
\end{figure*}

Following \citet{2024A&A...681L...8N}, we constrained the RC sample to the solar radius by requiring $8<R<9.5$ kpc. We only kept stars with $|Z|<0.15$ kpc because the metallicity/density bump is dominated by stars near the plane. In \citet{2024A&A...681L...8N}, the bimodal distribution first appears at [M/H]$\sim0.1$ and becomes very significant at [M/H]$>0.25$. Therefore, we divided the RC sample into three different types of $\mathrm{0.25<[M/H]<0.5}$, $\mathrm{0.1<[M/H]<0.25}$, and $\mathrm{[M/H]<0.1}$. In Figure~\ref{fig:guidingradius}, we show the normalized distributions of $R_g$ for these three types of stars in bins of $0^\circ<\phi<15^\circ$ and $-15^\circ<\phi<0^\circ$, respectively. The bimodality is particularly evident in the super metal-rich group within the range of $-15^\circ<\phi<0^\circ$, where the density bump is also apparent. In the bin of $0^\circ<\phi<15^\circ$, an inner $R_g$ peak is still visible, but its amplitude is much weaker than in the bin of $0-15^\circ<\phi<0^\circ$. This trend is consistent with the behavior of the density bump, which also becomes much less apparent at $\phi>0^\circ$. In agreement with \citet{2024A&A...681L...8N}, the inner $R_g$ peak is less prominent for stars in the range of $\mathrm{0.1<[M/H]<0.25}$ and disappears at $\mathrm{[M/H]<0.1}$. The $R_\mathrm{g}$ distribution of $\mathrm{[M/H]<0.1}$ is almost the same for both bins of $-15^\circ<\phi<0^\circ$ and $0^\circ<\phi<15^\circ$. If we interpret the inner $R_g$ peak as the accumulation of stars migrating outside from the density bump, it is natural that there are no obvious difference in $\mathrm{[M/H]<0.1}$ since the density bump is mainly related to super metal-rich stars of $\mathrm{[M/H]}>0.1$ as shown in Figure~\ref{fig:Metaldistribution}.

\begin{figure*}
\centering
\includegraphics[width = \textwidth]{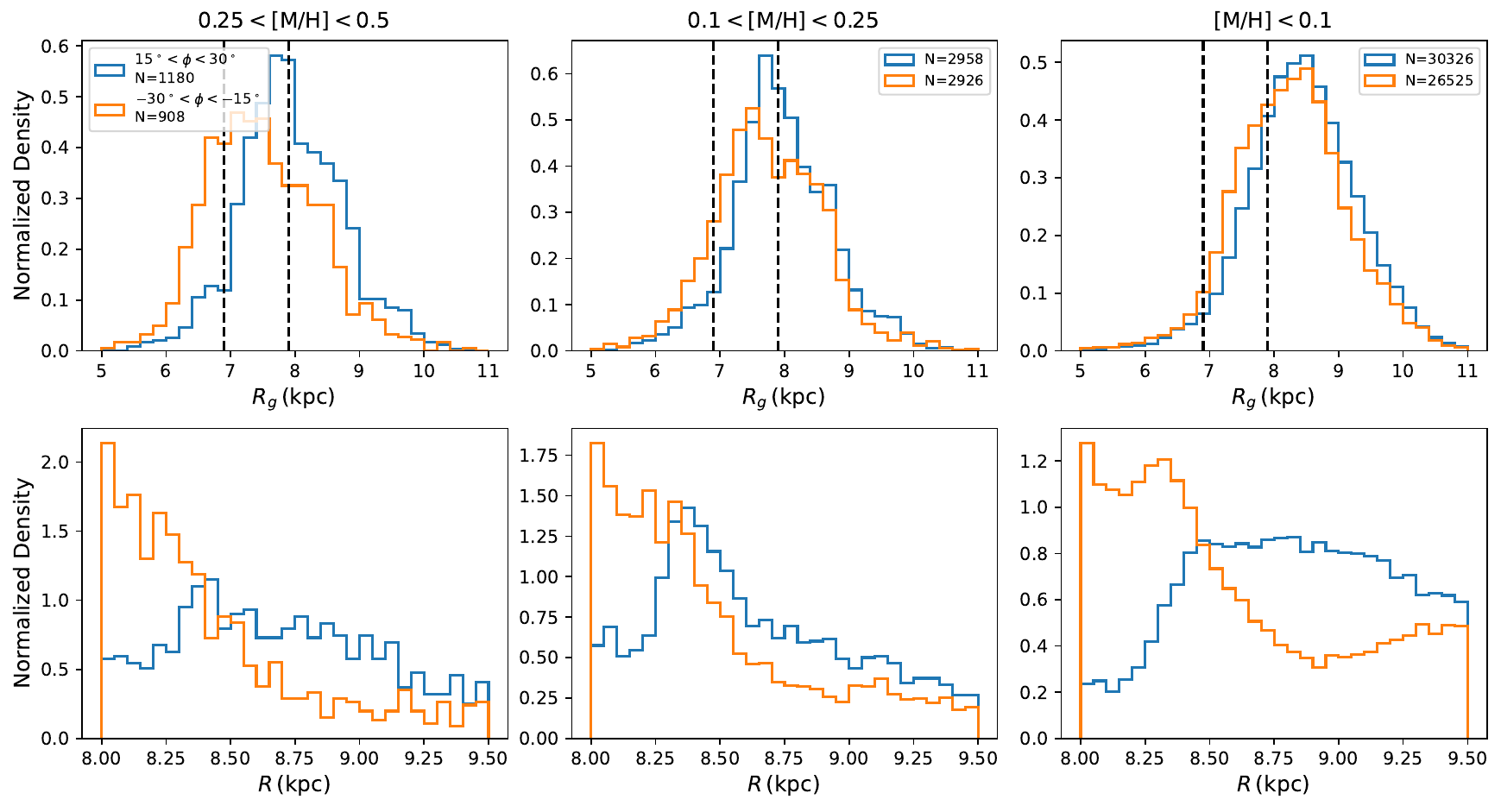}
\caption{Normalized density distributions of $R_g$ (top panel) and $R$ (bottom panel) for stars with $\mathrm{0.25<[M/H]<0.5}$ (left panel), $\mathrm{0.1<[M/H]<0.25}$ (middle panel), and $\mathrm{[M/H]<0.1}$ (right panel) in two different bins of $15^\circ<\phi<30^\circ$ (blue hist) and $-30^\circ<\phi<-15^\circ$ (orange hist). The difference in the $R$ distributions suggests that the selection effect is different in these two bins.}
\label{fig:guidingradius_15to30}
\end{figure*}

In Figure~\ref{fig:guidingradius_15to30}, we show the distribution of $R_g$ for another two bins of $15^\circ<\phi<30^\circ$ and $-30^\circ<\phi<-15^\circ$. Unlike Figure~\ref{fig:guidingradius}, the different distribution in $R$ suggests that stars in these two bins suffer from the influence of the selection function heavily and differently. In the bin $-30^\circ<\phi<-15^\circ$, the inner $R_g$ peak becomes so prominent that the bimodal distribution is much less obvious than in the bin $-15^\circ<\phi<0^\circ$. This prominence might be caused by the selection effect as we can see stars in $-30^\circ<\phi<-15^\circ$ are much more concentrated to $R<8.5$ kpc. Another possibility is that the metallicity/density bump is more evident in $-30^\circ<\phi<-15^\circ$ than in $-15^\circ<\phi<0^\circ$, causing a larger number of super metal-rich stars brought outside by radial migration. Although the selection effect works differently in the two $\phi$ bins of Figure~\ref{fig:guidingradius_15to30}, the $R_g$ distribution for stars of $\mathrm{[M/H]}<0.1$ only shows minimal variations, which indicates that the large difference in $\mathrm{0.25<[M/H]<0.5}$ is more likely an intrinsic feature related to the metallicity. In Figure~\ref{fig:allfitting}, we can see that not all $R$ bins within the range of $5<R<7$ kpc have a measurement of density inside $|Z|<0.15$ kpc, especially for bins of $\phi>15^\circ$. If we attribute the bimodal $R_g$ distribution to stars migrated outwards from the density bump, the lack of an inner $R_g$ peak will in turn support the absence of a density bump in $15^\circ<\phi<30^\circ$. 

A similar azimuthal variation is also seen in the local Herculus moving group, where its strength increases steadily towards the minor axis of the bar within $-15^\circ<\phi<15^\circ$ \citep{2024MNRAS.531L..14L}. This is expected if we assume a bar's co-rotation resonance origin for the Herculus \citep{2020ApJ...890..117D}. In this assumption, the Herculus members are trapped around the Galactic bar’s L4/L5 Lagrange points, and the orbits of these trapped stars are called Trojan orbits \citep{2020ApJ...890..117D,2025MNRAS.539.1595L}. A recent study shows a high fraction of super metal-rich stars in the Herculus, and this fraction becomes higher at smaller Galactocentric radius \citep{2025MNRAS.536..498L}. The location of the density bump is covered by the Trojan family orbits assuming a long and slow bar \citep[see in Figure 4 of ][]{2025MNRAS.539.1595L}. Therefore, a fraction of the super metal-rich Herculus members in the solar vicinity may come from the density bump through the co-rotation resonance, which gives some contributions to the its metal rich morphology.    

The connection between the density bump and the bimodal $R_g$ distribution may help us understand the effect of radial migration in reshaping the Galactic disk. However, a further study of it requires more information like stellar ages and a quantitative analysis of the radial migration, which is beyond the scope of this study.

\section{summary}\label{sec:summary}
In this study, we construct a comprehensive and full-sky covered catalog comprising approximately 8.4 million RC stars selected from the Gaia XP spectra. The selection is based on the $T_\mathrm{eff}-\log g-\mathrm{[M/H]}$ relationship and utilizes the previously identified RC sample of \citet{2018ApJ...858L...7T} as a reference. After correcting the selection function, we divide these RCs into different bins of R ranging from 2 to 15 kpc, and fit the vertical density profile with a two-component model consisting of a geometric thin and thick disk. Our results are summarized as follows:

1. The scale height $h_\mathrm{thin}$ of the thin disk presents a V-shaped pattern flaring both inside ($R<6.4$ kpc) and outside ($R>6.4$ kpc). The gradual increase in $h_\mathrm{thin}$ toward the Galactic center stops at $R\sim3$ kpc, where the bar's stars become important in the density fitting. For the inner flare, we speculate that the dynamical heating from the Galactic bar may scatter the near-plane stars to higher $|Z|$ positions and thus increase the disk thickness. For the thick disk, $h_\mathrm{thick}$ increases steadily from 0.45 at $R=2.5$ kpc to around 0.75 in the solar neighborhood.       

2. The radial structure of the mid-plane $\ln\,\nu_0$ and the surface $\ln\,\Sigma$ density profiles are composed of four components: a nearly exponential feature inside $R=3.5$ kpc, a nearly flat plateau at $3.5\leq R<6.2$ kpc, a typical exponential with $h_{R;\mathrm{inner}}=2.14$ kpc for $\ln\,\nu_0$ and $h_{R;\mathrm{inner}}=2.85$ kpc for $\ln\,\Sigma$ at $6.2\leq R<12.8$ kpc, and another steeper exponential with $h_{R;\mathrm{outer}}=1.05$ kpc for $\nu_0(R)$ and $h_{R;\mathrm{outer}}=1.45$ kpc for $\Sigma(R)$ at $13\leq R\leq15$ kpc.

3. By dividing the RC samples into different bins of azimuthal angle $\phi$, we find that the radial density profiles continue to exhibit a similar four components but demonstrate a clear dependence on $\phi$. The variation of the first break radius $R_\mathrm{br}$ is consistent with an elliptical bar model with a major axis length of 4.3 kpc and a minor axis length of 2.2 kpc. The second $R_\mathrm{br}$, which is the ending radii of the flat plateau, varies across different $\phi$ bins from 6 to 7 kpc. The scale length of the third component has a slight dependence on $\phi$ at $\phi<0^\circ$. An additional flat plateau between $R=$7.4 to 8.5 kpc is found near the solar neighborhood in the bin of $0^\circ<\phi<15^\circ$. In contrast to the third one, the fourth component shows greater symmetry, with its scale length $h_{R;\mathrm{outer}}$ only exhibiting minimal variations across all $\phi$ bins.

4. We find a density bump at $5<R<7$ kpc in $\phi$ bins of $-15^\circ<\phi<0^\circ$ and $-30^\circ<\phi<-15^\circ$. In the position of this density bump, we also identify a metallicity bump caused by a large amount of super metal-rich stars near the Galactic plane. The density bump disappears after removing data points of $|Z| < 0.15$ kpc from the fitting, further supporting its connection to stars near the Galactic plane. The consistency between the metallicity and the density bumps may suggest a possible origin of spiral perturbation according to previous studies \citep{2018A&A...611L...2K,2025MNRAS.537.1620D}. 

5. By analyzing the $R_g$ distribution of stars in the solar radius, the recent found inner $R_g$ peak in super metal-rich stars exhibits a similar characteristic in the azimuthal distribution to the density bump, which suggests that this peak might be related to stars migrated outwards from the density bump.           

Using RCs selected from the Gaia XP spectra, this study explores the variation of the density profile with $R$ and $\phi$ for the Galactic stellar disk. The large volume and wide spatial distribution of Gaia XP spectra enable us to explore the density profile of the stellar disk from the Galactic center to large Galactocentric distance. Our results reveal that the stellar disk is radially broken and azimuthally asymmetric. Compared to the outer disk, the inner disk is more asymmetric and exhibits several local features connected to stars near the Galactic plane. Although Gaia XP spectra span a large volume and full sky coverage, it is hard for us to extract a complete chemical abundance pattern and radial velocity from them due to the low-resolution and wiggles. The ongoing and upcoming data release of large spectroscopic surveys such as SDSS-\uppercase\expandafter{\romannumeral5} \citep{2025arXiv250706989K}, Gaia DR4 XP spectra \citep{2016A&A...595A...1G}, WEAVE \citep{2024MNRAS.530.2688J}, DESI\citep{2025arXiv250314745D}, and 4MOST \citep{2019Msngr.175....3D} will provide more extensive coverage of the inner Galaxy with higher quality spectroscopic data. Compared to the low-resolution Gaia DR3 XP spectra, these improved observations may bring a more complete and cleaner RC sample, which will allow us to construct a more clear picture of the inner disk morphology and the perturbations driven by bars or spirals. Our future work will focus on investigating the bar’s influence on the inner disk morphology, which may help explain the observed V-shaped pattern in the scale height of the geometric thin disk. 

\begin{acknowledgements}
This study is supported by the National Key R\&D Program of China under grant Nos. 2023YFE0107800, 2024YFA1611900, and the National Natural Science Foundation of China under grant Nos. 12588202, 12273055, 12373020. We thank the anonymous reviewer for the helpful comments and suggestions. This study is also supported by International Partnership Program of Chinese Academy of Sciences. Grant No. 178GJHZ2022040GC, and CAS Project for Young Scientists in Basic Research grant No. YSBR-062 and YSBR-092. We also thank the support by the China Manned Space Program with grant nos. CMS-CSST-2025-A11 and CMS-CSST-2025-A12. Xianhao Ye and Wenbo Wu acknowledge the support from the China Scholarship Council. DA, JIGH, and RRL acknowledge financial support from the Spanish Ministry of Science, Innovation and Universities (MICIU) projects PID2020-117493GB-I00 and PID2023-149982NB-I00. C. Allende Prieto acknowledges financial support from the
Spanish Ministry of Science, Innovation and Universities (MICIU
projects PID2020-117493GB-I00, PID2023-149982NB-I00 and
PID2023-146453NB-I00 (\textit{PLAtoSOnG}). MLC’s research is supported by the grant PID2021-129031NB-I00 of the Spanish Ministerio de Ciencia, Innovación y Universidades (MICINN). JL acknowledges support from National Natural Science Foundation of China (No. 12473021) and National Key R\&D Program of China (No. 2024YFA1611600). XY acknowledges support from National Natural Science Foundation of China (No. 12503024). This research made use of computing time available on the high-performance computing systems at the Instituto de Astrofisica de Canarias. This work presents results from the European Space Agency (ESA) space mission Gaia. Gaia data are being processed by the Gaia Data Processing and Analysis Consortium (DPAC). Funding for the DPAC is provided by national institutions, in particular the institutions participating in the Gaia MultiLateral Agreement (MLA). The Gaia mission website is https://www.cosmos.esa.int/gaia. The Gaia archive website is https://archives.esac.esa.int/gaia. This job has made use of the Python package GaiaXPy, developed and maintained by members of the Gaia Data Processing and Analysis Consortium (DPAC), and in particular, Coordination Unit 5 (CU5), and the Data Processing Centre located at the Institute of Astronomy, Cambridge, UK (DPCI). This publication makes use of data products from the Two Micron All Sky Survey, which is a joint project of the University of Massachusetts and the Infrared Processing and Analysis Center/California Institute of Technology, funded by the National Aeronautics and Space Administration and the National Science Foundation.

\end{acknowledgements} 

\bibliography{sample701}{}

@ARTICLE{2024ApJS..272....2L,
       author = {{Li}, Jiadong and {Wong}, Kaze W.~K. and {Hogg}, David W. and {Rix}, Hans-Walter and {Chandra}, Vedant},
        title = "{AspGap: Augmented Stellar Parameters and Abundances for 37 Million Red Giant Branch Stars from Gaia XP Low-resolution Spectra}",
      journal = {\apjs},
         year = 2024,
        month = may,
       volume = {272},
       number = {1},
          eid = {2},
        pages = {2},
          doi = {10.3847/1538-4365/ad2b4d},
archivePrefix = {arXiv},
       eprint = {2309.14294},
 primaryClass = {astro-ph.SR},
       adsurl = {https://ui.adsabs.harvard.edu/abs/2024ApJS..272....2L}
}

@ARTICLE{2014ApJ...790..127B,
       author = {{Bovy}, Jo and {Nidever}, David L. and {Rix}, Hans-Walter and {Girardi}, L{\'e}o and {Zasowski}, Gail and {Chojnowski}, S. Drew and {Holtzman}, Jon and {Epstein}, Courtney and {Frinchaboy}, Peter M. and {Hayden}, Michael R. and {Rodrigues}, Tha{\'\i}se S. and {Majewski}, Steven R. and {Johnson}, Jennifer A. and {Pinsonneault}, Marc H. and {Stello}, Dennis and {Allende Prieto}, Carlos and {Andrews}, Brett and {Basu}, Sarbani and {Beers}, Timothy C. and {Bizyaev}, Dmitry and {Burton}, Adam and {Chaplin}, William J. and {Cunha}, Katia and {Elsworth}, Yvonne and {Garc{\'\i}a}, Rafael A. and {Garc{\'\i}a-Her{\'n}andez}, Domingo A. and {Garc{\'\i}a P{\'e}rez}, Ana E. and {Hearty}, Fred R. and {Hekker}, Saskia and {Kallinger}, Thomas and {Kinemuchi}, Karen and {Koesterke}, Lars and {M{\'e}sz{\'a}ros}, Szabolcs and {Mosser}, Beno{\^\i}t and {O'Connell}, Robert W. and {Oravetz}, Daniel and {Pan}, Kaike and {Robin}, Annie C. and {Schiavon}, Ricardo P. and {Schneider}, Donald P. and {Schultheis}, Mathias and {Serenelli}, Aldo and {Shetrone}, Matthew and {Silva Aguirre}, Victor and {Simmons}, Audrey and {Skrutskie}, Michael and {Smith}, Verne V. and {Stassun}, Keivan and {Weinberg}, David H. and {Wilson}, John C. and {Zamora}, Olga},
        title = "{The APOGEE Red-clump Catalog: Precise Distances, Velocities, and High-resolution Elemental Abundances over a Large Area of the Milky Way's Disk}",
      journal = {\apj},
         year = 2014,
        month = aug,
       volume = {790},
       number = {2},
          eid = {127},
        pages = {127},
          doi = {10.1088/0004-637X/790/2/127},
archivePrefix = {arXiv},
       eprint = {1405.1032},
 primaryClass = {astro-ph.GA},
       adsurl = {https://ui.adsabs.harvard.edu/abs/2014ApJ...790..127B}
}

@ARTICLE{2022MNRAS.513.4130L,
       author = {{Lian}, Jianhui and {Zasowski}, Gail and {Mackereth}, Ted and {Imig}, Julie and {Holtzman}, Jon A. and {Beaton}, Rachael L. and {Bird}, Jonathan C. and {Cunha}, Katia and {Fern{\'a}ndez-Trincado}, Jos{\'e} G. and {Horta}, Danny and {Lane}, Richard R. and {Masters}, Karen L. and {Nitschelm}, Christian and {Roman-Lopes}, A.},
        title = "{The Milky Way tomography with APOGEE: intrinsic density distribution and structure of mono-abundance populations}",
      journal = {\mnras},
         year = 2022,
        month = jul,
       volume = {513},
       number = {3},
        pages = {4130-4151},
          doi = {10.1093/mnras/stac1151},
archivePrefix = {arXiv},
       eprint = {2204.10327},
 primaryClass = {astro-ph.GA},
       adsurl = {https://ui.adsabs.harvard.edu/abs/2022MNRAS.513.4130L}
}

@ARTICLE{2022MNRAS.512.1710H,
       author = {{He}, Xu-Jiang and {Luo}, A. -Li and {Chen}, Yu-Qin},
        title = "{Identification, mass, and age of primary red clump stars from spectral features derived with the LAMOST DR7}",
      journal = {\mnras},
         year = 2022,
        month = may,
       volume = {512},
       number = {2},
        pages = {1710-1721},
          doi = {10.1093/mnras/stac484},
archivePrefix = {arXiv},
       eprint = {2202.09185},
 primaryClass = {astro-ph.SR},
       adsurl = {https://ui.adsabs.harvard.edu/abs/2022MNRAS.512.1710H}
}

@ARTICLE{2017MNRAS.465.1621P,
	author = {{Portail}, Matthieu and {Gerhard}, Ortwin and {Wegg}, Christopher and {Ness}, Melissa},
	title = "{Dynamical modelling of the galactic bulge and bar: the Milky Way's pattern speed, stellar and dark matter mass distribution}",
	journal = {\mnras},
	year = 2017,
	month = feb,
	volume = {465},
	number = {2},
	pages = {1621-1644},
	doi = {10.1093/mnras/stw2819},
	archivePrefix = {arXiv},
	eprint = {1608.07954},
	primaryClass = {astro-ph.GA},
	adsurl = {https://ui.adsabs.harvard.edu/abs/2017MNRAS.465.1621P}
}

@ARTICLE{2021MNRAS.500.4710C,
	author = {{Chiba}, Rimpei and {Friske}, Jennifer K.~S. and {Sch{\"o}nrich}, Ralph},
	title = "{Resonance sweeping by a decelerating Galactic bar}",
	journal = {\mnras},
	year = 2021,
	month = jan,
	volume = {500},
	number = {4},
	pages = {4710-4729},
	doi = {10.1093/mnras/staa3585},
	archivePrefix = {arXiv},
	eprint = {1912.04304},
	primaryClass = {astro-ph.GA},
	adsurl = {https://ui.adsabs.harvard.edu/abs/2021MNRAS.500.4710C}
}

@ARTICLE{2019A&A...626A..41M,
	author = {{Monari}, G. and {Famaey}, B. and {Siebert}, A. and {Wegg}, C. and {Gerhard}, O.},
	title = "{Signatures of the resonances of a large Galactic bar in local velocity space}",
	journal = {\aap},
	year = 2019,
	month = jun,
	volume = {626},
	eid = {A41},
	pages = {A41},
	doi = {10.1051/0004-6361/201834820},
	archivePrefix = {arXiv},
	eprint = {1812.04151},
	primaryClass = {astro-ph.GA},
	adsurl = {https://ui.adsabs.harvard.edu/abs/2019A&A...626A..41M}
}

@ARTICLE{2020ApJS..249...29H,
       author = {{Huang}, Yang and {Sch{\"o}nrich}, Ralph and {Zhang}, Huawei and {Wu}, Yaqian and {Chen}, Bingqiu and {Wang}, Haifeng and {Xiang}, Maosheng and {Wang}, Chun and {Yuan}, Haibo and {Li}, Xinyi and {Sun}, Weixiang and {Li}, Ji and {Liu}, Xiaowei},
        title = "{Mapping the Galactic Disk with the LAMOST and Gaia Red Clump Sample. I. Precise Distances, Masses, Ages, and 3D Velocities of {\ensuremath{\sim}}140,000 Red Clump Stars}",
      journal = {\apjs},
         year = 2020,
        month = aug,
       volume = {249},
       number = {2},
          eid = {29},
        pages = {29},
          doi = {10.3847/1538-4365/ab994f},
archivePrefix = {arXiv},
       eprint = {2006.02686},
 primaryClass = {astro-ph.SR},
       adsurl = {https://ui.adsabs.harvard.edu/abs/2020ApJS..249...29H}
}

@ARTICLE{2021ApJ...912..106Y,
       author = {{Yu}, Zheng and {Li}, Ji and {Chen}, Bingqiu and {Huang}, Yang and {Jia}, Shuhua and {Xiang}, Maosheng and {Yuan}, Haibo and {Shi}, Jianrong and {Wang}, Chun and {Liu}, Xiaowei},
        title = "{Mapping the Galactic Disk with the LAMOST and Gaia Red Clump Sample. VII. The Stellar Disk Structure Revealed by the Mono-abundance Populations}",
      journal = {\apj},
         year = 2021,
        month = may,
       volume = {912},
       number = {2},
          eid = {106},
        pages = {106},
          doi = {10.3847/1538-4357/abf098},
archivePrefix = {arXiv},
       eprint = {2105.07151},
 primaryClass = {astro-ph.GA},
       adsurl = {https://ui.adsabs.harvard.edu/abs/2021ApJ...912..106Y}
}

@ARTICLE{2024MNRAS.531..495T,
       author = {{Tang}, Xi-Can and {Tian}, Hao and {Li}, Jing and {Chen}, Bing-qiu and {Chen}, Yi-Rong and {Liu}, Chao and {Qiu}, Dan},
        title = "{Detailed mapping of the Galactic disc structure in the solar neighbourhood through LAMOST K dwarfs}",
      journal = {\mnras},
         year = 2024,
        month = jun,
       volume = {531},
       number = {1},
        pages = {495-509},
          doi = {10.1093/mnras/stae1148},
archivePrefix = {arXiv},
       eprint = {2407.09312},
 primaryClass = {astro-ph.GA},
       adsurl = {https://ui.adsabs.harvard.edu/abs/2024MNRAS.531..495T}
}

@ARTICLE{2018ApJ...858L...7T,
       author = {{Ting}, Yuan-Sen and {Hawkins}, Keith and {Rix}, Hans-Walter},
        title = "{A Large and Pristine Sample of Standard Candles across the Milky Way: {\ensuremath{\sim}}100,000 Red Clump Stars with 3\% Contamination}",
      journal = {\apjl},
         year = 2018,
        month = may,
       volume = {858},
       number = {1},
          eid = {L7},
        pages = {L7},
          doi = {10.3847/2041-8213/aabf8e},
archivePrefix = {arXiv},
       eprint = {1803.06650},
 primaryClass = {astro-ph.SR},
       adsurl = {https://ui.adsabs.harvard.edu/abs/2018ApJ...858L...7T}
}

@ARTICLE{2013PASP..125..306F,
       author = {{Foreman-Mackey}, Daniel and {Hogg}, David W. and {Lang}, Dustin and {Goodman}, Jonathan},
        title = "{emcee: The MCMC Hammer}",
      journal = {\pasp},
         year = 2013,
        month = mar,
       volume = {125},
       number = {925},
        pages = {306},
          doi = {10.1086/670067},
archivePrefix = {arXiv},
       eprint = {1202.3665},
 primaryClass = {astro-ph.IM},
       adsurl = {https://ui.adsabs.harvard.edu/abs/2013PASP..125..306F}
}

@ARTICLE{2016ARA&A..54...95G,
       author = {{Girardi}, L{\'e}o},
        title = "{Red Clump Stars}",
      journal = {\araa},
         year = 2016,
        month = sep,
       volume = {54},
        pages = {95-133},
          doi = {10.1146/annurev-astro-081915-023354},
       adsurl = {https://ui.adsabs.harvard.edu/abs/2016ARA&A..54...95G}
}

@ARTICLE{2017MNRAS.471..722H,
       author = {{Hawkins}, Keith and {Leistedt}, Boris and {Bovy}, Jo and {Hogg}, David W.},
        title = "{Red clump stars and Gaia: calibration of the standard candle using a hierarchical probabilistic model}",
      journal = {\mnras},
         year = 2017,
        month = oct,
       volume = {471},
       number = {1},
        pages = {722-729},
          doi = {10.1093/mnras/stx1655},
archivePrefix = {arXiv},
       eprint = {1705.08988},
 primaryClass = {astro-ph.GA},
       adsurl = {https://ui.adsabs.harvard.edu/abs/2017MNRAS.471..722H}
}

@ARTICLE{2013ApJ...765L..41S,
       author = {{Stello}, Dennis and {Huber}, Daniel and {Bedding}, Timothy R. and {Benomar}, Othman and {Bildsten}, Lars and {Elsworth}, Yvonne P. and {Gilliland}, Ronald L. and {Mosser}, Beno{\^\i}t and {Paxton}, Bill and {White}, Timothy R.},
        title = "{Asteroseismic Classification of Stellar Populations among 13,000 Red Giants Observed by Kepler}",
      journal = {\apjl},
         year = 2013,
        month = mar,
       volume = {765},
       number = {2},
          eid = {L41},
        pages = {L41},
          doi = {10.1088/2041-8205/765/2/L41},
archivePrefix = {arXiv},
       eprint = {1302.0858},
 primaryClass = {astro-ph.SR},
       adsurl = {https://ui.adsabs.harvard.edu/abs/2013ApJ...765L..41S}
}

@ARTICLE{2025arXiv250314745D,
	author = {{DESI Collaboration} and {Karim}, M. Abdul and {Adame}, A.~G. and {Aguado}, D. and {Aguilar}, J. and {Ahlen}, S. and {Alam}, S. and {Aldering}, G. and {Alexander}, D.~M. and {Alfarsy}, R. and {Allen}, L. and {Allende Prieto}, C. and {Alves}, O. and {Anand}, A. and {Andrade}, U. and {Armengaud}, E. and {Avila}, S. and {Aviles}, A. and {Awan}, H. and {Bailey}, S. and {Baleato Lizancos}, A. and {Ballester}, O. and {Bault}, A. and {Bautista}, J. and {Bean}, R. and {Behera}, J. and {BenZvi}, S. and {Beraldo e Silva}, L. and {Bermejo-Climent}, J.~R. and {Beutler}, F. and {Bianchi}, D. and {Blake}, C. and {Blum}, R. and {Bolton}, A.~S. and {Bonici}, M. and {Brieden}, S. and {Brodzeller}, A. and {Brooks}, D. and {Buckley-Geer}, E. and {Burtin}, E. and {Bystr{\"o}m}, A. and {Canning}, R. and {Carnero Rosell}, A. and {Carr}, A. and {Carrilho}, P. and {Casas}, L. and {Castander}, F.~J. and {Cereskaite}, R. and {Cervantes-Cota}, J.~L. and {Chaussidon}, E. and {Chaves-Montero}, J. and {Chen}, S. and {Chen}, X. and {Circosta}, C. and {Claybaugh}, T. and {Cole}, S. and {Cooper}, A.~P. and {Cousinou}, M.-C. and {Cuceu}, A. and {Davis}, T.~M. and {Dawson}, K.~S. and {de Belsunce}, R. and {de la Cruz}, R. and {de la Macorra}, A. and {de Mattia}, A. and {Deiosso}, N. and {Della Costa}, J. and {Demina}, R. and {Demirbozan}, U. and {DeRose}, J. and {Dey}, A. and {Dey}, B. and {Ding}, J. and {Ding}, Z. and {Doel}, P. and {Douglass}, K. and {Dowicz}, M. and {Ebina}, H. and {Edelstein}, J. and {Eisenstein}, D.~J. and {Elbers}, W. and {Emas}, N. and {Escoffier}, S. and {Fagrelius}, P. and {Fan}, X. and {Fanning}, K. and {Favole}, G. and {Fawcett}, V.~A. and {Fern{\'a}ndez-Garc{\'\i}a}, E. and {Ferraro}, S. and {Findlay}, N. and {Font-Ribera}, A. and {Forero-Romero}, J.~E. and {Forero-S{\'a}nchez}, D. and {Frenk}, C.~S. and {G{\"a}nsicke}, B.~T. and {Galbany}, L. and {Garc{\'\i}a-Bellido}, J. and {Garcia-Quintero}, C. and {Garrison}, L.~H. and {Gazta{\~n}aga}, E. and {Gil-Mar{\'\i}n}, H. and {Gloudemans}, A. and {Gnedin}, O.~Y. and {Gontcho}, S. Gontcho A and {Gonzalez}, D. and {Gonzalez-Morales}, A.~X. and {Gonzalez-Perez}, V. and {Gordon}, C. and {Graur}, O. and {Green}, D. and {Gruen}, D. and {Gsponer}, R. and {Guandalin}, C. and {Gutierrez}, G. and {Guy}, J. and {Hahn}, C. and {Han}, J.~J. and {Han}, J. and {He}, S. and {Herrera-Alcantar}, H.~K. and {Heydenreich}, S. and {Honscheid}, K. and {Hou}, J. and {Howlett}, C. and {Huterer}, D. and {Ir{\v{s}}i{\v{c}}}, V. and {Ishak}, M. and {Jacques}, A. and {Jiang}, L. and {Jimenez}, J. and {Jing}, Y.~P. and {Joachimi}, B. and {Joudaki}, S. and {Joyce}, R. and {Jullo}, E. and {Juneau}, S. and {Kara{\c{c}}ayl{\i}}, N.~G. and {Karim}, T. and {Kehoe}, R. and {Kent}, S. and {Khederlarian}, A. and {Kirkby}, D. and {Kisner}, T. and {Kitaura}, F.-S. and {Kizhuprakkat}, N. and {Kong}, H. and {Koposov}, S.~E. and {Kremin}, A. and {Krolewski}, A. and {Lahav}, O. and {Lai}, Y. and {Lamman}, C. and {Lan}, T.-W. and {Landriau}, M. and {Lang}, D. and {Lange}, J.~U. and {Lasker}, J. and {Le Goff}, J.~M. and {Le Guillou}, L. and {Leauthaud}, A. and {Levi}, M.~E. and {Li}, S. and {Li}, T.~S. and {Liu}, W. and {Lodha}, K. and {Lokken}, M. and {Luo}, Y. and {Luo}, Y. and {Magneville}, C. and {Manera}, M. and {Manser}, C.~J. and {Margala}, D. and {Martini}, P. and {Maus}, M. and {McCullough}, J. and {McDonald}, P. and {Medina}, G.~E. and {Medina-Varela}, L. and {Meisner}, A. and {Mena-Fern{\'a}ndez}, J. and {Menegas}, A. and {Meneses-Rizo}, J. and {Mezcua}, M. and {Miquel}, R. and {Montero-Camacho}, P. and {Moon}, J. and {Moustakas}, J. and {Mu{\~n}oz-Guti{\'e}rrez}, A. and {Mu{\~n}oz-Santos}, D. and {Myers}, A.~D. and {Myles}, J. and {Nadathur}, S. and {Najita}, J. and {Napolitano}, L. and {Newman}, J.~A. and {Nikakhtar}, F. and {Nikutta}, R. and {Niz}, G. and {Noriega}, H.~E.},
	title = "{Data Release 1 of the Dark Energy Spectroscopic Instrument}",
	journal = {arXiv e-prints},
	year = 2025,
	month = mar,
	eid = {arXiv:2503.14745},
	pages = {arXiv:2503.14745},
	doi = {10.48550/arXiv.2503.14745},
	archivePrefix = {arXiv},
	eprint = {2503.14745},
	primaryClass = {astro-ph.CO},
	adsurl = {https://ui.adsabs.harvard.edu/abs/2025arXiv250314745D}
}

@ARTICLE{2016A&A...588A..87V,
       author = {{Vrard}, M. and {Mosser}, B. and {Samadi}, R.},
        title = "{Period spacings in red giants. II. Automated measurement}",
      journal = {\aap},
         year = 2016,
        month = apr,
       volume = {588},
          eid = {A87},
        pages = {A87},
          doi = {10.1051/0004-6361/201527259},
archivePrefix = {arXiv},
       eprint = {1602.04940},
 primaryClass = {astro-ph.SR},
       adsurl = {https://ui.adsabs.harvard.edu/abs/2016A&A...588A..87V}
}

@ARTICLE{2023ApJS..267....8A,
	author = {{Andrae}, Ren{\'e} and {Rix}, Hans-Walter and {Chandra}, Vedant},
	title = "{Robust Data-driven Metallicities for 175 Million Stars from Gaia XP Spectra}",
	journal = {\apjs},
	year = 2023,
	month = jul,
	volume = {267},
	number = {1},
	eid = {8},
	pages = {8},
	doi = {10.3847/1538-4365/acd53e},
	archivePrefix = {arXiv},
	eprint = {2302.02611},
	primaryClass = {astro-ph.SR},
	adsurl = {https://ui.adsabs.harvard.edu/abs/2023ApJS..267....8A}
}

@ARTICLE{2017A&A...607A.105M,
	author = {{Marrese}, P.~M. and {Marinoni}, S. and {Fabrizio}, M. and {Giuffrida}, G.},
	title = "{Gaia Data Release 1. Cross-match with external catalogues. Algorithm and results}",
	journal = {\aap},
	year = 2017,
	month = nov,
	volume = {607},
	eid = {A105},
	pages = {A105},
	doi = {10.1051/0004-6361/201730965},
	archivePrefix = {arXiv},
	eprint = {1710.06739},
	primaryClass = {astro-ph.SR},
	adsurl = {https://ui.adsabs.harvard.edu/abs/2017A&A...607A.105M}
}

@ARTICLE{2019A&A...621A.144M,
	author = {{Marrese}, P.~M. and {Marinoni}, S. and {Fabrizio}, M. and {Altavilla}, G.},
	title = "{Gaia Data Release 2. Cross-match with external catalogues: algorithms and results}",
	journal = {\aap},
	year = 2019,
	month = jan,
	volume = {621},
	eid = {A144},
	pages = {A144},
	doi = {10.1051/0004-6361/201834142},
	archivePrefix = {arXiv},
	eprint = {1808.09151},
	primaryClass = {astro-ph.SR},
	adsurl = {https://ui.adsabs.harvard.edu/abs/2019A&A...621A.144M}
}

@ARTICLE{2003A&A...409..205D,
	author = {{Drimmel}, R. and {Cabrera-Lavers}, A. and {L{\'o}pez-Corredoira}, M.},
	title = "{A three-dimensional Galactic extinction model}",
	journal = {\aap},
	year = 2003,
	month = oct,
	volume = {409},
	pages = {205-215},
	doi = {10.1051/0004-6361:20031070},
	archivePrefix = {arXiv},
	eprint = {astro-ph/0307273},
	primaryClass = {astro-ph},
	adsurl = {https://ui.adsabs.harvard.edu/abs/2003A&A...409..205D}
}

@ARTICLE{2016ApJ...818..130B,
	author = {{Bovy}, Jo and {Rix}, Hans-Walter and {Green}, Gregory M. and {Schlafly}, Edward F. and {Finkbeiner}, Douglas P.},
	title = "{On Galactic Density Modeling in the Presence of Dust Extinction}",
	journal = {\apj},
	year = 2016,
	month = feb,
	volume = {818},
	number = {2},
	eid = {130},
	pages = {130},
	doi = {10.3847/0004-637X/818/2/130},
	archivePrefix = {arXiv},
	eprint = {1509.06751},
	primaryClass = {astro-ph.GA},
	adsurl = {https://ui.adsabs.harvard.edu/abs/2016ApJ...818..130B}
}

@ARTICLE{2006A&A...453..635M,
	author = {{Marshall}, D.~J. and {Robin}, A.~C. and {Reyl{\'e}}, C. and {Schultheis}, M. and {Picaud}, S.},
	title = "{Modelling the Galactic interstellar extinction distribution in three dimensions}",
	journal = {\aap},
	year = 2006,
	month = jul,
	volume = {453},
	number = {2},
	pages = {635-651},
	doi = {10.1051/0004-6361:20053842},
	archivePrefix = {arXiv},
	eprint = {astro-ph/0604427},
	primaryClass = {astro-ph},
	adsurl = {https://ui.adsabs.harvard.edu/abs/2006A&A...453..635M}
}

@ARTICLE{2019ApJ...887...93G,
	author = {{Green}, Gregory M. and {Schlafly}, Edward and {Zucker}, Catherine and {Speagle}, Joshua S. and {Finkbeiner}, Douglas},
	title = "{A 3D Dust Map Based on Gaia, Pan-STARRS 1, and 2MASS}",
	journal = {\apj},
	year = 2019,
	month = dec,
	volume = {887},
	number = {1},
	eid = {93},
	pages = {93},
	doi = {10.3847/1538-4357/ab5362},
	archivePrefix = {arXiv},
	eprint = {1905.02734},
	primaryClass = {astro-ph.GA},
	adsurl = {https://ui.adsabs.harvard.edu/abs/2019ApJ...887...93G}
}

@ARTICLE{2018A&A...609A.116R,
	author = {{Ruiz-Dern}, L. and {Babusiaux}, C. and {Arenou}, F. and {Turon}, C. and {Lallement}, R.},
	title = "{Empirical photometric calibration of the Gaia red clump: Colours, effective temperature, and absolute magnitude}",
	journal = {\aap},
	year = 2018,
	month = jan,
	volume = {609},
	eid = {A116},
	pages = {A116},
	doi = {10.1051/0004-6361/201731572},
	archivePrefix = {arXiv},
	eprint = {1710.05803},
	primaryClass = {astro-ph.SR},
	adsurl = {https://ui.adsabs.harvard.edu/abs/2018A&A...609A.116R}
}

@ARTICLE{2000ApJ...539..732A,
	author = {{Alves}, David R.},
	title = "{K-Band Calibration of the Red Clump Luminosity}",
	journal = {\apj},
	year = 2000,
	month = aug,
	volume = {539},
	number = {2},
	pages = {732-741},
	doi = {10.1086/309278},
	archivePrefix = {arXiv},
	eprint = {astro-ph/0003329},
	primaryClass = {astro-ph},
	adsurl = {https://ui.adsabs.harvard.edu/abs/2000ApJ...539..732A}
}

@ARTICLE{2017ApJ...840...77C,
	author = {{Chen}, Y.~Q. and {Casagrande}, L. and {Zhao}, G. and {Bovy}, J. and {Silva Aguirre}, V. and {Zhao}, J.~K. and {Jia}, Y.~P.},
	title = "{Absolute Magnitudes of Seismic Red Clumps in the Kepler  Field and SAGA: The Age Dependency of the Distance Scale}",
	journal = {\apj},
	year = 2017,
	month = may,
	volume = {840},
	number = {2},
	eid = {77},
	pages = {77},
	doi = {10.3847/1538-4357/aa6d0f},
	archivePrefix = {arXiv},
	eprint = {1704.03903},
	primaryClass = {astro-ph.SR},
	adsurl = {https://ui.adsabs.harvard.edu/abs/2017ApJ...840...77C}
}

@ARTICLE{2022A&A...658A..91A,
       author = {{Anders}, F. and {Khalatyan}, A. and {Queiroz}, A.~B.~A. and {Chiappini}, C. and {Ard{\`e}vol}, J. and {Casamiquela}, L. and {Figueras}, F. and {Jim{\'e}nez-Arranz}, {\'O}. and {Jordi}, C. and {Mongui{\'o}}, M. and {Romero-G{\'o}mez}, M. and {Altamirano}, D. and {Antoja}, T. and {Assaad}, R. and {Cantat-Gaudin}, T. and {Castro-Ginard}, A. and {Enke}, H. and {Girardi}, L. and {Guiglion}, G. and {Khan}, S. and {Luri}, X. and {Miglio}, A. and {Minchev}, I. and {Ramos}, P. and {Santiago}, B.~X. and {Steinmetz}, M.},
        title = "{Photo-astrometric distances, extinctions, and astrophysical parameters for Gaia EDR3 stars brighter than G = 18.5}",
      journal = {\aap},
         year = 2022,
        month = feb,
       volume = {658},
          eid = {A91},
        pages = {A91},
          doi = {10.1051/0004-6361/202142369},
archivePrefix = {arXiv},
       eprint = {2111.01860},
 primaryClass = {astro-ph.GA},
       adsurl = {https://ui.adsabs.harvard.edu/abs/2022A&A...658A..91A}
}

@ARTICLE{2007A&A...463..559V,
	author = {{van Helshoecht}, V. and {Groenewegen}, M.~A.~T.},
	title = "{K-band magnitude of the red clump as a distance indicator}",
	journal = {\aap},
	year = 2007,
	month = feb,
	volume = {463},
	number = {2},
	pages = {559-565},
	doi = {10.1051/0004-6361:20052721},
	adsurl = {https://ui.adsabs.harvard.edu/abs/2007A&A...463..559V}
}

@ARTICLE{2019ApJ...872...95M,
	author = {{Mohammed}, Steven and {Schiminovich}, David and {Hawkins}, Keith and {Johnson}, Benjamin and {Wang}, Dun and {Hogg}, David W.},
	title = "{An Ultraviolet-Optical Color-Metallicity Relation for Red Clump Stars Using GALEX and Gaia}",
	journal = {\apj},
	year = 2019,
	month = feb,
	volume = {872},
	number = {1},
	eid = {95},
	pages = {95},
	doi = {10.3847/1538-4357/aaf236},
	archivePrefix = {arXiv},
	eprint = {1805.03236},
	primaryClass = {astro-ph.SR},
	adsurl = {https://ui.adsabs.harvard.edu/abs/2019ApJ...872...95M}
}

@ARTICLE{2019MNRAS.486.5600O,
	author = {{Onozato}, Hiroki and {Ita}, Yoshifusa and {Nakada}, Yoshikazu and {Nishiyama}, Shogo},
	title = "{The age and metallicity dependence of the near-infrared magnitudes of red clump stars}",
	journal = {\mnras},
	year = 2019,
	month = jul,
	volume = {486},
	number = {4},
	pages = {5600-5613},
	doi = {10.1093/mnras/stz1192},
	archivePrefix = {arXiv},
	eprint = {1904.12874},
	primaryClass = {astro-ph.SR},
	adsurl = {https://ui.adsabs.harvard.edu/abs/2019MNRAS.486.5600O}
}

@ARTICLE{2021ApJ...923..145W,
	author = {{Wang}, Shu and {Chen}, Xiaodian},
	title = "{3D Parameter Maps of Red Clump Stars in the Milky Way: Absolute Magnitudes and Intrinsic Colors}",
	journal = {\apj},
	year = 2021,
	month = dec,
	volume = {923},
	number = {2},
	eid = {145},
	pages = {145},
	doi = {10.3847/1538-4357/ac22a7},
	archivePrefix = {arXiv},
	eprint = {2108.13605},
	primaryClass = {astro-ph.GA},
	adsurl = {https://ui.adsabs.harvard.edu/abs/2021ApJ...923..145W}
}

@ARTICLE{2020ApJ...893..108P,
	author = {{Plevne}, Olcay and {{\"O}nal Ta{\c{s}}}, {\"O}zgecan and {Bilir}, Sel{\c{c}}uk and {Seabroke}, George M.},
	title = "{Multiwavelength Absolute Magnitudes and Colors of Red Clump Stars in the Gaia Era}",
	journal = {\apj},
	year = 2020,
	month = apr,
	volume = {893},
	number = {2},
	eid = {108},
	pages = {108},
	doi = {10.3847/1538-4357/ab80bb},
	archivePrefix = {arXiv},
	eprint = {2003.07887},
	primaryClass = {astro-ph.GA},
	adsurl = {https://ui.adsabs.harvard.edu/abs/2020ApJ...893..108P}
}

@ARTICLE{2021A&A...654A.107C,
	author = {{Culpan}, R. and {Pelisoli}, I. and {Geier}, S.},
	title = "{Clean catalogues of blue horizontal-branch stars using Gaia EDR3}",
	journal = {\aap},
	year = 2021,
	month = oct,
	volume = {654},
	eid = {A107},
	pages = {A107},
	doi = {10.1051/0004-6361/202040074},
	archivePrefix = {arXiv},
	eprint = {2108.05172},
	primaryClass = {astro-ph.SR},
	adsurl = {https://ui.adsabs.harvard.edu/abs/2021A&A...654A.107C}
}

@ARTICLE{2024MNRAS.531.2126F,
	author = {{Fallows}, Connor P. and {Sanders}, Jason L.},
	title = "{Stellar atmospheric parameters from Gaia BP/RP spectra using uncertain neural networks}",
	journal = {\mnras},
	year = 2024,
	month = jun,
	volume = {531},
	number = {1},
	pages = {2126-2147},
	doi = {10.1093/mnras/stae1303},
	archivePrefix = {arXiv},
	eprint = {2405.10699},
	primaryClass = {astro-ph.SR},
	adsurl = {https://ui.adsabs.harvard.edu/abs/2024MNRAS.531.2126F}
}

@ARTICLE{2017RAA....17...96L,
       author = {{Liu}, Chao and {Xu}, Yan and {Wan}, Jun-Chen and {Wang}, Hai-Feng and {Carlin}, Jeffrey L. and {Deng}, Li-Cai and {Newberg}, Heidi Jo and {Cao}, Zi-Huang and {Hou}, Yong-Hui and {Wang}, Yue-Fei and {Zhang}, Yong},
        title = "{Mapping the Milky Way with LAMOST I: method and overview}",
      journal = {Research in Astronomy and Astrophysics},
         year = 2017,
        month = sep,
       volume = {17},
       number = {9},
          eid = {096},
        pages = {096},
          doi = {10.1088/1674-4527/17/9/96},
archivePrefix = {arXiv},
       eprint = {1701.07831},
 primaryClass = {astro-ph.GA},
       adsurl = {https://ui.adsabs.harvard.edu/abs/2017RAA....17...96L}
}

@ARTICLE{2024MNRAS.531.1730T,
       author = {{Tian}, Hao and {Liu}, Chao and {Li}, Jiadong and {Zhang}, Bo},
        title = "{Mapping the Milky Way with LAMOST - IV. The large Galactic disc extending to 35 kpc}",
      journal = {\mnras},
         year = 2024,
        month = jun,
       volume = {531},
       number = {1},
        pages = {1730-1745},
          doi = {10.1093/mnras/stae1247},
       adsurl = {https://ui.adsabs.harvard.edu/abs/2024MNRAS.531.1730T}
}

@ARTICLE{2018A&A...612L...8L,
       author = {{L{\'o}pez-Corredoira}, M. and {Allende Prieto}, C. and {Garz{\'o}n}, F. and {Wang}, H. and {Liu}, C. and {Deng}, L.},
        title = "{Disk stars in the Milky Way detected beyond 25 kpc from its center}",
      journal = {\aap},
         year = 2018,
        month = apr,
       volume = {612},
          eid = {L8},
        pages = {L8},
          doi = {10.1051/0004-6361/201832880},
archivePrefix = {arXiv},
       eprint = {1804.03064},
 primaryClass = {astro-ph.GA},
       adsurl = {https://ui.adsabs.harvard.edu/abs/2018A&A...612L...8L}
}

@ARTICLE{2023A&A...669A..55C,
       author = {{Cantat-Gaudin}, Tristan and {Fouesneau}, Morgan and {Rix}, Hans-Walter and {Brown}, Anthony G.~A. and {Castro-Ginard}, Alfred and {Kostrzewa-Rutkowska}, Zuzanna and {Drimmel}, Ronald and {Hogg}, David W. and {Casey}, Andrew R. and {Khanna}, Shourya and {Oh}, Semyeong and {Price-Whelan}, Adrian M. and {Belokurov}, Vasily and {Saydjari}, Andrew K. and {Green}, G.},
        title = "{An empirical model of the Gaia DR3 selection function}",
      journal = {\aap},
         year = 2023,
        month = jan,
       volume = {669},
          eid = {A55},
        pages = {A55},
          doi = {10.1051/0004-6361/202244784},
archivePrefix = {arXiv},
       eprint = {2208.09335},
 primaryClass = {astro-ph.GA},
       adsurl = {https://ui.adsabs.harvard.edu/abs/2023A&A...669A..55C}
}

@ARTICLE{2024A&A...683A.128C,
       author = {{Cantat-Gaudin}, Tristan and {Fouesneau}, Morgan and {Rix}, Hans-Walter and {Brown}, Anthony G.~A. and {Drimmel}, Ronald and {Castro-Ginard}, Alfred and {Khanna}, Shourya and {Belokurov}, Vasily and {Casey}, Andrew R.},
        title = "{Uniting Gaia and APOGEE to unveil the cosmic chemistry of the Milky Way disc}",
      journal = {\aap},
         year = 2024,
        month = mar,
       volume = {683},
          eid = {A128},
        pages = {A128},
          doi = {10.1051/0004-6361/202348018},
archivePrefix = {arXiv},
       eprint = {2401.05023},
 primaryClass = {astro-ph.GA},
       adsurl = {https://ui.adsabs.harvard.edu/abs/2024A&A...683A.128C}
}

@ARTICLE{2023A&A...677A..37C,
       author = {{Castro-Ginard}, Alfred and {Brown}, Anthony G.~A. and {Kostrzewa-Rutkowska}, Zuzanna and {Cantat-Gaudin}, Tristan and {Drimmel}, Ronald and {Oh}, Semyeong and {Belokurov}, Vasily and {Casey}, Andrew R. and {Fouesneau}, Morgan and {Khanna}, Shourya and {Price-Whelan}, Adrian M. and {Rix}, Hans-Walter},
        title = "{Estimating the selection function of Gaia DR3 subsamples}",
      journal = {\aap},
         year = 2023,
        month = sep,
       volume = {677},
          eid = {A37},
        pages = {A37},
          doi = {10.1051/0004-6361/202346547},
archivePrefix = {arXiv},
       eprint = {2303.17738},
 primaryClass = {astro-ph.GA},
       adsurl = {https://ui.adsabs.harvard.edu/abs/2023A&A...677A..37C}
}

@ARTICLE{2018AJ....156..123A,
       author = {{Astropy Collaboration} and {Price-Whelan}, A.~M. and {Sip{\H{o}}cz}, B.~M. and {G{\"u}nther}, H.~M. and {Lim}, P.~L. and {Crawford}, S.~M. and {Conseil}, S. and {Shupe}, D.~L. and {Craig}, M.~W. and {Dencheva}, N. and {Ginsburg}, A. and {VanderPlas}, J.~T. and {Bradley}, L.~D. and {P{\'e}rez-Su{\'a}rez}, D. and {de Val-Borro}, M. and {Aldcroft}, T.~L. and {Cruz}, K.~L. and {Robitaille}, T.~P. and {Tollerud}, E.~J. and {Ardelean}, C. and {Babej}, T. and {Bach}, Y.~P. and {Bachetti}, M. and {Bakanov}, A.~V. and {Bamford}, S.~P. and {Barentsen}, G. and {Barmby}, P. and {Baumbach}, A. and {Berry}, K.~L. and {Biscani}, F. and {Boquien}, M. and {Bostroem}, K.~A. and {Bouma}, L.~G. and {Brammer}, G.~B. and {Bray}, E.~M. and {Breytenbach}, H. and {Buddelmeijer}, H. and {Burke}, D.~J. and {Calderone}, G. and {Cano Rodr{\'\i}guez}, J.~L. and {Cara}, M. and {Cardoso}, J.~V.~M. and {Cheedella}, S. and {Copin}, Y. and {Corrales}, L. and {Crichton}, D. and {D'Avella}, D. and {Deil}, C. and {Depagne}, {\'E}. and {Dietrich}, J.~P. and {Donath}, A. and {Droettboom}, M. and {Earl}, N. and {Erben}, T. and {Fabbro}, S. and {Ferreira}, L.~A. and {Finethy}, T. and {Fox}, R.~T. and {Garrison}, L.~H. and {Gibbons}, S.~L.~J. and {Goldstein}, D.~A. and {Gommers}, R. and {Greco}, J.~P. and {Greenfield}, P. and {Groener}, A.~M. and {Grollier}, F. and {Hagen}, A. and {Hirst}, P. and {Homeier}, D. and {Horton}, A.~J. and {Hosseinzadeh}, G. and {Hu}, L. and {Hunkeler}, J.~S. and {Ivezi{\'c}}, {\v{Z}}. and {Jain}, A. and {Jenness}, T. and {Kanarek}, G. and {Kendrew}, S. and {Kern}, N.~S. and {Kerzendorf}, W.~E. and {Khvalko}, A. and {King}, J. and {Kirkby}, D. and {Kulkarni}, A.~M. and {Kumar}, A. and {Lee}, A. and {Lenz}, D. and {Littlefair}, S.~P. and {Ma}, Z. and {Macleod}, D.~M. and {Mastropietro}, M. and {McCully}, C. and {Montagnac}, S. and {Morris}, B.~M. and {Mueller}, M. and {Mumford}, S.~J. and {Muna}, D. and {Murphy}, N.~A. and {Nelson}, S. and {Nguyen}, G.~H. and {Ninan}, J.~P. and {N{\"o}the}, M. and {Ogaz}, S. and {Oh}, S. and {Parejko}, J.~K. and {Parley}, N. and {Pascual}, S. and {Patil}, R. and {Patil}, A.~A. and {Plunkett}, A.~L. and {Prochaska}, J.~X. and {Rastogi}, T. and {Reddy Janga}, V. and {Sabater}, J. and {Sakurikar}, P. and {Seifert}, M. and {Sherbert}, L.~E. and {Sherwood-Taylor}, H. and {Shih}, A.~Y. and {Sick}, J. and {Silbiger}, M.~T. and {Singanamalla}, S. and {Singer}, L.~P. and {Sladen}, P.~H. and {Sooley}, K.~A. and {Sornarajah}, S. and {Streicher}, O. and {Teuben}, P. and {Thomas}, S.~W. and {Tremblay}, G.~R. and {Turner}, J.~E.~H. and {Terr{\'o}n}, V. and {van Kerkwijk}, M.~H. and {de la Vega}, A. and {Watkins}, L.~L. and {Weaver}, B.~A. and {Whitmore}, J.~B. and {Woillez}, J. and {Zabalza}, V. and {Astropy Contributors}},
        title = "{The Astropy Project: Building an Open-science Project and Status of the v2.0 Core Package}",
      journal = {\aj},
         year = 2018,
        month = sep,
       volume = {156},
       number = {3},
          eid = {123},
        pages = {123},
          doi = {10.3847/1538-3881/aabc4f},
archivePrefix = {arXiv},
       eprint = {1801.02634},
 primaryClass = {astro-ph.IM},
       adsurl = {https://ui.adsabs.harvard.edu/abs/2018AJ....156..123A}
}

@ARTICLE{2018A&A...615L..15G,
       author = {{GRAVITY Collaboration} and {Abuter}, R. and {Amorim}, A. and {Anugu}, N. and {Baub{\"o}ck}, M. and {Benisty}, M. and {Berger}, J.~P. and {Blind}, N. and {Bonnet}, H. and {Brandner}, W. and {Buron}, A. and {Collin}, C. and {Chapron}, F. and {Cl{\'e}net}, Y. and {Coud{\'e} Du Foresto}, V. and {de Zeeuw}, P.~T. and {Deen}, C. and {Delplancke-Str{\"o}bele}, F. and {Dembet}, R. and {Dexter}, J. and {Duvert}, G. and {Eckart}, A. and {Eisenhauer}, F. and {Finger}, G. and {F{\"o}rster Schreiber}, N.~M. and {F{\'e}dou}, P. and {Garcia}, P. and {Garcia Lopez}, R. and {Gao}, F. and {Gendron}, E. and {Genzel}, R. and {Gillessen}, S. and {Gordo}, P. and {Habibi}, M. and {Haubois}, X. and {Haug}, M. and {Hau{\ss}mann}, F. and {Henning}, Th. and {Hippler}, S. and {Horrobin}, M. and {Hubert}, Z. and {Hubin}, N. and {Jimenez Rosales}, A. and {Jochum}, L. and {Jocou}, K. and {Kaufer}, A. and {Kellner}, S. and {Kendrew}, S. and {Kervella}, P. and {Kok}, Y. and {Kulas}, M. and {Lacour}, S. and {Lapeyr{\`e}re}, V. and {Lazareff}, B. and {Le Bouquin}, J. -B. and {L{\'e}na}, P. and {Lippa}, M. and {Lenzen}, R. and {M{\'e}rand}, A. and {M{\"u}ler}, E. and {Neumann}, U. and {Ott}, T. and {Palanca}, L. and {Paumard}, T. and {Pasquini}, L. and {Perraut}, K. and {Perrin}, G. and {Pfuhl}, O. and {Plewa}, P.~M. and {Rabien}, S. and {Ram{\'\i}rez}, A. and {Ramos}, J. and {Rau}, C. and {Rodr{\'\i}guez-Coira}, G. and {Rohloff}, R. -R. and {Rousset}, G. and {Sanchez-Bermudez}, J. and {Scheithauer}, S. and {Sch{\"o}ller}, M. and {Schuler}, N. and {Spyromilio}, J. and {Straub}, O. and {Straubmeier}, C. and {Sturm}, E. and {Tacconi}, L.~J. and {Tristram}, K.~R.~W. and {Vincent}, F. and {von Fellenberg}, S. and {Wank}, I. and {Waisberg}, I. and {Widmann}, F. and {Wieprecht}, E. and {Wiest}, M. and {Wiezorrek}, E. and {Woillez}, J. and {Yazici}, S. and {Ziegler}, D. and {Zins}, G.},
        title = "{Detection of the gravitational redshift in the orbit of the star S2 near the Galactic centre massive black hole}",
      journal = {\aap},
         year = 2018,
        month = jul,
       volume = {615},
          eid = {L15},
        pages = {L15},
          doi = {10.1051/0004-6361/201833718},
archivePrefix = {arXiv},
       eprint = {1807.09409},
 primaryClass = {astro-ph.GA},
       adsurl = {https://ui.adsabs.harvard.edu/abs/2018A&A...615L..15G}
}

@ARTICLE{2019MNRAS.482.1417B,
       author = {{Bennett}, Morgan and {Bovy}, Jo},
        title = "{Vertical waves in the solar neighbourhood in Gaia DR2}",
      journal = {\mnras},
         year = 2019,
        month = jan,
       volume = {482},
       number = {1},
        pages = {1417-1425},
          doi = {10.1093/mnras/sty2813},
archivePrefix = {arXiv},
       eprint = {1809.03507},
 primaryClass = {astro-ph.GA},
       adsurl = {https://ui.adsabs.harvard.edu/abs/2019MNRAS.482.1417B}
}

@ARTICLE{2006AJ....131.1163S,
       author = {{Skrutskie}, M.~F. and {Cutri}, R.~M. and {Stiening}, R. and {Weinberg}, M.~D. and {Schneider}, S. and {Carpenter}, J.~M. and {Beichman}, C. and {Capps}, R. and {Chester}, T. and {Elias}, J. and {Huchra}, J. and {Liebert}, J. and {Lonsdale}, C. and {Monet}, D.~G. and {Price}, S. and {Seitzer}, P. and {Jarrett}, T. and {Kirkpatrick}, J.~D. and {Gizis}, J.~E. and {Howard}, E. and {Evans}, T. and {Fowler}, J. and {Fullmer}, L. and {Hurt}, R. and {Light}, R. and {Kopan}, E.~L. and {Marsh}, K.~A. and {McCallon}, H.~L. and {Tam}, R. and {Van Dyk}, S. and {Wheelock}, S.},
        title = "{The Two Micron All Sky Survey (2MASS)}",
      journal = {\aj},
         year = 2006,
        month = feb,
       volume = {131},
       number = {2},
        pages = {1163-1183},
          doi = {10.1086/498708},
       adsurl = {https://ui.adsabs.harvard.edu/abs/2006AJ....131.1163S}
}

@ARTICLE{2024ApJ...972..112C,
	author = {{Chandra}, Vedant and {Semenov}, Vadim A. and {Rix}, Hans-Walter and {Conroy}, Charlie and {Bonaca}, Ana and {Naidu}, Rohan P. and {Andrae}, Ren{\'e} and {Li}, Jiadong and {Hernquist}, Lars},
	title = "{The Three-phase Evolution of the Milky Way}",
	journal = {\apj},
	year = 2024,
	month = sep,
	volume = {972},
	number = {1},
	eid = {112},
	pages = {112},
	doi = {10.3847/1538-4357/ad5b60},
	archivePrefix = {arXiv},
	eprint = {2310.13050},
	primaryClass = {astro-ph.GA},
	adsurl = {https://ui.adsabs.harvard.edu/abs/2024ApJ...972..112C}
}

@ARTICLE{2024NatAs...8.1302L,
	author = {{Lian}, Jianhui and {Zasowski}, Gail and {Chen}, Bingqiu and {Imig}, Julie and {Wang}, Tao and {Boardman}, Nicholas and {Liu}, Xiaowei},
	title = "{The broken-exponential radial structure and larger size of the Milky Way galaxy}",
	journal = {Nature Astronomy},
	year = 2024,
	month = oct,
	volume = {8},
	number = {10},
	pages = {1302-1309},
	doi = {10.1038/s41550-024-02315-7},
	archivePrefix = {arXiv},
	eprint = {2406.05604},
	primaryClass = {astro-ph.GA},
	adsurl = {https://ui.adsabs.harvard.edu/abs/2024NatAs...8.1302L}
}

@ARTICLE{2024MNRAS.527.4863U,
	author = {{Uppal}, Namita and {Ganesh}, Shashikiran and {Schultheis}, Mathias},
	title = "{Warp and flare of the old Galactic disc as traced by the red clump stars}",
	journal = {\mnras},
	year = 2024,
	month = jan,
	volume = {527},
	number = {3},
	pages = {4863-4873},
	doi = {10.1093/mnras/stad3525},
	archivePrefix = {arXiv},
	eprint = {2311.09616},
	primaryClass = {astro-ph.GA},
	adsurl = {https://ui.adsabs.harvard.edu/abs/2024MNRAS.527.4863U}
}

@ARTICLE{2022A&A...664A..58C,
	author = {{Chrob{\'a}kov{\'a}}, {\v{Z}}. and {Nagy}, R. and {L{\'o}pez-Corredoira}, M.},
	title = "{Warp and flare of the Galactic disc revealed with supergiants by Gaia EDR3}",
	journal = {\aap},
	year = 2022,
	month = aug,
	volume = {664},
	eid = {A58},
	pages = {A58},
	doi = {10.1051/0004-6361/202243296},
	archivePrefix = {arXiv},
	eprint = {2206.08230},
	primaryClass = {astro-ph.GA},
	adsurl = {https://ui.adsabs.harvard.edu/abs/2022A&A...664A..58C}
}

@ARTICLE{2017AJ....154...94M,
	author = {{Majewski}, Steven R. and {Schiavon}, Ricardo P. and {Frinchaboy}, Peter M. and {Allende Prieto}, Carlos and {Barkhouser}, Robert and {Bizyaev}, Dmitry and {Blank}, Basil and {Brunner}, Sophia and {Burton}, Adam and {Carrera}, Ricardo and {Chojnowski}, S. Drew and {Cunha}, K{\'a}tia and {Epstein}, Courtney and {Fitzgerald}, Greg and {Garc{\'\i}a P{\'e}rez}, Ana E. and {Hearty}, Fred R. and {Henderson}, Chuck and {Holtzman}, Jon A. and {Johnson}, Jennifer A. and {Lam}, Charles R. and {Lawler}, James E. and {Maseman}, Paul and {M{\'e}sz{\'a}ros}, Szabolcs and {Nelson}, Matthew and {Nguyen}, Duy Coung and {Nidever}, David L. and {Pinsonneault}, Marc and {Shetrone}, Matthew and {Smee}, Stephen and {Smith}, Verne V. and {Stolberg}, Todd and {Skrutskie}, Michael F. and {Walker}, Eric and {Wilson}, John C. and {Zasowski}, Gail and {Anders}, Friedrich and {Basu}, Sarbani and {Beland}, Stephane and {Blanton}, Michael R. and {Bovy}, Jo and {Brownstein}, Joel R. and {Carlberg}, Joleen and {Chaplin}, William and {Chiappini}, Cristina and {Eisenstein}, Daniel J. and {Elsworth}, Yvonne and {Feuillet}, Diane and {Fleming}, Scott W. and {Galbraith-Frew}, Jessica and {Garc{\'\i}a}, Rafael A. and {Garc{\'\i}a-Hern{\'a}ndez}, D. An{\'\i}bal and {Gillespie}, Bruce A. and {Girardi}, L{\'e}o and {Gunn}, James E. and {Hasselquist}, Sten and {Hayden}, Michael R. and {Hekker}, Saskia and {Ivans}, Inese and {Kinemuchi}, Karen and {Klaene}, Mark and {Mahadevan}, Suvrath and {Mathur}, Savita and {Mosser}, Beno{\^\i}t and {Muna}, Demitri and {Munn}, Jeffrey A. and {Nichol}, Robert C. and {O'Connell}, Robert W. and {Parejko}, John K. and {Robin}, A.~C. and {Rocha-Pinto}, Helio and {Schultheis}, Matthias and {Serenelli}, Aldo M. and {Shane}, Neville and {Silva Aguirre}, Victor and {Sobeck}, Jennifer S. and {Thompson}, Benjamin and {Troup}, Nicholas W. and {Weinberg}, David H. and {Zamora}, Olga},
	title = "{The Apache Point Observatory Galactic Evolution Experiment (APOGEE)}",
	journal = {\aj},
	year = 2017,
	month = sep,
	volume = {154},
	number = {3},
	eid = {94},
	pages = {94},
	doi = {10.3847/1538-3881/aa784d},
	archivePrefix = {arXiv},
	eprint = {1509.05420},
	primaryClass = {astro-ph.IM},
	adsurl = {https://ui.adsabs.harvard.edu/abs/2017AJ....154...94M}
}

@ARTICLE{2008AJ....135...20E,
	author = {{Erwin}, Peter and {Pohlen}, Michael and {Beckman}, John E.},
	title = "{The Outer Disks of Early-Type Galaxies. I. Surface-Brightness Profiles of Barred Galaxies}",
	journal = {\aj},
	year = 2008,
	month = jan,
	volume = {135},
	number = {1},
	pages = {20-54},
	doi = {10.1088/0004-6256/135/1/20},
	archivePrefix = {arXiv},
	eprint = {0709.3505},
	primaryClass = {astro-ph},
	adsurl = {https://ui.adsabs.harvard.edu/abs/2008AJ....135...20E}
}

@ARTICLE{2011ApJ...733L..43M,
	author = {{Minniti}, D. and {Saito}, R.~K. and {Alonso-Garc{\'\i}a}, J. and {Lucas}, P.~W. and {Hempel}, M.},
	title = "{The Edge of the Milky Way Stellar Disk Revealed Using Clump Giant Stars as Distance Indicators}",
	journal = {\apjl},
	year = 2011,
	month = jun,
	volume = {733},
	number = {2},
	eid = {L43},
	pages = {L43},
	doi = {10.1088/2041-8205/733/2/L43},
	archivePrefix = {arXiv},
	eprint = {1105.3151},
	primaryClass = {astro-ph.GA},
	adsurl = {https://ui.adsabs.harvard.edu/abs/2011ApJ...733L..43M}
}

@ARTICLE{2020A&A...637A..96C,
	author = {{Chrob{\'a}kov{\'a}}, {\v{Z}}. and {Nagy}, R. and {L{\'o}pez-Corredoira}, M.},
	title = "{Structure of the outer Galactic disc with Gaia DR2}",
	journal = {\aap},
	year = 2020,
	month = may,
	volume = {637},
	eid = {A96},
	pages = {A96},
	doi = {10.1051/0004-6361/201937289},
	archivePrefix = {arXiv},
	eprint = {2004.03247},
	primaryClass = {astro-ph.GA},
	adsurl = {https://ui.adsabs.harvard.edu/abs/2020A&A...637A..96C}
}

@ARTICLE{2024RAA....24f5005L,
	author = {{Liu}, Xiaopeng and {Tian}, Hao and {Cui}, Wenyuan and {Li}, Linlin and {Liu}, Jiaming and {Huo}, Zhenyan and {Gao}, Yawei},
	title = "{An Asymmetric Galactic Stellar Disk Traced by OB-type Stars from LAMOST DR7}",
	journal = {Research in Astronomy and Astrophysics},
	year = 2024,
	month = jun,
	volume = {24},
	number = {6},
	eid = {065005},
	pages = {065005},
	doi = {10.1088/1674-4527/ad3dc2},
	adsurl = {https://ui.adsabs.harvard.edu/abs/2024RAA....24f5005L}
}

@ARTICLE{2018MNRAS.478.3367W,
	author = {{Wang}, Hai-Feng and {Liu}, Chao and {Xu}, Yan and {Wan}, Jun-Chen and {Deng}, Licai},
	title = "{Mapping the Milky Way with LAMOST- III. Complicated spatial structure in the outer disc}",
	journal = {\mnras},
	year = 2018,
	month = aug,
	volume = {478},
	number = {3},
	pages = {3367-3379},
	doi = {10.1093/mnras/sty1058},
	archivePrefix = {arXiv},
	eprint = {1804.10485},
	primaryClass = {astro-ph.GA},
	adsurl = {https://ui.adsabs.harvard.edu/abs/2018MNRAS.478.3367W}
}

@ARTICLE{1992ApJ...400L..25R,
	author = {{Robin}, Annie C. and {Creze}, Michel and {Mohan}, Vijay},
	title = "{The Edge of the Galactic Disk}",
	journal = {\apjl},
	year = 1992,
	month = nov,
	volume = {400},
	pages = {L25},
	doi = {10.1086/186640},
	archivePrefix = {arXiv},
	eprint = {astro-ph/9210001},
	primaryClass = {astro-ph},
	adsurl = {https://ui.adsabs.harvard.edu/abs/1992ApJ...400L..25R}
}

@ARTICLE{2014A&A...567A.106L,
	author = {{L{\'o}pez-Corredoira}, M. and {Molg{\'o}}, J.},
	title = "{Flare in the Galactic stellar outer disc detected in SDSS-SEGUE data}",
	journal = {\aap},
	year = 2014,
	month = jul,
	volume = {567},
	eid = {A106},
	pages = {A106},
	doi = {10.1051/0004-6361/201423706},
	archivePrefix = {arXiv},
	eprint = {1405.7649},
	primaryClass = {astro-ph.GA},
	adsurl = {https://ui.adsabs.harvard.edu/abs/2014A&A...567A.106L}
}

@ARTICLE{2016ApJ...823...30B,
	author = {{Bovy}, Jo and {Rix}, Hans-Walter and {Schlafly}, Edward F. and {Nidever}, David L. and {Holtzman}, Jon A. and {Shetrone}, Matthew and {Beers}, Timothy C.},
	title = "{The Stellar Population Structure of the Galactic Disk}",
	journal = {\apj},
	year = 2016,
	month = may,
	volume = {823},
	number = {1},
	eid = {30},
	pages = {30},
	doi = {10.3847/0004-637X/823/1/30},
	archivePrefix = {arXiv},
	eprint = {1509.05796},
	primaryClass = {astro-ph.GA},
	adsurl = {https://ui.adsabs.harvard.edu/abs/2016ApJ...823...30B}
}

@ARTICLE{2023A&A...674A...3M,
	author = {{Montegriffo}, P. and {De Angeli}, F. and {Andrae}, R. and {Riello}, M. and {Pancino}, E. and {Sanna}, N. and {Bellazzini}, M. and {Evans}, D.~W. and {Carrasco}, J.~M. and {Sordo}, R. and {Busso}, G. and {Cacciari}, C. and {Jordi}, C. and {van Leeuwen}, F. and {Vallenari}, A. and {Altavilla}, G. and {Barstow}, M.~A. and {Brown}, A.~G.~A. and {Burgess}, P.~W. and {Castellani}, M. and {Cowell}, S. and {Davidson}, M. and {De Luise}, F. and {Delchambre}, L. and {Diener}, C. and {Fabricius}, C. and {Fr{\'e}mat}, Y. and {Fouesneau}, M. and {Gilmore}, G. and {Giuffrida}, G. and {Hambly}, N.~C. and {Harrison}, D.~L. and {Hidalgo}, S. and {Hodgkin}, S.~T. and {Holland}, G. and {Marinoni}, S. and {Osborne}, P.~J. and {Pagani}, C. and {Palaversa}, L. and {Piersimoni}, A.~M. and {Pulone}, L. and {Ragaini}, S. and {Rainer}, M. and {Richards}, P.~J. and {Rowell}, N. and {Ruz-Mieres}, D. and {Sarro}, L.~M. and {Walton}, N.~A. and {Yoldas}, A.},
	title = "{Gaia Data Release 3. External calibration of BP/RP low-resolution spectroscopic data}",
	journal = {\aap},
	year = 2023,
	month = jun,
	volume = {674},
	eid = {A3},
	pages = {A3},
	doi = {10.1051/0004-6361/202243880},
	archivePrefix = {arXiv},
	eprint = {2206.06205},
	primaryClass = {astro-ph.IM},
	adsurl = {https://ui.adsabs.harvard.edu/abs/2023A&A...674A...3M}
}

@ARTICLE{2023A&A...674A...2D,
	author = {{De Angeli}, F. and {Weiler}, M. and {Montegriffo}, P. and {Evans}, D.~W. and {Riello}, M. and {Andrae}, R. and {Carrasco}, J.~M. and {Busso}, G. and {Burgess}, P.~W. and {Cacciari}, C. and {Davidson}, M. and {Harrison}, D.~L. and {Hodgkin}, S.~T. and {Jordi}, C. and {Osborne}, P.~J. and {Pancino}, E. and {Altavilla}, G. and {Barstow}, M.~A. and {Bailer-Jones}, C.~A.~L. and {Bellazzini}, M. and {Brown}, A.~G.~A. and {Castellani}, M. and {Cowell}, S. and {Delchambre}, L. and {De Luise}, F. and {Diener}, C. and {Fabricius}, C. and {Fouesneau}, M. and {Fr{\'e}mat}, Y. and {Gilmore}, G. and {Giuffrida}, G. and {Hambly}, N.~C. and {Hidalgo}, S. and {Holland}, G. and {Kostrzewa-Rutkowska}, Z. and {van Leeuwen}, F. and {Lobel}, A. and {Marinoni}, S. and {Miller}, N. and {Pagani}, C. and {Palaversa}, L. and {Piersimoni}, A.~M. and {Pulone}, L. and {Ragaini}, S. and {Rainer}, M. and {Richards}, P.~J. and {Rixon}, G.~T. and {Ruz-Mieres}, D. and {Sanna}, N. and {Sarro}, L.~M. and {Rowell}, N. and {Sordo}, R. and {Walton}, N.~A. and {Yoldas}, A.},
	title = "{Gaia Data Release 3. Processing and validation of BP/RP low-resolution spectral data}",
	journal = {\aap},
	year = 2023,
	month = jun,
	volume = {674},
	eid = {A2},
	pages = {A2},
	doi = {10.1051/0004-6361/202243680},
	archivePrefix = {arXiv},
	eprint = {2206.06143},
	primaryClass = {astro-ph.IM},
	adsurl = {https://ui.adsabs.harvard.edu/abs/2023A&A...674A...2D}
}

@ARTICLE{2021A&A...652A..86C,
	author = {{Carrasco}, J.~M. and {Weiler}, M. and {Jordi}, C. and {Fabricius}, C. and {De Angeli}, F. and {Evans}, D.~W. and {van Leeuwen}, F. and {Riello}, M. and {Montegriffo}, P.},
	title = "{Internal calibration of Gaia BP/RP low-resolution spectra}",
	journal = {\aap},
	year = 2021,
	month = aug,
	volume = {652},
	eid = {A86},
	pages = {A86},
	doi = {10.1051/0004-6361/202141249},
	archivePrefix = {arXiv},
	eprint = {2106.01752},
	primaryClass = {astro-ph.IM},
	adsurl = {https://ui.adsabs.harvard.edu/abs/2021A&A...652A..86C}
}

@ARTICLE{2025AJ....169...61Y,
	author = {{Yu}, Zheng and {Chen}, Bingqiu and {Lian}, Jianhui and {Wang}, Chun and {Liu}, Xiaowei},
	title = "{The Stellar Disk Structure Revealed by the Mono-age Populations of the LAMOST Red Clump Star Sample}",
	journal = {\aj},
	year = 2025,
	month = feb,
	volume = {169},
	number = {2},
	eid = {61},
	pages = {61},
	doi = {10.3847/1538-3881/ad9582},
	archivePrefix = {arXiv},
	eprint = {2412.14743},
	primaryClass = {astro-ph.GA},
	adsurl = {https://ui.adsabs.harvard.edu/abs/2025AJ....169...61Y}
}

@ARTICLE{1979A&AS...38...15V,
	author = {{van der Kruit}, P.~C.},
	title = "{Optical surface photometry of eight spiral galaxies studied in Westerbork.}",
	journal = {\aaps},
	year = 1979,
	month = oct,
	volume = {38},
	pages = {15-38},
	adsurl = {https://ui.adsabs.harvard.edu/abs/1979A&AS...38...15V}
}

@ARTICLE{2002A&A...392..807P,
	author = {{Pohlen}, M. and {Dettmar}, R. -J. and {L{\"u}tticke}, R. and {Aronica}, G.},
	title = "{Outer edges of face-on spiral galaxies. Deep optical imaging of NGC 5923, UGC 9837 and NGC 5434}",
	journal = {\aap},
	year = 2002,
	month = sep,
	volume = {392},
	pages = {807-816},
	doi = {10.1051/0004-6361:20020994},
	adsurl = {https://ui.adsabs.harvard.edu/abs/2002A&A...392..807P}
}

@ARTICLE{2006A&A...454..759P,
	author = {{Pohlen}, M. and {Trujillo}, I.},
	title = "{The structure of galactic disks. Studying late-type spiral galaxies using SDSS}",
	journal = {\aap},
	year = 2006,
	month = aug,
	volume = {454},
	number = {3},
	pages = {759-772},
	doi = {10.1051/0004-6361:20064883},
	archivePrefix = {arXiv},
	eprint = {astro-ph/0603682},
	primaryClass = {astro-ph},
	adsurl = {https://ui.adsabs.harvard.edu/abs/2006A&A...454..759P}
}

@ARTICLE{2016A&A...596A..25L,
	author = {{Laine}, Jarkko and {Laurikainen}, Eija and {Salo}, Heikki},
	title = "{Influence of galaxy stellar mass and observed wavelength on disc breaks in S$^{4}$G, NIRS0S, and SDSS data}",
	journal = {\aap},
	year = 2016,
	month = nov,
	volume = {596},
	eid = {A25},
	pages = {A25},
	doi = {10.1051/0004-6361/201628397},
	archivePrefix = {arXiv},
	eprint = {1610.00610},
	primaryClass = {astro-ph.GA},
	adsurl = {https://ui.adsabs.harvard.edu/abs/2016A&A...596A..25L}
}

@ARTICLE{2024A&A...682L..17X,
	author = {{Xu}, Dewang and {Yu}, Si-Yue},
	title = "{JWST reveals a high fraction of disk breaks at 1 {\ensuremath{\leq}} z {\ensuremath{\leq}} 3}",
	journal = {\aap},
	year = 2024,
	month = feb,
	volume = {682},
	eid = {L17},
	pages = {L17},
	doi = {10.1051/0004-6361/202449252},
	archivePrefix = {arXiv},
	eprint = {2402.04233},
	primaryClass = {astro-ph.GA},
	adsurl = {https://ui.adsabs.harvard.edu/abs/2024A&A...682L..17X}
}

@ARTICLE{2014ApJ...782...64K,
	author = {{Kim}, Taehyun and {Gadotti}, Dimitri A. and {Sheth}, Kartik and {Athanassoula}, E. and {Bosma}, Albert and {Lee}, Myung Gyoon and {Madore}, Barry F. and {Elmegreen}, Bruce and {Knapen}, Johan H. and {Zaritsky}, Dennis and {Ho}, Luis C. and {Comer{\'o}n}, S{\'e}bastien and {Holwerda}, Benne and {Hinz}, Joannah L. and {Mu{\~n}oz-Mateos}, Juan-Carlos and {Cisternas}, Mauricio and {Erroz-Ferrer}, Santiago and {Buta}, Ron and {Laurikainen}, Eija and {Salo}, Heikki and {Laine}, Jarkko and {Men{\'e}ndez-Delmestre}, Kar{\'\i}n and {Regan}, Michael W. and {de Swardt}, Bonita and {Gil de Paz}, Armando and {Seibert}, Mark and {Mizusawa}, Trisha},
	title = "{Unveiling the Structure of Barred Galaxies at 3.6 {\ensuremath{\mu}}m with the Spitzer Survey of Stellar Structure in Galaxies (S$^{4}$G). I. Disk Breaks}",
	journal = {\apj},
	year = 2014,
	month = feb,
	volume = {782},
	number = {2},
	eid = {64},
	pages = {64},
	doi = {10.1088/0004-637X/782/2/64},
	archivePrefix = {arXiv},
	eprint = {1312.3384},
	primaryClass = {astro-ph.GA},
	adsurl = {https://ui.adsabs.harvard.edu/abs/2014ApJ...782...64K}
}

@ARTICLE{2000AJ....120.1579Y,
	author = {{York}, Donald G. and {Adelman}, J. and {Anderson}, John E., Jr. and {Anderson}, Scott F. and {Annis}, James and {Bahcall}, Neta A. and {Bakken}, J.~A. and {Barkhouser}, Robert and {Bastian}, Steven and {Berman}, Eileen and {Boroski}, William N. and {Bracker}, Steve and {Briegel}, Charlie and {Briggs}, John W. and {Brinkmann}, J. and {Brunner}, Robert and {Burles}, Scott and {Carey}, Larry and {Carr}, Michael A. and {Castander}, Francisco J. and {Chen}, Bing and {Colestock}, Patrick L. and {Connolly}, A.~J. and {Crocker}, J.~H. and {Csabai}, Istv{\'a}n and {Czarapata}, Paul C. and {Davis}, John Eric and {Doi}, Mamoru and {Dombeck}, Tom and {Eisenstein}, Daniel and {Ellman}, Nancy and {Elms}, Brian R. and {Evans}, Michael L. and {Fan}, Xiaohui and {Federwitz}, Glenn R. and {Fiscelli}, Larry and {Friedman}, Scott and {Frieman}, Joshua A. and {Fukugita}, Masataka and {Gillespie}, Bruce and {Gunn}, James E. and {Gurbani}, Vijay K. and {de Haas}, Ernst and {Haldeman}, Merle and {Harris}, Frederick H. and {Hayes}, J. and {Heckman}, Timothy M. and {Hennessy}, G.~S. and {Hindsley}, Robert B. and {Holm}, Scott and {Holmgren}, Donald J. and {Huang}, Chi-hao and {Hull}, Charles and {Husby}, Don and {Ichikawa}, Shin-Ichi and {Ichikawa}, Takashi and {Ivezi{\'c}}, {\v{Z}}eljko and {Kent}, Stephen and {Kim}, Rita S.~J. and {Kinney}, E. and {Klaene}, Mark and {Kleinman}, A.~N. and {Kleinman}, S. and {Knapp}, G.~R. and {Korienek}, John and {Kron}, Richard G. and {Kunszt}, Peter Z. and {Lamb}, D.~Q. and {Lee}, B. and {Leger}, R. French and {Limmongkol}, Siriluk and {Lindenmeyer}, Carl and {Long}, Daniel C. and {Loomis}, Craig and {Loveday}, Jon and {Lucinio}, Rich and {Lupton}, Robert H. and {MacKinnon}, Bryan and {Mannery}, Edward J. and {Mantsch}, P.~M. and {Margon}, Bruce and {McGehee}, Peregrine and {McKay}, Timothy A. and {Meiksin}, Avery and {Merelli}, Aronne and {Monet}, David G. and {Munn}, Jeffrey A. and {Narayanan}, Vijay K. and {Nash}, Thomas and {Neilsen}, Eric and {Neswold}, Rich and {Newberg}, Heidi Jo and {Nichol}, R.~C. and {Nicinski}, Tom and {Nonino}, Mario and {Okada}, Norio and {Okamura}, Sadanori and {Ostriker}, Jeremiah P. and {Owen}, Russell and {Pauls}, A. George and {Peoples}, John and {Peterson}, R.~L. and {Petravick}, Donald and {Pier}, Jeffrey R. and {Pope}, Adrian and {Pordes}, Ruth and {Prosapio}, Angela and {Rechenmacher}, Ron and {Quinn}, Thomas R. and {Richards}, Gordon T. and {Richmond}, Michael W. and {Rivetta}, Claudio H. and {Rockosi}, Constance M. and {Ruthmansdorfer}, Kurt and {Sandford}, Dale and {Schlegel}, David J. and {Schneider}, Donald P. and {Sekiguchi}, Maki and {Sergey}, Gary and {Shimasaku}, Kazuhiro and {Siegmund}, Walter A. and {Smee}, Stephen and {Smith}, J. Allyn and {Snedden}, S. and {Stone}, R. and {Stoughton}, Chris and {Strauss}, Michael A. and {Stubbs}, Christopher and {SubbaRao}, Mark and {Szalay}, Alexander S. and {Szapudi}, Istvan and {Szokoly}, Gyula P. and {Thakar}, Anirudda R. and {Tremonti}, Christy and {Tucker}, Douglas L. and {Uomoto}, Alan and {Vanden Berk}, Dan and {Vogeley}, Michael S. and {Waddell}, Patrick and {Wang}, Shu-i. and {Watanabe}, Masaru and {Weinberg}, David H. and {Yanny}, Brian and {Yasuda}, Naoki and {SDSS Collaboration}},
	title = "{The Sloan Digital Sky Survey: Technical Summary}",
	journal = {\aj},
	year = 2000,
	month = sep,
	volume = {120},
	number = {3},
	pages = {1579-1587},
	doi = {10.1086/301513},
	archivePrefix = {arXiv},
	eprint = {astro-ph/0006396},
	primaryClass = {astro-ph},
	adsurl = {https://ui.adsabs.harvard.edu/abs/2000AJ....120.1579Y}
}

@ARTICLE{2004AJ....128..502A,
	author = {{Abazajian}, Kevork and {Adelman-McCarthy}, Jennifer K. and {Ag{\"u}eros}, Marcel A. and {Allam}, Sahar S. and {Anderson}, Kurt and {Anderson}, Scott F. and {Annis}, James and {Bahcall}, Neta A. and {Baldry}, Ivan K. and {Bastian}, Steven and {Berlind}, Andreas and {Bernardi}, Mariangela and {Blanton}, Michael R. and {Bochanski}, John J., Jr. and {Boroski}, William N. and {Briggs}, John W. and {Brinkmann}, J. and {Brunner}, Robert J. and {Budav{\'a}ri}, Tam{\'a}s and {Carey}, Larry N. and {Carliles}, Samuel and {Castander}, Francisco J. and {Connolly}, A.~J. and {Csabai}, Istv{\'a}n and {Doi}, Mamoru and {Dong}, Feng and {Eisenstein}, Daniel J. and {Evans}, Michael L. and {Fan}, Xiaohui and {Finkbeiner}, Douglas P. and {Friedman}, Scott D. and {Frieman}, Joshua A. and {Fukugita}, Masataka and {Gal}, Roy R. and {Gillespie}, Bruce and {Glazebrook}, Karl and {Gray}, Jim and {Grebel}, Eva K. and {Gunn}, James E. and {Gurbani}, Vijay K. and {Hall}, Patrick B. and {Hamabe}, Masaru and {Harris}, Frederick H. and {Harris}, Hugh C. and {Harvanek}, Michael and {Heckman}, Timothy M. and {Hendry}, John S. and {Hennessy}, Gregory S. and {Hindsley}, Robert B. and {Hogan}, Craig J. and {Hogg}, David W. and {Holmgren}, Donald J. and {Ichikawa}, Shin-ichi and {Ichikawa}, Takashi and {Ivezi{\'c}}, {\v{Z}}eljko and {Jester}, Sebastian and {Johnston}, David E. and {Jorgensen}, Anders M. and {Kent}, Stephen M. and {Kleinman}, S.~J. and {Knapp}, G.~R. and {Kniazev}, Alexei Yu. and {Kron}, Richard G. and {Krzesinski}, Jurek and {Kunszt}, Peter Z. and {Kuropatkin}, Nickolai and {Lamb}, Donald Q. and {Lampeitl}, Hubert and {Lee}, Brian C. and {Leger}, R. French and {Li}, Nolan and {Lin}, Huan and {Loh}, Yeong-Shang and {Long}, Daniel C. and {Loveday}, Jon and {Lupton}, Robert H. and {Malik}, Tanu and {Margon}, Bruce and {Matsubara}, Takahiko and {McGehee}, Peregrine M. and {McKay}, Timothy A. and {Meiksin}, Avery and {Munn}, Jeffrey A. and {Nakajima}, Reiko and {Nash}, Thomas and {Neilsen}, Eric H., Jr. and {Newberg}, Heidi Jo and {Newman}, Peter R. and {Nichol}, Robert C. and {Nicinski}, Tom and {Nieto-Santisteban}, Maria and {Nitta}, Atsuko and {Okamura}, Sadanori and {O'Mullane}, William and {Ostriker}, Jeremiah P. and {Owen}, Russell and {Padmanabhan}, Nikhil and {Peoples}, John and {Pier}, Jeffrey R. and {Pope}, Adrian C. and {Quinn}, Thomas R. and {Richards}, Gordon T. and {Richmond}, Michael W. and {Rix}, Hans-Walter and {Rockosi}, Constance M. and {Schlegel}, David J. and {Schneider}, Donald P. and {Scranton}, Ryan and {Sekiguchi}, Maki and {Seljak}, Uros and {Sergey}, Gary and {Sesar}, Branimir and {Sheldon}, Erin and {Shimasaku}, Kazu and {Siegmund}, Walter A. and {Silvestri}, Nicole M. and {Smith}, J. Allyn and {Smol{\v{c}}i{\'c}}, Vernesa and {Snedden}, Stephanie A. and {Stebbins}, Albert and {Stoughton}, Chris and {Strauss}, Michael A. and {SubbaRao}, Mark and {Szalay}, Alexander S. and {Szapudi}, Istv{\'a}n and {Szkody}, Paula and {Szokoly}, Gyula P. and {Tegmark}, Max and {Teodoro}, Luis and {Thakar}, Aniruddha R. and {Tremonti}, Christy and {Tucker}, Douglas L. and {Uomoto}, Alan and {Vanden Berk}, Daniel E. and {Vandenberg}, Jan and {Vogeley}, Michael S. and {Voges}, Wolfgang and {Vogt}, Nicole P. and {Walkowicz}, Lucianne M. and {Wang}, Shu-i. and {Weinberg}, David H. and {West}, Andrew A. and {White}, Simon D.~M. and {Wilhite}, Brian C. and {Xu}, Yongzhong and {Yanny}, Brian and {Yasuda}, Naoki and {Yip}, Ching-Wa and {Yocum}, D.~R. and {York}, Donald G. and {Zehavi}, Idit and {Zibetti}, Stefano and {Zucker}, Daniel B.},
	title = "{The Second Data Release of the Sloan Digital Sky Survey}",
	journal = {\aj},
	year = 2004,
	month = jul,
	volume = {128},
	number = {1},
	pages = {502-512},
	doi = {10.1086/421365},
	archivePrefix = {arXiv},
	eprint = {astro-ph/0403325},
	primaryClass = {astro-ph},
	adsurl = {https://ui.adsabs.harvard.edu/abs/2004AJ....128..502A}
}

@ARTICLE{2011ApJS..193...29A,
	author = {{Aihara}, Hiroaki and {Allende Prieto}, Carlos and {An}, Deokkeun and {Anderson}, Scott F. and {Aubourg}, {\'E}ric and {Balbinot}, Eduardo and {Beers}, Timothy C. and {Berlind}, Andreas A. and {Bickerton}, Steven J. and {Bizyaev}, Dmitry and {Blanton}, Michael R. and {Bochanski}, John J. and {Bolton}, Adam S. and {Bovy}, Jo and {Brandt}, W.~N. and {Brinkmann}, J. and {Brown}, Peter J. and {Brownstein}, Joel R. and {Busca}, Nicolas G. and {Campbell}, Heather and {Carr}, Michael A. and {Chen}, Yanmei and {Chiappini}, Cristina and {Comparat}, Johan and {Connolly}, Natalia and {Cortes}, Marina and {Croft}, Rupert A.~C. and {Cuesta}, Antonio J. and {da Costa}, Luiz N. and {Davenport}, James R.~A. and {Dawson}, Kyle and {Dhital}, Saurav and {Ealet}, Anne and {Ebelke}, Garrett L. and {Edmondson}, Edward M. and {Eisenstein}, Daniel J. and {Escoffier}, Stephanie and {Esposito}, Massimiliano and {Evans}, Michael L. and {Fan}, Xiaohui and {Femen{\'\i}a Castell{\'a}}, Bruno and {Font-Ribera}, Andreu and {Frinchaboy}, Peter M. and {Ge}, Jian and {Gillespie}, Bruce A. and {Gilmore}, G. and {Gonz{\'a}lez Hern{\'a}ndez}, Jonay I. and {Gott}, J. Richard and {Gould}, Andrew and {Grebel}, Eva K. and {Gunn}, James E. and {Hamilton}, Jean-Christophe and {Harding}, Paul and {Harris}, David W. and {Hawley}, Suzanne L. and {Hearty}, Frederick R. and {Ho}, Shirley and {Hogg}, David W. and {Holtzman}, Jon A. and {Honscheid}, Klaus and {Inada}, Naohisa and {Ivans}, Inese I. and {Jiang}, Linhua and {Johnson}, Jennifer A. and {Jordan}, Cathy and {Jordan}, Wendell P. and {Kazin}, Eyal A. and {Kirkby}, David and {Klaene}, Mark A. and {Knapp}, G.~R. and {Kneib}, Jean-Paul and {Kochanek}, C.~S. and {Koesterke}, Lars and {Kollmeier}, Juna A. and {Kron}, Richard G. and {Lampeitl}, Hubert and {Lang}, Dustin and {Le Goff}, Jean-Marc and {Lee}, Young Sun and {Lin}, Yen-Ting and {Long}, Daniel C. and {Loomis}, Craig P. and {Lucatello}, Sara and {Lundgren}, Britt and {Lupton}, Robert H. and {Ma}, Zhibo and {MacDonald}, Nicholas and {Mahadevan}, Suvrath and {Maia}, Marcio A.~G. and {Makler}, Martin and {Malanushenko}, Elena and {Malanushenko}, Viktor and {Mandelbaum}, Rachel and {Maraston}, Claudia and {Margala}, Daniel and {Masters}, Karen L. and {McBride}, Cameron K. and {McGehee}, Peregrine M. and {McGreer}, Ian D. and {M{\'e}nard}, Brice and {Miralda-Escud{\'e}}, Jordi and {Morrison}, Heather L. and {Mullally}, F. and {Muna}, Demitri and {Munn}, Jeffrey A. and {Murayama}, Hitoshi and {Myers}, Adam D. and {Naugle}, Tracy and {Neto}, Angelo Fausti and {Nguyen}, Duy Cuong and {Nichol}, Robert C. and {O'Connell}, Robert W. and {Ogando}, Ricardo L.~C. and {Olmstead}, Matthew D. and {Oravetz}, Daniel J. and {Padmanabhan}, Nikhil and {Palanque-Delabrouille}, Nathalie and {Pan}, Kaike and {Pandey}, Parul and {P{\^a}ris}, Isabelle and {Percival}, Will J. and {Petitjean}, Patrick and {Pfaffenberger}, Robert and {Pforr}, Janine and {Phleps}, Stefanie and {Pichon}, Christophe and {Pieri}, Matthew M. and {Prada}, Francisco and {Price-Whelan}, Adrian M. and {Raddick}, M. Jordan and {Ramos}, Beatriz H.~F. and {Reyl{\'e}}, C{\'e}line and {Rich}, James and {Richards}, Gordon T. and {Rix}, Hans-Walter and {Robin}, Annie C. and {Rocha-Pinto}, Helio J. and {Rockosi}, Constance M. and {Roe}, Natalie A. and {Rollinde}, Emmanuel and {Ross}, Ashley J. and {Ross}, Nicholas P. and {Rossetto}, Bruno M. and {S{\'a}nchez}, Ariel G. and {Sayres}, Conor and {Schlegel}, David J. and {Schlesinger}, Katharine J. and {Schmidt}, Sarah J. and {Schneider}, Donald P. and {Sheldon}, Erin and {Shu}, Yiping and {Simmerer}, Jennifer and {Simmons}, Audrey E. and {Sivarani}, Thirupathi and {Snedden}, Stephanie A. and {Sobeck}, Jennifer S. and {Steinmetz}, Matthias and {Strauss}, Michael A. and {Szalay}, Alexander S. and {Tanaka}, Masayuki and {Thakar}, Aniruddha R. and {Thomas}, Daniel and {Tinker}, Jeremy L. and {Tofflemire}, Benjamin M. and {Tojeiro}, Rita and {Tremonti}, Christy A. and {Vandenberg}, Jan and {Vargas Maga{\~n}a}, M. and {Verde}, Licia and {Vogt}, Nicole P. and {Wake}, David A. and {Wang}, Ji and {Weaver}, Benjamin A. and {Weinberg}, David H. and {White}, Martin and {White}, Simon D.~M. and {Yanny}, Brian and {Yasuda}, Naoki and {Yeche}, Christophe and {Zehavi}, Idit},
	title = "{The Eighth Data Release of the Sloan Digital Sky Survey: First Data from SDSS-III}",
	journal = {\apjs},
	year = 2011,
	month = apr,
	volume = {193},
	number = {2},
	eid = {29},
	pages = {29},
	doi = {10.1088/0067-0049/193/2/29},
	archivePrefix = {arXiv},
	eprint = {1101.1559},
	primaryClass = {astro-ph.IM},
	adsurl = {https://ui.adsabs.harvard.edu/abs/2011ApJS..193...29A}
}

@ARTICLE{2012RAA....12.1197C,
	author = {{Cui}, Xiang-Qun and {Zhao}, Yong-Heng and {Chu}, Yao-Quan and {Li}, Guo-Ping and {Li}, Qi and {Zhang}, Li-Ping and {Su}, Hong-Jun and {Yao}, Zheng-Qiu and {Wang}, Ya-Nan and {Xing}, Xiao-Zheng and {Li}, Xin-Nan and {Zhu}, Yong-Tian and {Wang}, Gang and {Gu}, Bo-Zhong and {Luo}, A. -Li and {Xu}, Xin-Qi and {Zhang}, Zhen-Chao and {Liu}, Gen-Rong and {Zhang}, Hao-Tong and {Yang}, De-Hua and {Cao}, Shu-Yun and {Chen}, Hai-Yuan and {Chen}, Jian-Jun and {Chen}, Kun-Xin and {Chen}, Ying and {Chu}, Jia-Ru and {Feng}, Lei and {Gong}, Xue-Fei and {Hou}, Yong-Hui and {Hu}, Hong-Zhuan and {Hu}, Ning-Sheng and {Hu}, Zhong-Wen and {Jia}, Lei and {Jiang}, Fang-Hua and {Jiang}, Xiang and {Jiang}, Zi-Bo and {Jin}, Ge and {Li}, Ai-Hua and {Li}, Yan and {Li}, Ye-Ping and {Liu}, Guan-Qun and {Liu}, Zhi-Gang and {Lu}, Wen-Zhi and {Mao}, Yin-Dun and {Men}, Li and {Qi}, Yong-Jun and {Qi}, Zhao-Xiang and {Shi}, Huo-Ming and {Tang}, Zheng-Hong and {Tao}, Qing-Sheng and {Wang}, Da-Qi and {Wang}, Dan and {Wang}, Guo-Min and {Wang}, Hai and {Wang}, Jia-Ning and {Wang}, Jian and {Wang}, Jian-Ling and {Wang}, Jian-Ping and {Wang}, Lei and {Wang}, Shu-Qing and {Wang}, You and {Wang}, Yue-Fei and {Xu}, Ling-Zhe and {Xu}, Yan and {Yang}, Shi-Hai and {Yu}, Yong and {Yuan}, Hui and {Yuan}, Xiang-Yan and {Zhai}, Chao and {Zhang}, Jing and {Zhang}, Yan-Xia and {Zhang}, Yong and {Zhao}, Ming and {Zhou}, Fang and {Zhou}, Guo-Hua and {Zhu}, Jie and {Zou}, Si-Cheng},
	title = "{The Large Sky Area Multi-Object Fiber Spectroscopic Telescope (LAMOST)}",
	journal = {Research in Astronomy and Astrophysics},
	year = 2012,
	month = sep,
	volume = {12},
	number = {9},
	pages = {1197-1242},
	doi = {10.1088/1674-4527/12/9/003},
	adsurl = {https://ui.adsabs.harvard.edu/abs/2012RAA....12.1197C}
}

@ARTICLE{2012RAA....12.1243L,
	author = {{Luo}, A. -Li and {Zhang}, Hao-Tong and {Zhao}, Yong-Heng and {Zhao}, Gang and {Cui}, Xiang-Qun and {Li}, Guo-Ping and {Chu}, Yao-Quan and {Shi}, Jian-Rong and {Wang}, Gang and {Zhang}, Jian-Nan and {Bai}, Zhong-Rui and {Chen}, Xiao-Yan and {Wang}, Feng-Fei and {Guo}, Yan-Xin and {Chen}, Jian-Jun and {Du}, Bing and {Kong}, Xiao and {Lei}, Ya-Juan and {Li}, Yin-Bi and {Song}, Yi-Han and {Wu}, Yue and {Zhang}, Yan-Xia and {Zhou}, Xin-Lin and {Zuo}, Fang and {Du}, Peng and {He}, Lin and {Hou}, Wen and {Dong}, Yi-Qiao and {Li}, Jian and {Li}, Guang-Wei and {Li}, Shuang and {Song}, Jing and {Tian}, Yuan and {Wang}, Meng-Xin and {Wu}, Ke-Fei and {Yang}, Hui-Qin and {Yuan}, Hai-Long and {Cao}, Shu-Yun and {Chen}, Hai-Yuan and {Chen}, Kun-Xin and {Chen}, Ying and {Chu}, Jia-Ru and {Feng}, Lei and {Gong}, Xue-Fei and {Gu}, Bo-Zhong and {Hou}, Yong-Hui and {Huo}, Zhi-Ying and {Hu}, Hong-Zhuan and {Hu}, Ning-Sheng and {Hu}, Zhong-Wen and {Jia}, Lei and {Jiang}, Fang-Hua and {Jiang}, Xiang and {Jiang}, Zi-Bo and {Jin}, Ge and {Li}, Ai-Hua and {Li}, Qi and {Li}, Xin-Nan and {Li}, Yan and {Li}, Ye-Ping and {Liu}, Gen-Rong and {Liu}, Guan-Qun and {Liu}, Zhi-Gang and {Lu}, Qi-Shuai and {Lu}, Wen-Zhi and {Luo}, Yu and {Mao}, Yin-Dun and {Men}, Li and {Ni}, Ji-Jun and {Qi}, Yong-Jun and {Qi}, Zhao-Xiang and {Shi}, Huo-Ming and {Su}, Ding-Qiang and {Sun}, Shi-Wei and {Su}, Hong-Jun and {Tang}, Zheng-Hong and {Tao}, Qing-Sheng and {Tu}, Liang-Ping and {Wang}, Da-Qing and {Wang}, Dan and {Wang}, Guo-Min and {Wang}, Hai and {Wang}, Jia-Ning and {Wang}, Jian and {Wang}, Jian-Ling and {Wang}, Jian-Ping and {Wang}, Lei and {Wang}, Shou-Guan and {Wang}, Shu-Qing and {Wang}, Ya-Nan and {Wang}, You and {Wang}, Yue-Fei and {Wei}, Ming-Zhi and {Xue}, Xiang-Xiang and {Xing}, Xiao-Zheng and {Xu}, Ling-Zhe and {Xu}, Xin-Qi and {Xu}, Yan and {Yang}, De-Hua and {Yang}, Shi-Hai and {Yao}, Zheng-Qiu and {Yu}, Yong and {Yuan}, Hui and {Zhai}, Chao and {Zhang}, En-Peng and {Zhang}, Jing and {Zhang}, Li-Ping and {Zhang}, Wei and {Zhang}, Yong and {Zhang}, Zhen-Chao and {Zhao}, Ming and {Zhou}, Fang and {Zhu}, Yong-Tian and {Zhu}, Jie and {Zou}, Si-Cheng},
	title = "{Data release of the LAMOST pilot survey}",
	journal = {Research in Astronomy and Astrophysics},
	year = 2012,
	month = sep,
	volume = {12},
	number = {9},
	pages = {1243-1246},
	doi = {10.1088/1674-4527/12/9/004},
	adsurl = {https://ui.adsabs.harvard.edu/abs/2012RAA....12.1243L}
}

@ARTICLE{2006ChJAA...6..265Z,
	author = {{Zhao}, Gang and {Chen}, Yu-Qin and {Shi}, Jian-Rong and {Liang}, Yan-Chun and {Hou}, Jin-Liang and {Chen}, Li and {Zhang}, Hua-Wei and {Li}, Ai-Gen},
	title = "{Stellar Abundance and Galactic Chemical Evolution through LAMOST Spectroscopic Survey}",
	journal = {\cjaa},
	year = 2006,
	month = jun,
	volume = {6},
	number = {3},
	pages = {265-280},
	doi = {10.1088/1009-9271/6/3/01},
	adsurl = {https://ui.adsabs.harvard.edu/abs/2006ChJAA...6..265Z}
}

@ARTICLE{2012RAA....12..723Z,
	author = {{Zhao}, Gang and {Zhao}, Yong-Heng and {Chu}, Yao-Quan and {Jing}, Yi-Peng and {Deng}, Li-Cai},
	title = "{LAMOST spectral survey {\textemdash} An overview}",
	journal = {Research in Astronomy and Astrophysics},
	year = 2012,
	month = jul,
	volume = {12},
	number = {7},
	pages = {723-734},
	doi = {10.1088/1674-4527/12/7/002},
	adsurl = {https://ui.adsabs.harvard.edu/abs/2012RAA....12..723Z}
}

@ARTICLE{2015MNRAS.449.2604D,
	author = {{De Silva}, G.~M. and {Freeman}, K.~C. and {Bland-Hawthorn}, J. and {Martell}, S. and {de Boer}, E. Wylie and {Asplund}, M. and {Keller}, S. and {Sharma}, S. and {Zucker}, D.~B. and {Zwitter}, T. and {Anguiano}, B. and {Bacigalupo}, C. and {Bayliss}, D. and {Beavis}, M.~A. and {Bergemann}, M. and {Campbell}, S. and {Cannon}, R. and {Carollo}, D. and {Casagrande}, L. and {Casey}, A.~R. and {Da Costa}, G. and {D'Orazi}, V. and {Dotter}, A. and {Duong}, L. and {Heger}, A. and {Ireland}, M.~J. and {Kafle}, P.~R. and {Kos}, J. and {Lattanzio}, J. and {Lewis}, G.~F. and {Lin}, J. and {Lind}, K. and {Munari}, U. and {Nataf}, D.~M. and {O'Toole}, S. and {Parker}, Q. and {Reid}, W. and {Schlesinger}, K.~J. and {Sheinis}, A. and {Simpson}, J.~D. and {Stello}, D. and {Ting}, Y. -S. and {Traven}, G. and {Watson}, F. and {Wittenmyer}, R. and {Yong}, D. and {{\v{Z}}erjal}, M.},
	title = "{The GALAH survey: scientific motivation}",
	journal = {\mnras},
	year = 2015,
	month = may,
	volume = {449},
	number = {3},
	pages = {2604-2617},
	doi = {10.1093/mnras/stv327},
	archivePrefix = {arXiv},
	eprint = {1502.04767},
	primaryClass = {astro-ph.GA},
	adsurl = {https://ui.adsabs.harvard.edu/abs/2015MNRAS.449.2604D}
}

@ARTICLE{2021MNRAS.506..150B,
	author = {{Buder}, Sven and {Sharma}, Sanjib and {Kos}, Janez and {Amarsi}, Anish M. and {Nordlander}, Thomas and {Lind}, Karin and {Martell}, Sarah L. and {Asplund}, Martin and {Bland-Hawthorn}, Joss and {Casey}, Andrew R. and {de Silva}, Gayandhi M. and {D'Orazi}, Valentina and {Freeman}, Ken C. and {Hayden}, Michael R. and {Lewis}, Geraint F. and {Lin}, Jane and {Schlesinger}, Katharine J. and {Simpson}, Jeffrey D. and {Stello}, Dennis and {Zucker}, Daniel B. and {Zwitter}, Toma{\v{z}} and {Beeson}, Kevin L. and {Buck}, Tobias and {Casagrande}, Luca and {Clark}, Jake T. and {{\v{C}}otar}, Klemen and {da Costa}, Gary S. and {de Grijs}, Richard and {Feuillet}, Diane and {Horner}, Jonathan and {Kafle}, Prajwal R. and {Khanna}, Shourya and {Kobayashi}, Chiaki and {Liu}, Fan and {Montet}, Benjamin T. and {Nandakumar}, Govind and {Nataf}, David M. and {Ness}, Melissa K. and {Spina}, Lorenzo and {Tepper-Garc{\'\i}a}, Thor and {Ting}, Yuan-Sen and {Traven}, Gregor and {Vogrin{\v{c}}i{\v{c}}}, Rok and {Wittenmyer}, Robert A. and {Wyse}, Rosemary F.~G. and {{\v{Z}}erjal}, Maru{\v{s}}a and {Galah Collaboration}},
	title = "{The GALAH+ survey: Third data release}",
	journal = {\mnras},
	year = 2021,
	month = sep,
	volume = {506},
	number = {1},
	pages = {150-201},
	doi = {10.1093/mnras/stab1242},
	archivePrefix = {arXiv},
	eprint = {2011.02505},
	primaryClass = {astro-ph.GA},
	adsurl = {https://ui.adsabs.harvard.edu/abs/2021MNRAS.506..150B}
}

@ARTICLE{2016arXiv161205560C,
	author = {{Chambers}, K.~C. and {Magnier}, E.~A. and {Metcalfe}, N. and {Flewelling}, H.~A. and {Huber}, M.~E. and {Waters}, C.~Z. and {Denneau}, L. and {Draper}, P.~W. and {Farrow}, D. and {Finkbeiner}, D.~P. and {Holmberg}, C. and {Koppenhoefer}, J. and {Price}, P.~A. and {Rest}, A. and {Saglia}, R.~P. and {Schlafly}, E.~F. and {Smartt}, S.~J. and {Sweeney}, W. and {Wainscoat}, R.~J. and {Burgett}, W.~S. and {Chastel}, S. and {Grav}, T. and {Heasley}, J.~N. and {Hodapp}, K.~W. and {Jedicke}, R. and {Kaiser}, N. and {Kudritzki}, R. -P. and {Luppino}, G.~A. and {Lupton}, R.~H. and {Monet}, D.~G. and {Morgan}, J.~S. and {Onaka}, P.~M. and {Shiao}, B. and {Stubbs}, C.~W. and {Tonry}, J.~L. and {White}, R. and {Ba{\~n}ados}, E. and {Bell}, E.~F. and {Bender}, R. and {Bernard}, E.~J. and {Boegner}, M. and {Boffi}, F. and {Botticella}, M.~T. and {Calamida}, A. and {Casertano}, S. and {Chen}, W. -P. and {Chen}, X. and {Cole}, S. and {Deacon}, N. and {Frenk}, C. and {Fitzsimmons}, A. and {Gezari}, S. and {Gibbs}, V. and {Goessl}, C. and {Goggia}, T. and {Gourgue}, R. and {Goldman}, B. and {Grant}, P. and {Grebel}, E.~K. and {Hambly}, N.~C. and {Hasinger}, G. and {Heavens}, A.~F. and {Heckman}, T.~M. and {Henderson}, R. and {Henning}, T. and {Holman}, M. and {Hopp}, U. and {Ip}, W. -H. and {Isani}, S. and {Jackson}, M. and {Keyes}, C.~D. and {Koekemoer}, A.~M. and {Kotak}, R. and {Le}, D. and {Liska}, D. and {Long}, K.~S. and {Lucey}, J.~R. and {Liu}, M. and {Martin}, N.~F. and {Masci}, G. and {McLean}, B. and {Mindel}, E. and {Misra}, P. and {Morganson}, E. and {Murphy}, D.~N.~A. and {Obaika}, A. and {Narayan}, G. and {Nieto-Santisteban}, M.~A. and {Norberg}, P. and {Peacock}, J.~A. and {Pier}, E.~A. and {Postman}, M. and {Primak}, N. and {Rae}, C. and {Rai}, A. and {Riess}, A. and {Riffeser}, A. and {Rix}, H.~W. and {R{\"o}ser}, S. and {Russel}, R. and {Rutz}, L. and {Schilbach}, E. and {Schultz}, A.~S.~B. and {Scolnic}, D. and {Strolger}, L. and {Szalay}, A. and {Seitz}, S. and {Small}, E. and {Smith}, K.~W. and {Soderblom}, D.~R. and {Taylor}, P. and {Thomson}, R. and {Taylor}, A.~N. and {Thakar}, A.~R. and {Thiel}, J. and {Thilker}, D. and {Unger}, D. and {Urata}, Y. and {Valenti}, J. and {Wagner}, J. and {Walder}, T. and {Walter}, F. and {Watters}, S.~P. and {Werner}, S. and {Wood-Vasey}, W.~M. and {Wyse}, R.},
	title = "{The Pan-STARRS1 Surveys}",
	journal = {arXiv e-prints},
	year = 2016,
	month = dec,
	eid = {arXiv:1612.05560},
	pages = {arXiv:1612.05560},
	doi = {10.48550/arXiv.1612.05560},
	archivePrefix = {arXiv},
	eprint = {1612.05560},
	primaryClass = {astro-ph.IM},
	adsurl = {https://ui.adsabs.harvard.edu/abs/2016arXiv161205560C}
}

@ARTICLE{2009AJ....137.4377Y,
	author = {{Yanny}, Brian and {Rockosi}, Constance and {Newberg}, Heidi Jo and {Knapp}, Gillian R. and {Adelman-McCarthy}, Jennifer K. and {Alcorn}, Bonnie and {Allam}, Sahar and {Allende Prieto}, Carlos and {An}, Deokkeun and {Anderson}, Kurt S.~J. and {Anderson}, Scott and {Bailer-Jones}, Coryn A.~L. and {Bastian}, Steve and {Beers}, Timothy C. and {Bell}, Eric and {Belokurov}, Vasily and {Bizyaev}, Dmitry and {Blythe}, Norm and {Bochanski}, John J. and {Boroski}, William N. and {Brinchmann}, Jarle and {Brinkmann}, J. and {Brewington}, Howard and {Carey}, Larry and {Cudworth}, Kyle M. and {Evans}, Michael and {Evans}, N.~W. and {Gates}, Evalyn and {G{\"a}nsicke}, B.~T. and {Gillespie}, Bruce and {Gilmore}, Gerald and {Nebot Gomez-Moran}, Ada and {Grebel}, Eva K. and {Greenwell}, Jim and {Gunn}, James E. and {Jordan}, Cathy and {Jordan}, Wendell and {Harding}, Paul and {Harris}, Hugh and {Hendry}, John S. and {Holder}, Diana and {Ivans}, Inese I. and {Ivezi{\v{c}}}, {\v{Z}}eljko and {Jester}, Sebastian and {Johnson}, Jennifer A. and {Kent}, Stephen M. and {Kleinman}, Scot and {Kniazev}, Alexei and {Krzesinski}, Jurek and {Kron}, Richard and {Kuropatkin}, Nikolay and {Lebedeva}, Svetlana and {Lee}, Young Sun and {French Leger}, R. and {L{\'e}pine}, S{\'e}bastien and {Levine}, Steve and {Lin}, Huan and {Long}, Daniel C. and {Loomis}, Craig and {Lupton}, Robert and {Malanushenko}, Olena and {Malanushenko}, Viktor and {Margon}, Bruce and {Martinez-Delgado}, David and {McGehee}, Peregrine and {Monet}, Dave and {Morrison}, Heather L. and {Munn}, Jeffrey A. and {Neilsen}, Jr., Eric H. and {Nitta}, Atsuko and {Norris}, John E. and {Oravetz}, Dan and {Owen}, Russell and {Padmanabhan}, Nikhil and {Pan}, Kaike and {Peterson}, R.~S. and {Pier}, Jeffrey R. and {Platson}, Jared and {Re Fiorentin}, Paola and {Richards}, Gordon T. and {Rix}, Hans-Walter and {Schlegel}, David J. and {Schneider}, Donald P. and {Schreiber}, Matthias R. and {Schwope}, Axel and {Sibley}, Valena and {Simmons}, Audrey and {Snedden}, Stephanie A. and {Allyn Smith}, J. and {Stark}, Larry and {Stauffer}, Fritz and {Steinmetz}, M. and {Stoughton}, C. and {SubbaRao}, Mark and {Szalay}, Alex and {Szkody}, Paula and {Thakar}, Aniruddha R. and {Sivarani}, Thirupathi and {Tucker}, Douglas and {Uomoto}, Alan and {Vanden Berk}, Dan and {Vidrih}, Simon and {Wadadekar}, Yogesh and {Watters}, Shannon and {Wilhelm}, Ron and {Wyse}, Rosemary F.~G. and {Yarger}, Jean and {Zucker}, Dan},
	title = "{SEGUE: A Spectroscopic Survey of 240,000 Stars with g = 14-20}",
	journal = {\aj},
	year = 2009,
	month = may,
	volume = {137},
	number = {5},
	pages = {4377-4399},
	doi = {10.1088/0004-6256/137/5/4377},
	archivePrefix = {arXiv},
	eprint = {0902.1781},
	primaryClass = {astro-ph.GA},
	adsurl = {https://ui.adsabs.harvard.edu/abs/2009AJ....137.4377Y}
}

@ARTICLE{2016A&A...595A...2G,
	author = {{Gaia Collaboration} and {Brown}, A.~G.~A. and {Vallenari}, A. and {Prusti}, T. and {de Bruijne}, J.~H.~J. and {Mignard}, F. and {Drimmel}, R. and {Babusiaux}, C. and {Bailer-Jones}, C.~A.~L. and {Bastian}, U. and {Biermann}, M. and {Evans}, D.~W. and {Eyer}, L. and {Jansen}, F. and {Jordi}, C. and {Katz}, D. and {Klioner}, S.~A. and {Lammers}, U. and {Lindegren}, L. and {Luri}, X. and {O'Mullane}, W. and {Panem}, C. and {Pourbaix}, D. and {Randich}, S. and {Sartoretti}, P. and {Siddiqui}, H.~I. and {Soubiran}, C. and {Valette}, V. and {van Leeuwen}, F. and {Walton}, N.~A. and {Aerts}, C. and {Arenou}, F. and {Cropper}, M. and {H{\o}g}, E. and {Lattanzi}, M.~G. and {Grebel}, E.~K. and {Holland}, A.~D. and {Huc}, C. and {Passot}, X. and {Perryman}, M. and {Bramante}, L. and {Cacciari}, C. and {Casta{\~n}eda}, J. and {Chaoul}, L. and {Cheek}, N. and {De Angeli}, F. and {Fabricius}, C. and {Guerra}, R. and {Hern{\'a}ndez}, J. and {Jean-Antoine-Piccolo}, A. and {Masana}, E. and {Messineo}, R. and {Mowlavi}, N. and {Nienartowicz}, K. and {Ord{\'o}{\~n}ez-Blanco}, D. and {Panuzzo}, P. and {Portell}, J. and {Richards}, P.~J. and {Riello}, M. and {Seabroke}, G.~M. and {Tanga}, P. and {Th{\'e}venin}, F. and {Torra}, J. and {Els}, S.~G. and {Gracia-Abril}, G. and {Comoretto}, G. and {Garcia-Reinaldos}, M. and {Lock}, T. and {Mercier}, E. and {Altmann}, M. and {Andrae}, R. and {Astraatmadja}, T.~L. and {Bellas-Velidis}, I. and {Benson}, K. and {Berthier}, J. and {Blomme}, R. and {Busso}, G. and {Carry}, B. and {Cellino}, A. and {Clementini}, G. and {Cowell}, S. and {Creevey}, O. and {Cuypers}, J. and {Davidson}, M. and {De Ridder}, J. and {de Torres}, A. and {Delchambre}, L. and {Dell'Oro}, A. and {Ducourant}, C. and {Fr{\'e}mat}, Y. and {Garc{\'\i}a-Torres}, M. and {Gosset}, E. and {Halbwachs}, J. -L. and {Hambly}, N.~C. and {Harrison}, D.~L. and {Hauser}, M. and {Hestroffer}, D. and {Hodgkin}, S.~T. and {Huckle}, H.~E. and {Hutton}, A. and {Jasniewicz}, G. and {Jordan}, S. and {Kontizas}, M. and {Korn}, A.~J. and {Lanzafame}, A.~C. and {Manteiga}, M. and {Moitinho}, A. and {Muinonen}, K. and {Osinde}, J. and {Pancino}, E. and {Pauwels}, T. and {Petit}, J. -M. and {Recio-Blanco}, A. and {Robin}, A.~C. and {Sarro}, L.~M. and {Siopis}, C. and {Smith}, M. and {Smith}, K.~W. and {Sozzetti}, A. and {Thuillot}, W. and {van Reeven}, W. and {Viala}, Y. and {Abbas}, U. and {Abreu Aramburu}, A. and {Accart}, S. and {Aguado}, J.~J. and {Allan}, P.~M. and {Allasia}, W. and {Altavilla}, G. and {{\'A}lvarez}, M.~A. and {Alves}, J. and {Anderson}, R.~I. and {Andrei}, A.~H. and {Anglada Varela}, E. and {Antiche}, E. and {Antoja}, T. and {Ant{\'o}n}, S. and {Arcay}, B. and {Bach}, N. and {Baker}, S.~G. and {Balaguer-N{\'u}{\~n}ez}, L. and {Barache}, C. and {Barata}, C. and {Barbier}, A. and {Barblan}, F. and {Barrado y Navascu{\'e}s}, D. and {Barros}, M. and {Barstow}, M.~A. and {Becciani}, U. and {Bellazzini}, M. and {Bello Garc{\'\i}a}, A. and {Belokurov}, V. and {Bendjoya}, P. and {Berihuete}, A. and {Bianchi}, L. and {Bienaym{\'e}}, O. and {Billebaud}, F. and {Blagorodnova}, N. and {Blanco-Cuaresma}, S. and {Boch}, T. and {Bombrun}, A. and {Borrachero}, R. and {Bouquillon}, S. and {Bourda}, G. and {Bouy}, H. and {Bragaglia}, A. and {Breddels}, M.~A. and {Brouillet}, N. and {Br{\"u}semeister}, T. and {Bucciarelli}, B. and {Burgess}, P. and {Burgon}, R. and {Burlacu}, A. and {Busonero}, D. and {Buzzi}, R. and {Caffau}, E. and {Cambras}, J. and {Campbell}, H. and {Cancelliere}, R. and {Cantat-Gaudin}, T. and {Carlucci}, T. and {Carrasco}, J.~M. and {Castellani}, M. and {Charlot}, P. and {Charnas}, J. and {Chiavassa}, A. and {Clotet}, M. and {Cocozza}, G. and {Collins}, R.~S. and {Costigan}, G. and {Crifo}, F. and {Cross}, N.~J.~G. and {Crosta}, M. and {Crowley}, C. and {Dafonte}, C. and {Damerdji}, Y. and {Dapergolas}, A. and {David}, P. and {David}, M. and {De Cat}, P.},
	title = "{Gaia Data Release 1. Summary of the astrometric, photometric, and survey properties}",
	journal = {\aap},
	year = 2016,
	month = nov,
	volume = {595},
	eid = {A2},
	pages = {A2},
	doi = {10.1051/0004-6361/201629512},
	archivePrefix = {arXiv},
	eprint = {1609.04172},
	primaryClass = {astro-ph.IM},
	adsurl = {https://ui.adsabs.harvard.edu/abs/2016A&A...595A...2G}
}

@ARTICLE{2023A&A...674A...1G,
	author = {{Gaia Collaboration} and {Vallenari}, A. and {Brown}, A.~G.~A. and {Prusti}, T. and {de Bruijne}, J.~H.~J. and {Arenou}, F. and {Babusiaux}, C. and {Biermann}, M. and {Creevey}, O.~L. and {Ducourant}, C. and {Evans}, D.~W. and {Eyer}, L. and {Guerra}, R. and {Hutton}, A. and {Jordi}, C. and {Klioner}, S.~A. and {Lammers}, U.~L. and {Lindegren}, L. and {Luri}, X. and {Mignard}, F. and {Panem}, C. and {Pourbaix}, D. and {Randich}, S. and {Sartoretti}, P. and {Soubiran}, C. and {Tanga}, P. and {Walton}, N.~A. and {Bailer-Jones}, C.~A.~L. and {Bastian}, U. and {Drimmel}, R. and {Jansen}, F. and {Katz}, D. and {Lattanzi}, M.~G. and {van Leeuwen}, F. and {Bakker}, J. and {Cacciari}, C. and {Casta{\~n}eda}, J. and {De Angeli}, F. and {Fabricius}, C. and {Fouesneau}, M. and {Fr{\'e}mat}, Y. and {Galluccio}, L. and {Guerrier}, A. and {Heiter}, U. and {Masana}, E. and {Messineo}, R. and {Mowlavi}, N. and {Nicolas}, C. and {Nienartowicz}, K. and {Pailler}, F. and {Panuzzo}, P. and {Riclet}, F. and {Roux}, W. and {Seabroke}, G.~M. and {Sordo}, R. and {Th{\'e}venin}, F. and {Gracia-Abril}, G. and {Portell}, J. and {Teyssier}, D. and {Altmann}, M. and {Andrae}, R. and {Audard}, M. and {Bellas-Velidis}, I. and {Benson}, K. and {Berthier}, J. and {Blomme}, R. and {Burgess}, P.~W. and {Busonero}, D. and {Busso}, G. and {C{\'a}novas}, H. and {Carry}, B. and {Cellino}, A. and {Cheek}, N. and {Clementini}, G. and {Damerdji}, Y. and {Davidson}, M. and {de Teodoro}, P. and {Nu{\~n}ez Campos}, M. and {Delchambre}, L. and {Dell'Oro}, A. and {Esquej}, P. and {Fern{\'a}ndez-Hern{\'a}ndez}, J. and {Fraile}, E. and {Garabato}, D. and {Garc{\'\i}a-Lario}, P. and {Gosset}, E. and {Haigron}, R. and {Halbwachs}, J. -L. and {Hambly}, N.~C. and {Harrison}, D.~L. and {Hern{\'a}ndez}, J. and {Hestroffer}, D. and {Hodgkin}, S.~T. and {Holl}, B. and {Jan{\ss}en}, K. and {Jevardat de Fombelle}, G. and {Jordan}, S. and {Krone-Martins}, A. and {Lanzafame}, A.~C. and {L{\"o}ffler}, W. and {Marchal}, O. and {Marrese}, P.~M. and {Moitinho}, A. and {Muinonen}, K. and {Osborne}, P. and {Pancino}, E. and {Pauwels}, T. and {Recio-Blanco}, A. and {Reyl{\'e}}, C. and {Riello}, M. and {Rimoldini}, L. and {Roegiers}, T. and {Rybizki}, J. and {Sarro}, L.~M. and {Siopis}, C. and {Smith}, M. and {Sozzetti}, A. and {Utrilla}, E. and {van Leeuwen}, M. and {Abbas}, U. and {{\'A}brah{\'a}m}, P. and {Abreu Aramburu}, A. and {Aerts}, C. and {Aguado}, J.~J. and {Ajaj}, M. and {Aldea-Montero}, F. and {Altavilla}, G. and {{\'A}lvarez}, M.~A. and {Alves}, J. and {Anders}, F. and {Anderson}, R.~I. and {Anglada Varela}, E. and {Antoja}, T. and {Baines}, D. and {Baker}, S.~G. and {Balaguer-N{\'u}{\~n}ez}, L. and {Balbinot}, E. and {Balog}, Z. and {Barache}, C. and {Barbato}, D. and {Barros}, M. and {Barstow}, M.~A. and {Bartolom{\'e}}, S. and {Bassilana}, J. -L. and {Bauchet}, N. and {Becciani}, U. and {Bellazzini}, M. and {Berihuete}, A. and {Bernet}, M. and {Bertone}, S. and {Bianchi}, L. and {Binnenfeld}, A. and {Blanco-Cuaresma}, S. and {Blazere}, A. and {Boch}, T. and {Bombrun}, A. and {Bossini}, D. and {Bouquillon}, S. and {Bragaglia}, A. and {Bramante}, L. and {Breedt}, E. and {Bressan}, A. and {Brouillet}, N. and {Brugaletta}, E. and {Bucciarelli}, B. and {Burlacu}, A. and {Butkevich}, A.~G. and {Buzzi}, R. and {Caffau}, E. and {Cancelliere}, R. and {Cantat-Gaudin}, T. and {Carballo}, R. and {Carlucci}, T. and {Carnerero}, M.~I. and {Carrasco}, J.~M. and {Casamiquela}, L. and {Castellani}, M. and {Castro-Ginard}, A. and {Chaoul}, L. and {Charlot}, P. and {Chemin}, L. and {Chiaramida}, V. and {Chiavassa}, A. and {Chornay}, N. and {Comoretto}, G. and {Contursi}, G. and {Cooper}, W.~J. and {Cornez}, T. and {Cowell}, S. and {Crifo}, F. and {Cropper}, M. and {Crosta}, M. and {Crowley}, C. and {Dafonte}, C. and {Dapergolas}, A. and {David}, M. and {David}, P. and {de Laverny}, P. and {De Luise}, F. and {De March}, R.},
	title = "{Gaia Data Release 3. Summary of the content and survey properties}",
	journal = {\aap},
	year = 2023,
	month = jun,
	volume = {674},
	eid = {A1},
	pages = {A1},
	doi = {10.1051/0004-6361/202243940},
	archivePrefix = {arXiv},
	eprint = {2208.00211},
	primaryClass = {astro-ph.GA},
	adsurl = {https://ui.adsabs.harvard.edu/abs/2023A&A...674A...1G}
}

@ARTICLE{2018A&A...616A...1G,
	author = {{Gaia Collaboration} and {Brown}, A.~G.~A. and {Vallenari}, A. and {Prusti}, T. and {de Bruijne}, J.~H.~J. and {Babusiaux}, C. and {Bailer-Jones}, C.~A.~L. and {Biermann}, M. and {Evans}, D.~W. and {Eyer}, L. and {Jansen}, F. and {Jordi}, C. and {Klioner}, S.~A. and {Lammers}, U. and {Lindegren}, L. and {Luri}, X. and {Mignard}, F. and {Panem}, C. and {Pourbaix}, D. and {Randich}, S. and {Sartoretti}, P. and {Siddiqui}, H.~I. and {Soubiran}, C. and {van Leeuwen}, F. and {Walton}, N.~A. and {Arenou}, F. and {Bastian}, U. and {Cropper}, M. and {Drimmel}, R. and {Katz}, D. and {Lattanzi}, M.~G. and {Bakker}, J. and {Cacciari}, C. and {Casta{\~n}eda}, J. and {Chaoul}, L. and {Cheek}, N. and {De Angeli}, F. and {Fabricius}, C. and {Guerra}, R. and {Holl}, B. and {Masana}, E. and {Messineo}, R. and {Mowlavi}, N. and {Nienartowicz}, K. and {Panuzzo}, P. and {Portell}, J. and {Riello}, M. and {Seabroke}, G.~M. and {Tanga}, P. and {Th{\'e}venin}, F. and {Gracia-Abril}, G. and {Comoretto}, G. and {Garcia-Reinaldos}, M. and {Teyssier}, D. and {Altmann}, M. and {Andrae}, R. and {Audard}, M. and {Bellas-Velidis}, I. and {Benson}, K. and {Berthier}, J. and {Blomme}, R. and {Burgess}, P. and {Busso}, G. and {Carry}, B. and {Cellino}, A. and {Clementini}, G. and {Clotet}, M. and {Creevey}, O. and {Davidson}, M. and {De Ridder}, J. and {Delchambre}, L. and {Dell'Oro}, A. and {Ducourant}, C. and {Fern{\'a}ndez-Hern{\'a}ndez}, J. and {Fouesneau}, M. and {Fr{\'e}mat}, Y. and {Galluccio}, L. and {Garc{\'\i}a-Torres}, M. and {Gonz{\'a}lez-N{\'u}{\~n}ez}, J. and {Gonz{\'a}lez-Vidal}, J.~J. and {Gosset}, E. and {Guy}, L.~P. and {Halbwachs}, J. -L. and {Hambly}, N.~C. and {Harrison}, D.~L. and {Hern{\'a}ndez}, J. and {Hestroffer}, D. and {Hodgkin}, S.~T. and {Hutton}, A. and {Jasniewicz}, G. and {Jean-Antoine-Piccolo}, A. and {Jordan}, S. and {Korn}, A.~J. and {Krone-Martins}, A. and {Lanzafame}, A.~C. and {Lebzelter}, T. and {L{\"o}ffler}, W. and {Manteiga}, M. and {Marrese}, P.~M. and {Mart{\'\i}n-Fleitas}, J.~M. and {Moitinho}, A. and {Mora}, A. and {Muinonen}, K. and {Osinde}, J. and {Pancino}, E. and {Pauwels}, T. and {Petit}, J. -M. and {Recio-Blanco}, A. and {Richards}, P.~J. and {Rimoldini}, L. and {Robin}, A.~C. and {Sarro}, L.~M. and {Siopis}, C. and {Smith}, M. and {Sozzetti}, A. and {S{\"u}veges}, M. and {Torra}, J. and {van Reeven}, W. and {Abbas}, U. and {Abreu Aramburu}, A. and {Accart}, S. and {Aerts}, C. and {Altavilla}, G. and {{\'A}lvarez}, M.~A. and {Alvarez}, R. and {Alves}, J. and {Anderson}, R.~I. and {Andrei}, A.~H. and {Anglada Varela}, E. and {Antiche}, E. and {Antoja}, T. and {Arcay}, B. and {Astraatmadja}, T.~L. and {Bach}, N. and {Baker}, S.~G. and {Balaguer-N{\'u}{\~n}ez}, L. and {Balm}, P. and {Barache}, C. and {Barata}, C. and {Barbato}, D. and {Barblan}, F. and {Barklem}, P.~S. and {Barrado}, D. and {Barros}, M. and {Barstow}, M.~A. and {Bartholom{\'e} Mu{\~n}oz}, S. and {Bassilana}, J. -L. and {Becciani}, U. and {Bellazzini}, M. and {Berihuete}, A. and {Bertone}, S. and {Bianchi}, L. and {Bienaym{\'e}}, O. and {Blanco-Cuaresma}, S. and {Boch}, T. and {Boeche}, C. and {Bombrun}, A. and {Borrachero}, R. and {Bossini}, D. and {Bouquillon}, S. and {Bourda}, G. and {Bragaglia}, A. and {Bramante}, L. and {Breddels}, M.~A. and {Bressan}, A. and {Brouillet}, N. and {Br{\"u}semeister}, T. and {Brugaletta}, E. and {Bucciarelli}, B. and {Burlacu}, A. and {Busonero}, D. and {Butkevich}, A.~G. and {Buzzi}, R. and {Caffau}, E. and {Cancelliere}, R. and {Cannizzaro}, G. and {Cantat-Gaudin}, T. and {Carballo}, R. and {Carlucci}, T. and {Carrasco}, J.~M. and {Casamiquela}, L. and {Castellani}, M. and {Castro-Ginard}, A. and {Charlot}, P. and {Chemin}, L. and {Chiavassa}, A. and {Cocozza}, G. and {Costigan}, G. and {Cowell}, S. and {Crifo}, F. and {Crosta}, M. and {Crowley}, C. and {Cuypers}, J. and {Dafonte}, C. and {Damerdji}, Y. and {Dapergolas}, A. and {David}, P. and {David}, M. and {de Laverny}, P. and {De Luise}, F.},
	title = "{Gaia Data Release 2. Summary of the contents and survey properties}",
	journal = {\aap},
	year = 2018,
	month = aug,
	volume = {616},
	eid = {A1},
	pages = {A1},
	doi = {10.1051/0004-6361/201833051},
	archivePrefix = {arXiv},
	eprint = {1804.09365},
	primaryClass = {astro-ph.GA},
	adsurl = {https://ui.adsabs.harvard.edu/abs/2018A&A...616A...1G}
}

@ARTICLE{1983MNRAS.202.1025G,
	author = {{Gilmore}, G. and {Reid}, N.},
	title = "{New light on faint stars - III. Galactic structure towards the South Pole and the Galactic thick disc.}",
	journal = {\mnras},
	year = 1983,
	month = mar,
	volume = {202},
	pages = {1025-1047},
	doi = {10.1093/mnras/202.4.1025},
	adsurl = {https://ui.adsabs.harvard.edu/abs/1983MNRAS.202.1025G}
}

@ARTICLE{1982PASJ...34..365Y,
	author = {{Yoshii}, Y.},
	title = "{Density distribution of faint stars in the direction of the north galactic pole.}",
	journal = {\pasj},
	year = 1982,
	month = jan,
	volume = {34},
	pages = {365-379},
	adsurl = {https://ui.adsabs.harvard.edu/abs/1982PASJ...34..365Y}
}

@ARTICLE{2008ApJ...673..864J,
	author = {{Juri{\'c}}, Mario and {Ivezi{\'c}}, {\v{Z}}eljko and {Brooks}, Alyson and {Lupton}, Robert H. and {Schlegel}, David and {Finkbeiner}, Douglas and {Padmanabhan}, Nikhil and {Bond}, Nicholas and {Sesar}, Branimir and {Rockosi}, Constance M. and {Knapp}, Gillian R. and {Gunn}, James E. and {Sumi}, Takahiro and {Schneider}, Donald P. and {Barentine}, J.~C. and {Brewington}, Howard J. and {Brinkmann}, J. and {Fukugita}, Masataka and {Harvanek}, Michael and {Kleinman}, S.~J. and {Krzesinski}, Jurek and {Long}, Dan and {Neilsen}, Jr., Eric H. and {Nitta}, Atsuko and {Snedden}, Stephanie A. and {York}, Donald G.},
	title = "{The Milky Way Tomography with SDSS. I. Stellar Number Density Distribution}",
	journal = {\apj},
	year = 2008,
	month = feb,
	volume = {673},
	number = {2},
	pages = {864-914},
	doi = {10.1086/523619},
	archivePrefix = {arXiv},
	eprint = {astro-ph/0510520},
	primaryClass = {astro-ph},
	adsurl = {https://ui.adsabs.harvard.edu/abs/2008ApJ...673..864J}
}

@ARTICLE{2001ApJ...553..184C,
	author = {{Chen}, Bing and {Stoughton}, Chris and {Smith}, J. Allyn and {Uomoto}, Alan and {Pier}, Jeffrey R. and {Yanny}, Brian and {Ivezi{\'c}}, {\v{Z}}eljko and {York}, Donald G. and {Anderson}, John E. and {Annis}, James and {Brinkmann}, Jon and {Csabai}, Istv{\'a}n and {Fukugita}, Masataka and {Hindsley}, Robert and {Lupton}, Robert and {Munn}, Jeffrey A. and {SDSS Collaboration}},
	title = "{Stellar Population Studies with the SDSS. I. The Vertical Distribution of Stars in the Milky Way}",
	journal = {\apj},
	year = 2001,
	month = may,
	volume = {553},
	number = {1},
	pages = {184-197},
	doi = {10.1086/320647},
	adsurl = {https://ui.adsabs.harvard.edu/abs/2001ApJ...553..184C}
}

@ARTICLE{2016ARA&A..54..529B,
	author = {{Bland-Hawthorn}, Joss and {Gerhard}, Ortwin},
	title = "{The Galaxy in Context: Structural, Kinematic, and Integrated Properties}",
	journal = {\araa},
	year = 2016,
	month = sep,
	volume = {54},
	pages = {529-596},
	doi = {10.1146/annurev-astro-081915-023441},
	archivePrefix = {arXiv},
	eprint = {1602.07702},
	primaryClass = {astro-ph.GA},
	adsurl = {https://ui.adsabs.harvard.edu/abs/2016ARA&A..54..529B}
}

@ARTICLE{2018ApJS..237...33X,
	author = {{Xiang}, Maosheng and {Shi}, Jianrong and {Liu}, Xiaowei and {Yuan}, Haibo and {Chen}, Bingqiu and {Huang}, Yang and {Wang}, Chun and {Wu}, Yaqian and {Tian}, Zhijia and {Huo}, Zhiying and {Zhang}, Huawei and {Zhang}, Meng},
	title = "{Stellar Mass Distribution and Star Formation History of the Galactic Disk Revealed by Mono-age Stellar Populations from LAMOST}",
	journal = {\apjs},
	year = 2018,
	month = aug,
	volume = {237},
	number = {2},
	eid = {33},
	pages = {33},
	doi = {10.3847/1538-4365/aad237},
	archivePrefix = {arXiv},
	eprint = {1807.04592},
	primaryClass = {astro-ph.GA},
	adsurl = {https://ui.adsabs.harvard.edu/abs/2018ApJS..237...33X}
}

@ARTICLE{2019ApJ...871..208L,
	author = {{Li}, Chengdong and {Zhao}, Gang and {Jia}, Yunpeng and {Liao}, Shilong and {Yang}, Chengqun and {Wang}, Qixun},
	title = "{Flare and Warp of Galactic Disk with OB Stars from Gaia DR2}",
	journal = {\apj},
	year = 2019,
	month = feb,
	volume = {871},
	number = {2},
	eid = {208},
	pages = {208},
	doi = {10.3847/1538-4357/aafa17},
	adsurl = {https://ui.adsabs.harvard.edu/abs/2019ApJ...871..208L}
}

@ARTICLE{2014Natur.509..342F,
	author = {{Feast}, Michael W. and {Menzies}, John W. and {Matsunaga}, Noriyuki and {Whitelock}, Patricia A.},
	title = "{Cepheid variables in the flared outer disk of our galaxy}",
	journal = {\nat},
	year = 2014,
	month = may,
	volume = {509},
	number = {7500},
	pages = {342-344},
	doi = {10.1038/nature13246},
	archivePrefix = {arXiv},
	eprint = {1406.7660},
	primaryClass = {astro-ph.GA},
	adsurl = {https://ui.adsabs.harvard.edu/abs/2014Natur.509..342F}
}

@ARTICLE{1957AJ.....62...90B,
	author = {{Burke}, Bernard F.},
	title = "{Systematic distortion of the outer regions of the galaxy.}",
	journal = {\aj},
	year = 1957,
	month = may,
	volume = {62},
	pages = {90},
	doi = {10.1086/107463},
	adsurl = {https://ui.adsabs.harvard.edu/abs/1957AJ.....62...90B}
}

@ARTICLE{1958MNRAS.118..379O,
	author = {{Oort}, J.~H. and {Kerr}, F.~J. and {Westerhout}, G.},
	title = "{The galactic system as a spiral nebula (Council Note)}",
	journal = {\mnras},
	year = 1958,
	month = jan,
	volume = {118},
	pages = {379},
	doi = {10.1093/mnras/118.4.379},
	adsurl = {https://ui.adsabs.harvard.edu/abs/1958MNRAS.118..379O}
}

@ARTICLE{1993AJ....105.2127C,
	author = {{Carney}, Bruce W. and {Seitzer}, Patrick},
	title = "{Optical detection of the Galaxy's Southern Stellar Warp and Outer Disk}",
	journal = {\aj},
	year = 1993,
	month = jun,
	volume = {105},
	pages = {2127},
	doi = {10.1086/116591},
	adsurl = {https://ui.adsabs.harvard.edu/abs/1993AJ....105.2127C}
}

@ARTICLE{2002A&A...386..169L,
	author = {{L{\'o}pez-Corredoira}, M. and {Betancort-Rijo}, J. and {Beckman}, J.~E.},
	title = "{Generation of galactic disc warps due to intergalactic accretion flows onto the disc}",
	journal = {\aap},
	year = 2002,
	month = apr,
	volume = {386},
	pages = {169-186},
	doi = {10.1051/0004-6361:20020229},
	archivePrefix = {arXiv},
	eprint = {astro-ph/0202156},
	primaryClass = {astro-ph},
	adsurl = {https://ui.adsabs.harvard.edu/abs/2002A&A...386..169L}
}

@ARTICLE{2002A&A...394..883L,
	author = {{L{\'o}pez-Corredoira}, M. and {Cabrera-Lavers}, A. and {Garz{\'o}n}, F. and {Hammersley}, P.~L.},
	title = "{Old stellar Galactic disc in near-plane regions according to 2MASS: Scales, cut-off, flare and warp}",
	journal = {\aap},
	year = 2002,
	month = nov,
	volume = {394},
	pages = {883-899},
	doi = {10.1051/0004-6361:20021175},
	archivePrefix = {arXiv},
	eprint = {astro-ph/0208236},
	primaryClass = {astro-ph},
	adsurl = {https://ui.adsabs.harvard.edu/abs/2002A&A...394..883L}
}

@ARTICLE{2009A&A...495..819R,
	author = {{Reyl{\'e}}, C. and {Marshall}, D.~J. and {Robin}, A.~C. and {Schultheis}, M.},
	title = "{The Milky Way's external disc constrained by 2MASS star counts}",
	journal = {\aap},
	year = 2009,
	month = mar,
	volume = {495},
	number = {3},
	pages = {819-826},
	doi = {10.1051/0004-6361/200811341},
	archivePrefix = {arXiv},
	eprint = {0812.3739},
	primaryClass = {astro-ph},
	adsurl = {https://ui.adsabs.harvard.edu/abs/2009A&A...495..819R}
}

@ARTICLE{2017A&A...602A..67A,
	author = {{Am{\^o}res}, E.~B. and {Robin}, A.~C. and {Reyl{\'e}}, C.},
	title = "{Evolution over time of the Milky Way's disc shape}",
	journal = {\aap},
	year = 2017,
	month = jun,
	volume = {602},
	eid = {A67},
	pages = {A67},
	doi = {10.1051/0004-6361/201628461},
	archivePrefix = {arXiv},
	eprint = {1701.00475},
	primaryClass = {astro-ph.GA},
	adsurl = {https://ui.adsabs.harvard.edu/abs/2017A&A...602A..67A}
}

@ARTICLE{2019NatAs...3..320C,
	author = {{Chen}, Xiaodian and {Wang}, Shu and {Deng}, Licai and {de Grijs}, Richard and {Liu}, Chao and {Tian}, Hao},
	title = "{An intuitive 3D map of the Galactic warp's precession traced by classical Cepheids}",
	journal = {Nature Astronomy},
	year = 2019,
	month = feb,
	volume = {3},
	pages = {320-325},
	doi = {10.1038/s41550-018-0686-7},
	archivePrefix = {arXiv},
	eprint = {1902.00998},
	primaryClass = {astro-ph.GA},
	adsurl = {https://ui.adsabs.harvard.edu/abs/2019NatAs...3..320C}
}

@ARTICLE{1998AJ....115.2384D,
	author = {{Dehnen}, Walter},
	title = "{The Distribution of Nearby Stars in Velocity Space Inferred from HIPPARCOS Data}",
	journal = {\aj},
	year = 1998,
	month = jun,
	volume = {115},
	number = {6},
	pages = {2384-2396},
	doi = {10.1086/300364},
	archivePrefix = {arXiv},
	eprint = {astro-ph/9803110},
	primaryClass = {astro-ph},
	adsurl = {https://ui.adsabs.harvard.edu/abs/1998AJ....115.2384D}
}

@ARTICLE{2018MNRAS.478.3809S,
	author = {{Sch{\"o}nrich}, Ralph and {Dehnen}, Walter},
	title = "{Warp, waves, and wrinkles in the Milky Way}",
	journal = {\mnras},
	year = 2018,
	month = aug,
	volume = {478},
	number = {3},
	pages = {3809-3824},
	doi = {10.1093/mnras/sty1256},
	archivePrefix = {arXiv},
	eprint = {1712.06616},
	primaryClass = {astro-ph.GA},
	adsurl = {https://ui.adsabs.harvard.edu/abs/2018MNRAS.478.3809S}
}

@ARTICLE{2024MNRAS.535.1898B,
	author = {{Binney}, James},
	title = "{Disc distortion revisited}",
	journal = {\mnras},
	year = 2024,
	month = dec,
	volume = {535},
	number = {2},
	pages = {1898-1912},
	doi = {10.1093/mnras/stae2481},
	archivePrefix = {arXiv},
	eprint = {2411.04879},
	primaryClass = {astro-ph.GA},
	adsurl = {https://ui.adsabs.harvard.edu/abs/2024MNRAS.535.1898B}
}

@ARTICLE{1989MNRAS.237..785O,
	author = {{Ostriker}, E.~C. and {Binney}, J.~J.},
	title = "{Warped and tilted galactic discs}",
	journal = {\mnras},
	year = 1989,
	month = apr,
	volume = {237},
	pages = {785-798},
	doi = {10.1093/mnras/237.3.785},
	adsurl = {https://ui.adsabs.harvard.edu/abs/1989MNRAS.237..785O}
}

@ARTICLE{2023ApJ...957L..24H,
	author = {{Han}, Jiwon Jesse and {Semenov}, Vadim and {Conroy}, Charlie and {Hernquist}, Lars},
	title = "{Tilted Dark Halos Are Common and Long-lived, and Can Warp Galactic Disks}",
	journal = {\apjl},
	year = 2023,
	month = nov,
	volume = {957},
	number = {2},
	eid = {L24},
	pages = {L24},
	doi = {10.3847/2041-8213/ad0641},
	archivePrefix = {arXiv},
	eprint = {2309.07208},
	primaryClass = {astro-ph.GA},
	adsurl = {https://ui.adsabs.harvard.edu/abs/2023ApJ...957L..24H}
}

@ARTICLE{2023NatAs...7.1481H,
	author = {{Han}, Jiwon Jesse and {Conroy}, Charlie and {Hernquist}, Lars},
	title = "{A tilted dark halo origin of the Galactic disk warp and flare}",
	journal = {Nature Astronomy},
	year = 2023,
	month = dec,
	volume = {7},
	pages = {1481-1485},
	doi = {10.1038/s41550-023-02076-9},
	archivePrefix = {arXiv},
	eprint = {2309.07209},
	primaryClass = {astro-ph.GA},
	adsurl = {https://ui.adsabs.harvard.edu/abs/2023NatAs...7.1481H}
}

@ARTICLE{2024ApJ...975...28D,
	author = {{Deng}, Mingji and {Du}, Cuihua and {Yang}, Yanbin and {Liao}, Jiwei and {Ye}, Dashuang},
	title = "{A Potential Dynamical Origin of the Galactic Disk Warp: The Gaia{\textendash}Sausage{\textendash}Enceladus Major Merger}",
	journal = {\apj},
	year = 2024,
	month = nov,
	volume = {975},
	number = {1},
	eid = {28},
	pages = {28},
	doi = {10.3847/1538-4357/ad7799},
	archivePrefix = {arXiv},
	eprint = {2409.03264},
	primaryClass = {astro-ph.GA},
	adsurl = {https://ui.adsabs.harvard.edu/abs/2024ApJ...975...28D}
}

@ARTICLE{2024arXiv241022250K,
	author = {{Kurbatov}, Evgeny P. and {Belokurov}, Vasily and {Koposov}, Sergey and {Kravtsov}, Andrey and {Davies}, Elliot Y. and {Brown}, Anthony G.~A. and {Cantat-Gaudin}, Tristan and {Castro-Ginard}, Alfred and {Casey}, Andrew R. and {Drimmel}, Ronald and {Fouesneau}, Morgan and {Khanna}, Shourya and {Rix}, Hans-Walter and {Wallace}, Alex},
	title = "{The realm of Aurora. Density distribution of metal-poor giants in the heart of the Galaxy}",
	journal = {arXiv e-prints},
	year = 2024,
	month = oct,
	eid = {arXiv:2410.22250},
	pages = {arXiv:2410.22250},
	doi = {10.48550/arXiv.2410.22250},
	archivePrefix = {arXiv},
	eprint = {2410.22250},
	primaryClass = {astro-ph.GA},
	adsurl = {https://ui.adsabs.harvard.edu/abs/2024arXiv241022250K}
}

@ARTICLE{2024MNRAS.533.2997W,
	author = {{Wille}, A. and {Machado}, R.~E.~G.},
	title = "{Warps induced by satellites on barred and non-barred galaxies}",
	journal = {\mnras},
	year = 2024,
	month = sep,
	volume = {533},
	number = {3},
	pages = {2997-3007},
	doi = {10.1093/mnras/stae2004},
	archivePrefix = {arXiv},
	eprint = {2408.09932},
	primaryClass = {astro-ph.GA},
	adsurl = {https://ui.adsabs.harvard.edu/abs/2024MNRAS.533.2997W}
}

@ARTICLE{2025arXiv250114089C,
	author = {{Chen}, Boquan and {Orkney}, Matthew and {Ting}, Yuan-Sen and {Hayden}, Michael},
	title = "{Discovery of A Starburst in the Early Milky Way at [Fe/H] $< -2$}",
	journal = {arXiv e-prints},
	year = 2025,
	month = jan,
	eid = {arXiv:2501.14089},
	pages = {arXiv:2501.14089},
	doi = {10.48550/arXiv.2501.14089},
	archivePrefix = {arXiv},
	eprint = {2501.14089},
	primaryClass = {astro-ph.GA},
	adsurl = {https://ui.adsabs.harvard.edu/abs/2025arXiv250114089C}
}
\bibliographystyle{aasjournalv7}

\end{document}